\documentclass[aps,twocolumn,prd,superscriptaddress,nofootinbib,floatfix]{revtex4-2}

\usepackage{mathtools}
\usepackage{amsfonts}
\usepackage{amssymb}
\usepackage{mathrsfs}
\usepackage{bbm}
\usepackage{slashed}
\usepackage{amsmath}
\usepackage{bm}
\usepackage{bbold}

\usepackage{graphicx}
\usepackage{color}
\usepackage{colortbl}
\usepackage{array}

\usepackage{float}
\usepackage{placeins}
\usepackage{booktabs}
\usepackage[caption=false]{subfig}
\usepackage{makecell}
\usepackage{tabstackengine}

\usepackage{xspace}
\usepackage{hyperref}
\usepackage[nameinlink]{cleveref}
\usepackage{bookmark}
\usepackage{siunitx}

\usepackage{xifthen}
\usepackage{xcolor}
\hypersetup{
	colorlinks,
	linkcolor={red!75!black},
	citecolor={blue!75!black},
	urlcolor={blue!75!black}
}

\usepackage[utf8]{inputenc}
\allowdisplaybreaks[4]

\newcommand{\tr}{\text{tr}}
\newcommand{\Tr}{{\text{Tr}}}

\newcommand{\feyn}[1]{
	\setbox0=\hbox{\ensuremath{#1}}
	\hbox to\wd0{\hbox to0pt{\hbox to\wd0{\hss/\hss}\hss}\box0}}

\def\Eq#1{Eq.~\labelcref{#1}}

\def\Fig#1{Fig.~\labelcref{#1}}
\def\Tab#1{Tab.~\labelcref{#1}}
\def\sec#1{Sec.~\labelcref{#1}}
\def\app#1{App.~\labelcref{#1}}

\setkeys{Gin}{width=0.48\textwidth}
\graphicspath{{./figures/}}

\newcommand{\lVA}{\bar \lambda_{{V-A}}}
\newcommand{\lVpA}{\bar \lambda_{{V+A}}}
\newcommand{\lVAadj}{\bar \lambda_{({V-A})^{\text{adj}}}}
\newcommand{\leta}{\bar \lambda_{\eta}}
\newcommand{\la}{\bar \lambda_{{a}}}
\newcommand{\lSm}{\bar \lambda_{({S-P})_{-}^{\text{adj}}}}
\newcommand{\lSp}{\bar \lambda_{({S+P})_{-}^{\text{adj}}}}
\newcommand{\lSpp}{\bar \lambda_{({S+P})_{+}^{\text{adj}}}}

\newcommand{\gettitle}{}

\newcommand{\getDalianAffiliation}{\affiliation{School of Physics, Dalian University of Technology, Dalian, 116024, P.R. China}}

\newcommand{\getGiessenAffiliation}{\affiliation{Institut f\"ur Theoretische Physik, Justus-Liebig-Universit\"at Giessen, 35392 Giessen, Germany}}

\newcommand{\getHFHFAffiliation}{\affiliation{Helmholtz Research Academy Hesse for FAIR (HFHF), Campus Giessen, Giessen, Germany}}

\newcommand{\getHeidelbergAffiliation}{\affiliation{Institut f\"ur Theoretische Physik, Universit\"at Heidelberg, Philosophenweg 16, 69120 Heidelberg, Germany}}

\newcommand{\getYanTaiAffiliation}{\affiliation{Department of Physics, Yantai University, YanTai, 264005,  P.R. China}}

\hypersetup{
	colorlinks,
	linkcolor={red!75!black},
	citecolor={blue!75!black},
	urlcolor={blue!75!black},
	pdftitle={\gettitle},
	pdfauthor={Zelle},
	pdfkeywords={effective theory} {analytic continuation}
	{correlations functions} {hadronization}
	{functional renormalisation group}
	{real time} {bound states}
	bookmarksopen=true,
	bookmarksopenlevel=2,
	bookmarksnumbered=true
}

\begin{document}

\title{Fierz-complete four-quark interactions and the QCD phase diagram}

\author{Zi-ning Wang}
\getDalianAffiliation

\author{Li-jun Zhou}
\getDalianAffiliation

\author{Chuang Huang}
\getHeidelbergAffiliation

\author{Rui Wen}
\getYanTaiAffiliation

\author{Shi Yin}
\getGiessenAffiliation
\getHFHFAffiliation

\author{Wei-jie Fu}
\email{wjfu@dlut.edu.cn}
\getDalianAffiliation

\begin{abstract} 

The dynamics of Fierz-complete four-quark interactions and its influence on the QCD phase diagram have been investigated within the functional renormalization group approach to QCD at finite temperature and densities. It is found that in the vacuum the pion and sigma channels play the overwhelmingly dominant role, and all the other channels are negligible. However, when it is near the critical end point (CEP), the magnitude of four-quark couplings in other channels increases sizably and they become more and more important. In comparison to the single scalar-pseudoscalar channel of four-quark interactions, the dynamics of Fierz-complete four-quark interactions increases a bit the curvature of the phase boundary, and moves the CEP to location of larger baryon chemical potential and smaller temperature. 

\end{abstract}
	
\maketitle

\section{Introduction}
\label{sec:int} 

QCD phase diagram at finite temperature and densities provides us with an exceptional window to study the properties of strong interactions, which plays a pivotal role not only in the studies of strongly interacting matter, e.g., the quark-gluon plasma (QGP), produced in relativistic heavy-ion collisions, but also in understanding the evolution of early universe and properties of compact stars, such as the neutron star \cite{Stephanov:2007fk, Luo:2017faz, Bzdak:2019pkr, Fu:2022gou, Fukushima:2025ujk, Fischer:2026uni}. One of the most prominent characteristics of QCD phase diagram in the plane of temperature and baryon chemical potential is the critical end point (CEP), which connects the continuous chiral crossover in the regime of low baryon density to the first order phase transition in the regime of high density.

However, the existence of CEP, and if it indeed exists, the location of CEP, are still open questions. Our knowledge about the QCD phase diagram, especially at high baryon chemical potential, is quite limited. This is partly attributed to the fact that direct simulations of QCD on lattice at finite baryon chemical potential are hindered by the sign problem \cite{Bazavov:2018mes, Borsanyi:2020fev}. Even though, recent years have already seen significant progress in the studies of CEP from both theoretical and experimental endeavors. An obvious deviation from non-critical-point model calculations is found in the kurtosis of net-proton number fluctuations at the Relativistic Heavy Ion Collider (RHIC) by the STAR collaboration \cite{STAR:2025zdq}, see also \cite{STAR:2020tga, STAR:2021fge, STAR:2022vlo,STAR:2022etb}. Furthermore, a non-monotonic dependence on the collision energy in the variance of the mean transverse momentum fluctuations of charged particles is also found by the STAR collaboration \cite{STAR:2026vjv}. Both the fluctuation measurements in the net-proton number and the mean transverse momentum could be potentially related to the critical fluctuations of CEP, see, e.g., \cite{Stephanov:2008qz, Stephanov:2011pb, Fu:2016tey, Fu:2021oaw, Fu:2023lcm, Lu:2026ezr} and \cite{Chen:2025vwl} for the former and latter, respectively.

Theoretical estimate of the location CEP in the phase diagram from the first-principles QCD at finite temperature and densities started from the QCD calculations within the functional renormalization group (fRG) approach \cite{Fu:2019hdw}, where it was found that the CEP is located in the region of relatively large baryon chemical potential, with the baryon chemical potential of CEP, $\mu_{B_{\mathrm{CEP}}} \gtrsim 600$ MeV. This was subsequently corroborated by two independent QCD calculations within the Dyson-Schwinger equation (DSE) \cite{Gao:2020fbl, Gunkel:2021oya}. Since then, more and more calculations from different approaches, including lattice QCD extrapolation based on Yang-Lee edge singularities \cite{Basar:2023nkp, Clarke:2024ugt, Adam:2025phc} or contours of constant entropy density \cite{Shah:2024img, Borsanyi:2025dyp}, functional QCD from fRG \cite{Braun:2023qak, Fu:2024rto, Pawlowski:2025jpg, Fu:2026qnl} and DSE \cite{Lu:2025cls, Lu:2026ezr}, Bayesian holography \cite{Cai:2022omk, Hippert:2023bel, Zhu:2025gxo}, etc., have indicated that the critical end point is located in the region of $\mu_B/T \gtrsim 4 \sim 5$, or no CEP is found. 

In this work, we would like to improve on the computation of fRG approach to QCD at finite temperature and densities in \cite{Fu:2019hdw}, by extending the single scalar-pseudoscalar channel of four-quark interactions to the Fierz-complete four-quark basis, which allows us not only to remove the ambiguity arising from the projection of different tensor structures for the four-quark vertices on the level of technology, but more important, to investigate the influence of four-quark vertices of different tensor structures on the calculated results of phase boundary and the location of CEP. Although the scalar-pseudoscalar channel dominates over all other channels and plays the overwhelming role in the vacuum, other channels of four-quark interactions, e.g., the vector or the diquark channel, become more and more relevant with the increase of the baryon chemical potential \cite{Braun:2017srn, Braun:2018bik, Braun:2019aow}. Therefore, studies in this work would contribute to the analysis of systematics for the fRG approach to the first-principles QCD at finite temperature and densities. Moreover, it is found the four-quark interactions embedded in QCD within the fRG approach are very suited for the studies of dynamical chiral symmetry breaking \cite{Fu:2022uow, Fu:2024ysj, Fu:2025hcm} and parton distribution functions of hadrons \cite{Zhang:2025ofc, Cui:2026bod}.

This paper is organized as follows: In \sec{sec:FRG-QCD}, the fRG approach to QCD is recapitulated, followed by the discussions of correlation function in \sec{sec:cor-fun}, including the four-quark, quark-meson, and quark-gluon functions. Results of QCD phase transition and phase diagram are presented in \sec{sec:phase-diagram}. We conclude and summarize in \sec{sec:conclusion}. The details about the calculations and some further results are presented in Appendices, including QCD at finite temperature and densities within the fRG approach in \app{app:action-QCD}, four-quark interactions of Fierz-complete tensors for the light quarks in \app{app:action-4quark}, flow equation of the effective action with dynamical hadronization in \app{app:flow-action}, numerical setup and parameters in \app{app:parameters}, some further numerical results in \app{app:further-nums}, flow equations of the Yukawa couplings in \app{app:Yukawa-flows}, flow equations of the four-quark interactions in \app{app:flows-4quark}.

\section{Functional renormalization group approach to QCD}
\label{sec:FRG-QCD}

%
\begin{figure}[t]
\includegraphics[width=0.45\textwidth]{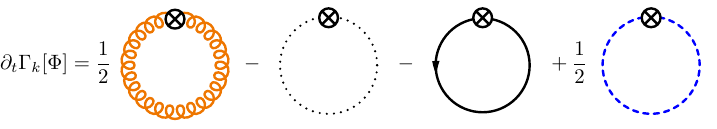}
\caption{Diagrammatic representation of the flow equation of the QCD effective action, where the four loops represent the contributions from the gluon, ghost, quark, and meson, respectively. Here $t=\ln(k/\Lambda)$ denotes the RG time with some ultraviolet cutoff $\Lambda$. The crossed circles denote the infrared regulators.}
\label{fig:QCD_equation}
\end{figure}
%

The fRG approach to QCD at finite temperature and densities has been developed in \cite{Fu:2019hdw}. In this approach, quantum fluctuations of QCD as well as the related thermodynamic fluctuations are included successively as the renormalization group (RG) scale is evolved from the ultraviolet perturbative to the infrared nonperturbative regimes, by integrating the flow equation of effective action as shown in \Fig{fig:QCD_equation}, where the different lines denote the nonperturbative propagators of gluons, ghosts, quarks, and mesons, respectively. The crossed circles stand for the infrared regulators, which suppress fluctuations of momentum modes below the infrared cutoff or the RG scale $k$. As the RG scale $k$ decreases, and when it is in the regime of low energy, say $k \lesssim 0.5 \sim1$ GeV, the gluon decouples from the system due to the emergence of a finite gluon mass gap related to the confinement of QCD \cite{vonSmekal:1997ohs, vonSmekal:1997ern, Fischer:2008uz, Cyrol:2016tym, Fu:2025hcm}; furthermore, the quark is also gapped since the chiral symmetry is broken dynamically in the low energy QCD, and thus a dynamical quark mass is produced \cite{Mitter:2014wpa, Braun:2014ata, Cyrol:2017ewj, Corell:2018yil, Fu:2019hdw, Fu:2022uow, Fu:2024ysj}. Therefore, neither the gluon or the quark are the relevant degrees of freedom in the phenomena of low energy QCD, e.g., bound states, QCD phase transitions, etc. On the contrary, the relevant degrees of freedom of low energy QCD are the composite, collective degrees of freedom in QCD, e.g., the pion meson. The pion, as the Goldstone boson of the chiral symmetry breaking, is the lightest hadron and its dynamics plays an important role in the low energy QCD. For example, the pion accounts for the dynamics of soft modes in a large region on the QCD phase diagram near the critical end point \cite{Braun:2023qak}. Moreover, the scalar sigma mode also plays an important role in the QCD phase transitions since it is the critical mode of the CEP and thus is massless at CEP \cite{Tan:2025bsv}.

The infrared dynamics of four-quark resonant scatterings is found to be closely related to the dynamical chiral symmetry breaking and the emergence of mesonic bound states \cite{Fu:2022uow, Fu:2024ysj, Fu:2025hcm}. In the coupled flow equations for the quark mass gap and the four-quark vertices, the significantly enhanced four-quark couplings in the regime of low energy would results in the increase of the dynamical quark mass, which in turn slows down the resonant enhancement of the four-quark couplings, and then a balance is obtained \cite{Fu:2022uow}. In the vicinity of a resonance, the four-quark vertex is dominated by the exchange of a bound state and the resonant pole just corresponds to the pole mass of the bound state, where all the properties of bound state, e.g., the Bethe-Salpeter amplitude, are encoded completely in the four-quark vertices, see \cite{Fu:2024ysj, Fu:2025hcm} for more details.

The QCD effective action used in this work is presented in \Eq{eq:QCDaction} in \app{app:action-QCD}. In comparison to the previous paper \cite{Fu:2019hdw}, where only the scalar-pseudoscalar channel for the four-quark scatterings is taken into account, here we have extended the truncation to include the Fierz-complete four-quark interactions for the $u$ and $d$ light quarks, as that done in \cite{Fu:2022uow, Fu:2024ysj} for the physics of bound states within the fRG approach. The four-quark sector of the QCD effective action reads
\begin{align} 
    \Gamma_{4q} =&-\int_{x}\sum_{\alpha \in \mathcal{F}} \lambda_{\alpha}\,{\mathcal{T}}^{(\alpha)}_{ijmn}\,\bar {q_l}_i {q_l}_j \bar {q_l}_m {q_l}_n\,, \label{eq:Gamma4q}
\end{align}
with $\int_{x}=\int_0^{1/T}d x_0 \int d^3 x$ and the temperature $T$, where $q_l$ represents the $u$ and $d$ light quarks, $q_l=(q_u, q_d)$, and a summation is assumed for the repeated quark indices $i,\,j,\,m,\,n$. The $\lambda_{\alpha}$ denotes the four-quark coupling strength for different channels. The Fierz-complete four-quark basis set of light quarks, here denoted by $\mathcal{F}$, includes ten tensors of different channels, i.e.,
\begin{align}
    \mathcal{F} = \Big\{&\sigma,\, \,\pi, \,\, a, \,\, \eta, \,\, (V\pm A)\,,\, (V-A)^{\mathrm{adj}}\,,\nonumber\\[2ex] &\,(S\pm P)^{\mathrm{adj}}_-,\,\, (S+ P)^{\mathrm{adj}}_+\,\Big\} \,.\label{eq:set-alpha10tensors}
\end{align}
Their explicit expressions are shown in \app{app:action-4quark}, see, e.g., \cite{Fu:2022uow, Fu:2024ysj, Fu:2025hcm} for more details. 

%
\begin{figure*}[t]
\includegraphics[width=1.0\textwidth]{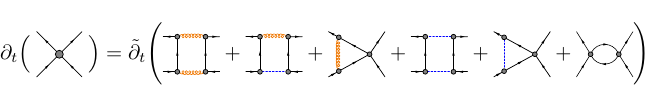}
\caption{Diagrammatic representation of the flow equation of the four-quark vertices, where $\tilde \partial_t$ only hits the RG-scale dependence through the regulator in propagators, whose implementation would lead to the insertion of a regulator for each inner line of diagrams on the right side of the equation. Here only one representative diagram out of one class of diagrams with the same vertices is shown and prefactors of each diagrams are not shown for the sake of brevity.}
\label{fig:4quark-eqn}
\end{figure*}
%

The scalar and pseudoscalar channels are directly related to the chiral symmetry breaking, since the chiral condensate $\langle \bar q_i q_i \rangle$ ($i=u,\,d,\,s$) with $\langle \cdots \rangle$ denoting the expectation value serves as the order parameter for the chiral symmetry breaking, and the pseudoscalar channel corresponds to the (pseudo-)Goldstone bosons of the chiral symmetry breaking. These two channels play a dominant role in the region of QCD phase diagram, where the baryon  chemical potential is not too high \cite{Braun:2019aow}. Therefore, the scalar and pseudoscalar channels, i.e., the $\sigma$ and $\pi$ channels out of the ten channels in \Eq{eq:set-alpha10tensors}, are dynamically hadronized in the work, such that the relevant four-quark couplings $\lambda_{\sigma}$ and $\lambda_{\pi}$ are replaced with the exchanges of the $\sigma$ and $\pi$ mesons, respectively, while those of the other eight channels are left unchanged. The $\sigma$ and $\pi$ mesons interact with the quarks through the Yukawa couplings, to wit, 
\begin{align}
    \Gamma_{\bar{q}q \phi} =h_{\sigma}\bar{q}_l T^0\sigma q_l+h_{\pi} \bar{q}_l i \gamma_5\bm{T}\cdot\bm{\pi}q_l\,, \label{eq:qqphi}
\end{align}
see also \Eq{eq:QCDaction} and the relevant discussions in \app{app:action-QCD}. Note that in the previous computation in \cite{Fu:2019hdw} the Yukawa couplings $h_{\sigma}$ and $h_{\pi}$ are assumed to be identical, this is a reasonable truncation given the constraint without the Fierz-complete four-quark tensor structures in \Eq{eq:set-alpha10tensors}. Consequently, it is natural to improve on this truncation and distinguish the $\sigma$ and $\pi$ Yukawa couplings in this work. We will compute the flows of both $h_{\sigma}$ and $h_{\pi}$ by means of the dynamical hadronization for the scalar and pseudoscalar channels, respectively.

In the sections as follows, we present the flow equations for the four-quark interactions and discuss the dynamical hadronization for the scalar and pseudoscalar channels. Furthermore, the flow equations of the quark-gluon vertex are slightly modified due to the presence of the four-quark interactions, and the corresponding flows are also provided.

\section{Correlation functions}
\label{sec:cor-fun}

In this section we discuss different correlation functions, mainly focusing on the effects arising from the Fierz-complete four-quark interactions in comparison to the calculations in \cite{Fu:2019hdw}.

\subsection{Four-quark correlation functions}
\label{subsec:4quark-cor-fun} 

%
\begin{figure}[t]
\includegraphics[width=0.45\textwidth]{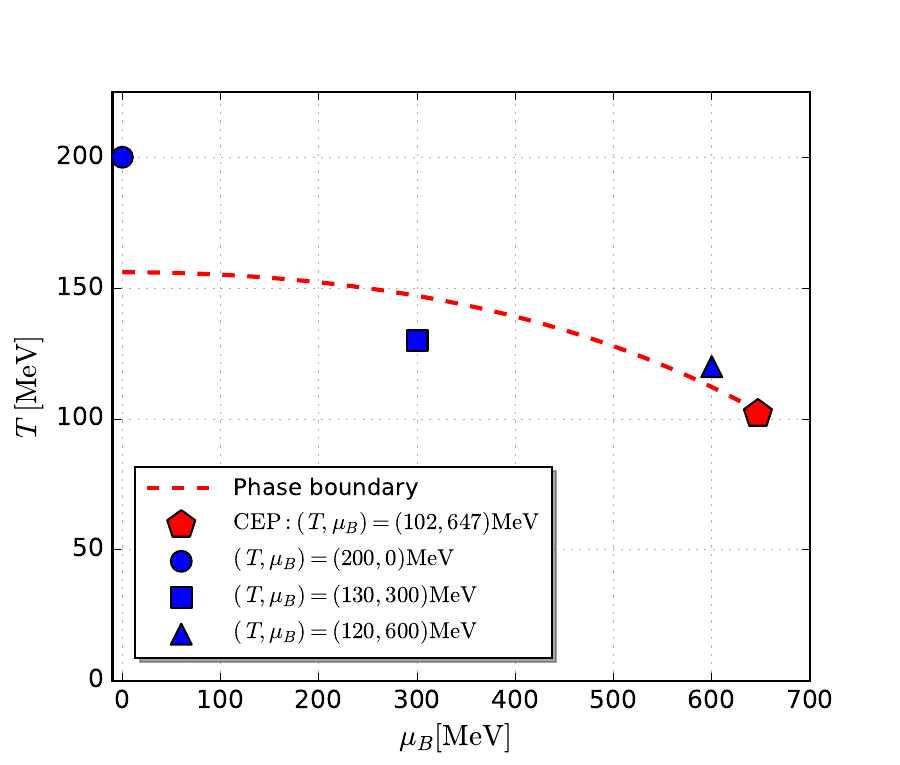}
\caption{Three representative points ($T, \mu_B$) in the QCD phase diagram, where the phase boundary and the critical end point (red pentagon) obtained in our calculations in this work are also shown to guide the eyes.}\label{fig:schematic}
\end{figure}
%

%
\begin{figure*}[t]
\includegraphics[width=0.45\textwidth]{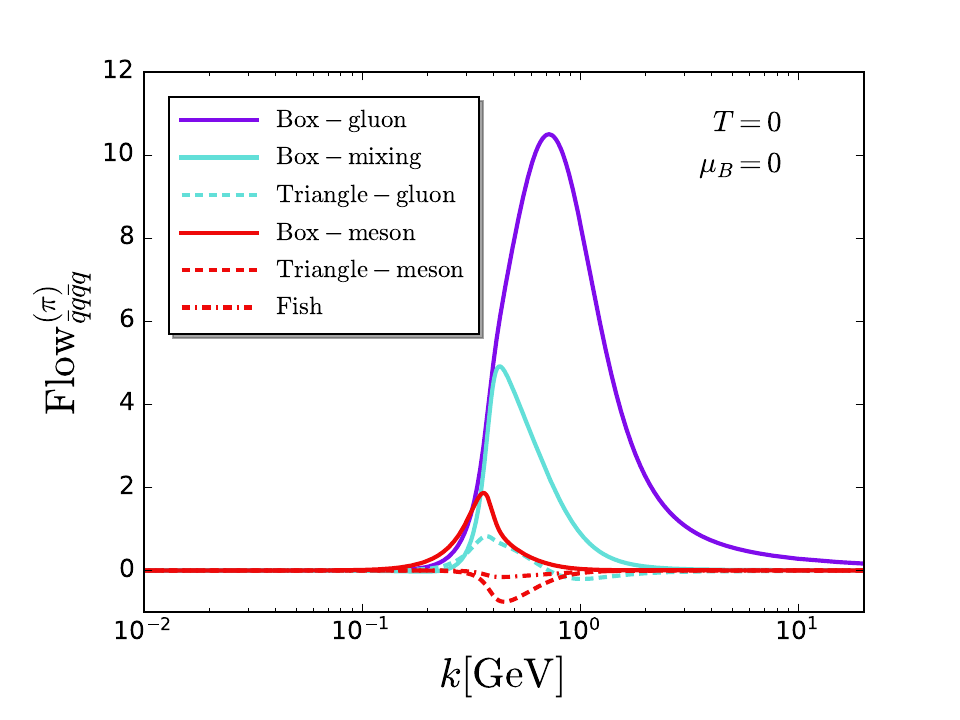}\hspace{0.3cm}
\includegraphics[width=0.45\textwidth]{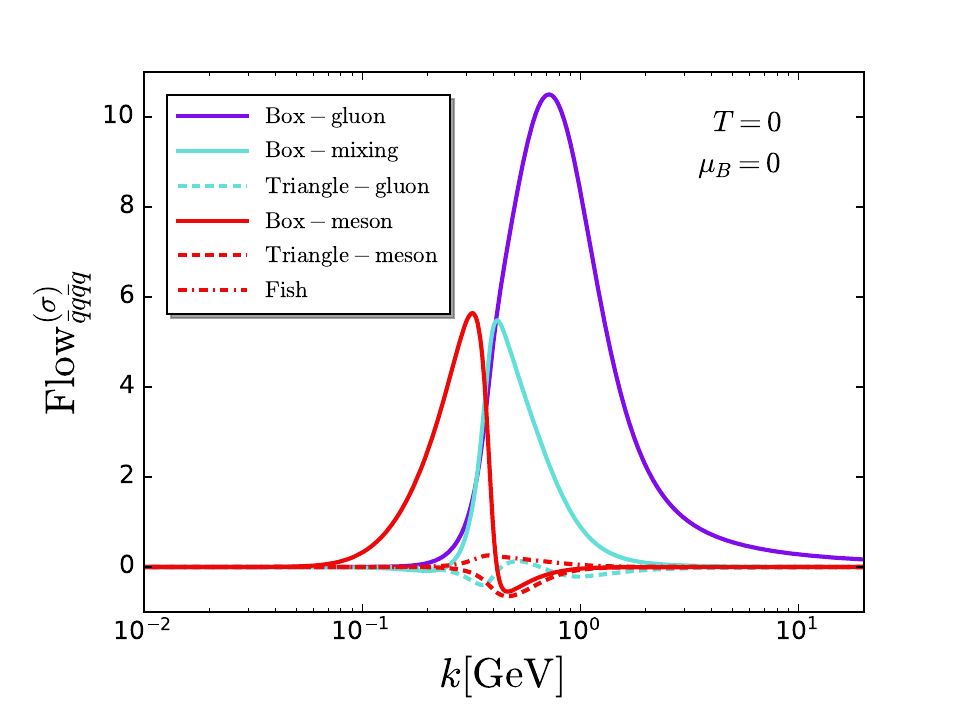}
\caption{Four-quark flows for the pion (left panel) and sigma (right panel) channels as functions of the RG scale in the vacuum, where different lines in the legend correspond one by one to the six loop diagrams on the right side of the flow equation in \Fig{fig:4quark-eqn}.}\label{fig:flow4pi-sigma}
\end{figure*}
%

%
\begin{figure*}[t]
\includegraphics[width=0.45\textwidth]{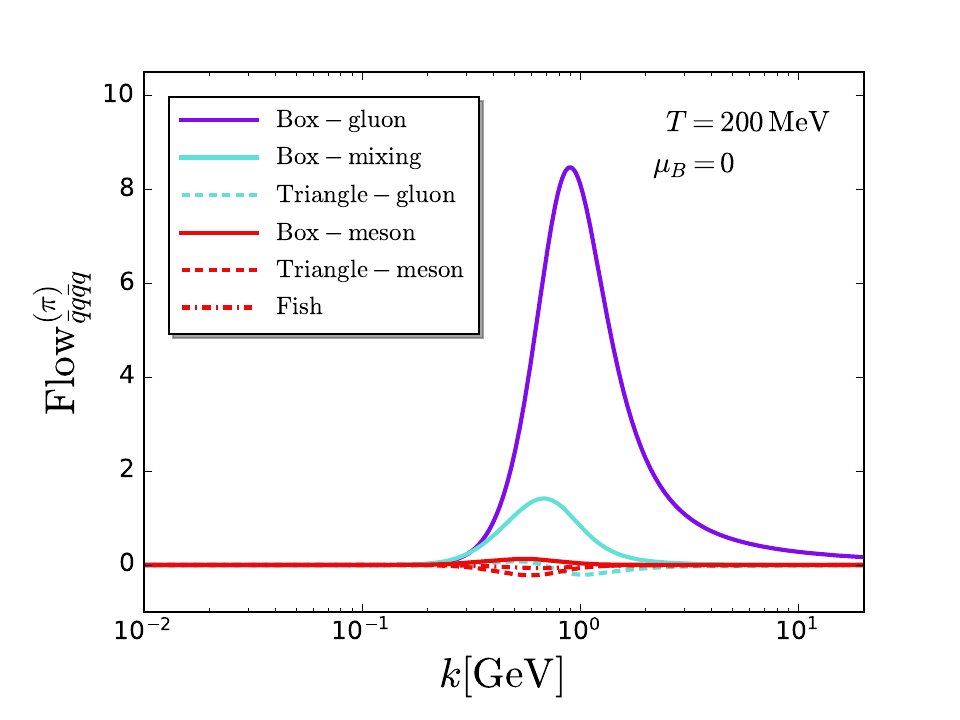}\hspace{0.3cm}
\includegraphics[width=0.45\textwidth]{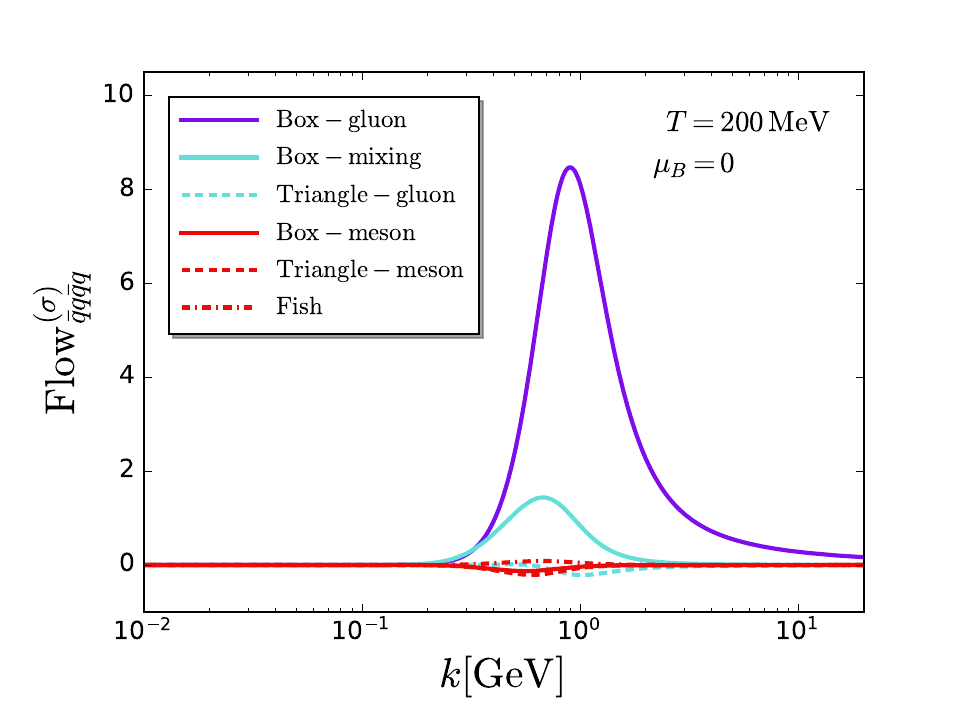}
\caption{Four-quark flows for the pion (left panel) and sigma (right panel) channels as functions of the RG scale with $T=200$ MeV and $\mu_B=0$, where different lines in the legend correspond one by one to the six loop diagrams on the right side of the flow equation in \Fig{fig:4quark-eqn}.}\label{fig:flow4pi-sigmaT200}
\end{figure*}
%

%
\begin{figure*}[t]
\includegraphics[width=0.45\textwidth]{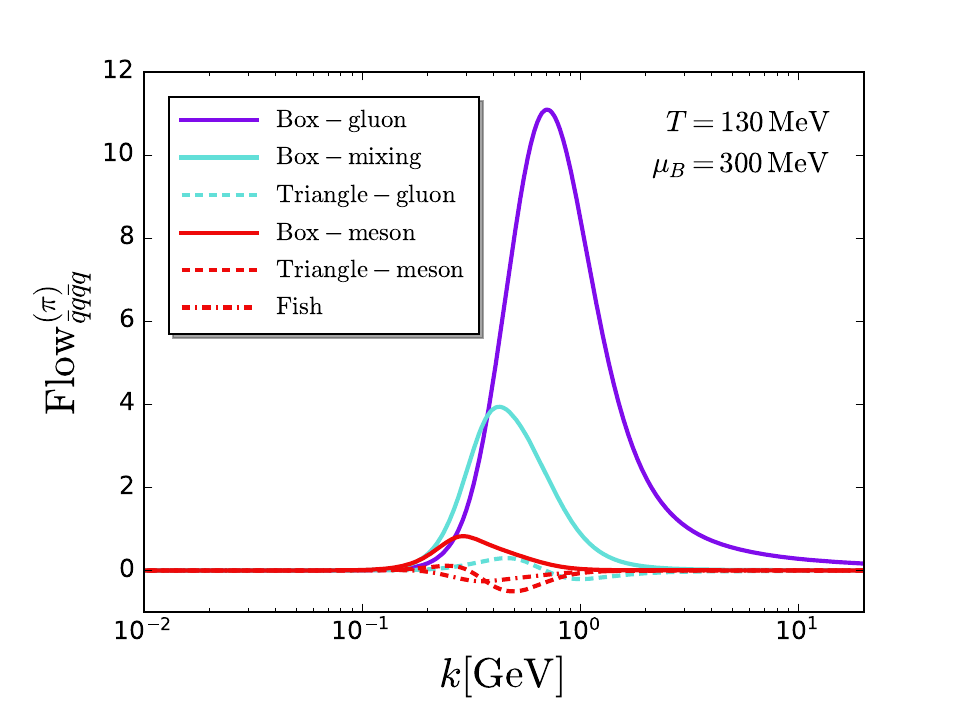}\hspace{0.3cm}
\includegraphics[width=0.45\textwidth]{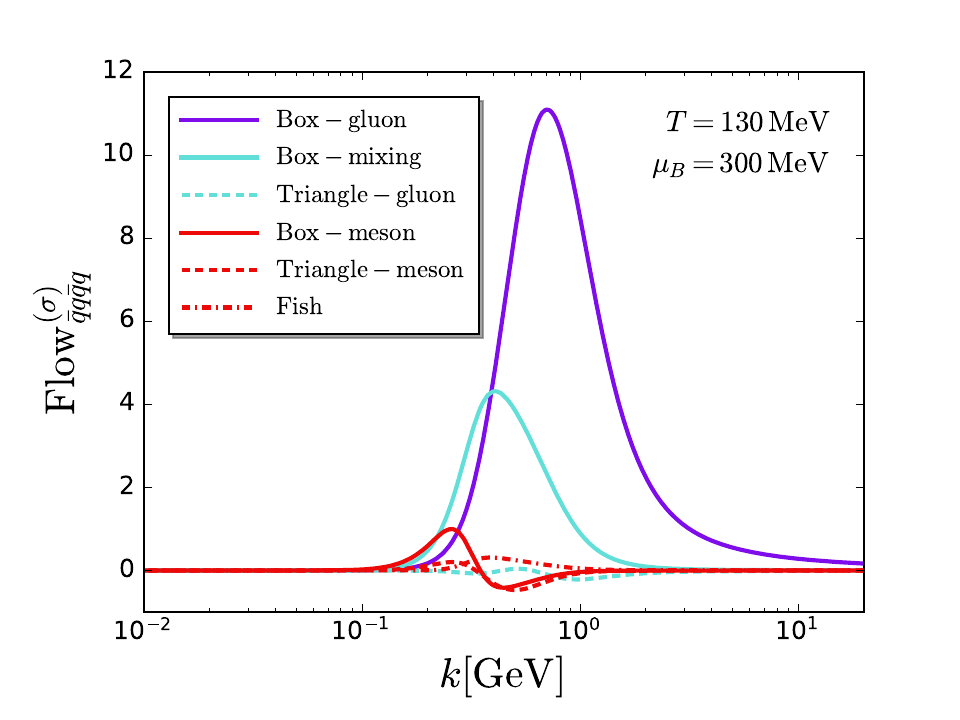}
\caption{Four-quark flows for the pion (left panel) and sigma (right panel) channels as functions of the RG scale with $T=130$ MeV and $\mu_B=300$ MeV, where different lines in the legend correspond one by one to the six loop diagrams on the right side of the flow equation in \Fig{fig:4quark-eqn}.}\label{fig:flow4pi-sigmaT130mub300}
\end{figure*}
%

%
\begin{figure*}[t]
\includegraphics[width=0.45\textwidth]{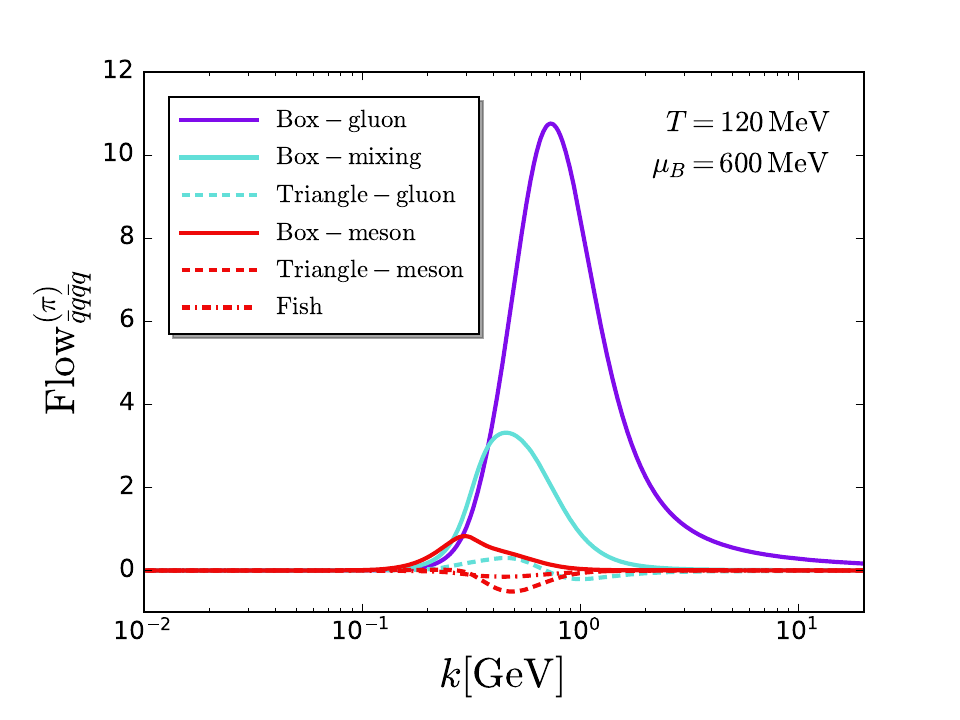}\hspace{0.3cm}
\includegraphics[width=0.45\textwidth]{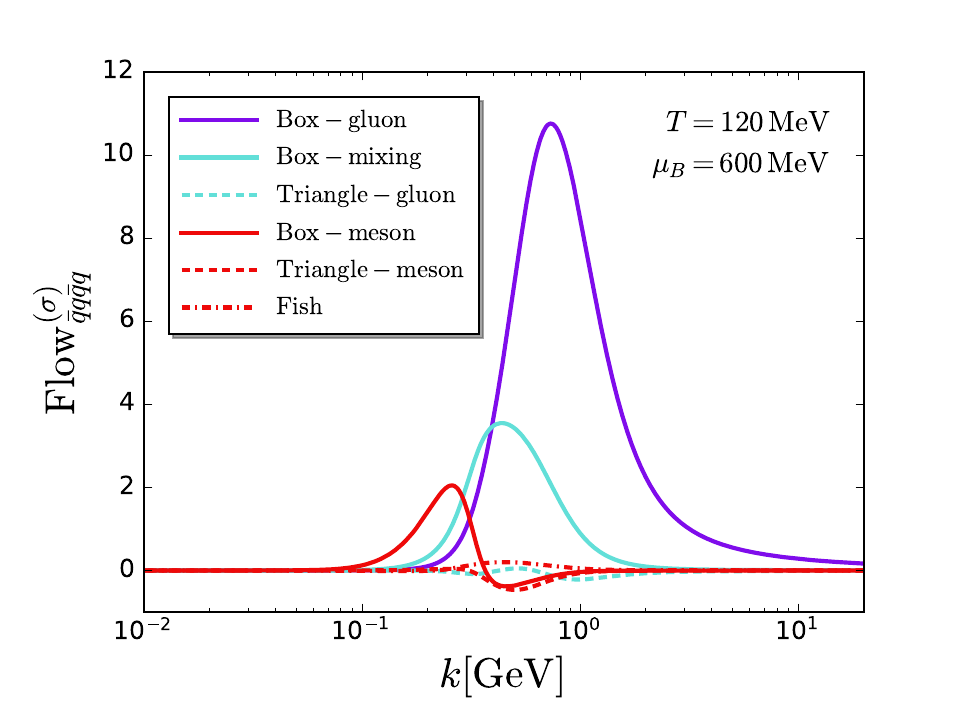}
\caption{Four-quark flows for the pion (left panel) and sigma (right panel) channels as functions of the RG scale with $T=120$ MeV and $\mu_B=600$ MeV, where different lines in the legend correspond one by one to the six loop diagrams on the right side of the flow equation in \Fig{fig:4quark-eqn}.}\label{fig:flow4pi-sigmaT120mub600}
\end{figure*}
%

The flow equation of the four-quark correlation functions are shown diagrammatically in \Fig{fig:4quark-eqn}. On the r.h.s. of the flow equation the orange curly and black solid lines stand for the full gluon and quark propagators, respectively. The blue dashed line denotes the full propagators for both the sigma and pion mesons. The four quark vertices on the right side of the equation represent those of the Fierz-complete basis in \Eq{eq:set-alpha10tensors}, except the $\sigma$ and $\pi$ channels, since they are vanishing after the dynamical hadronization is implemented, see the discussions in the following. According to the energy scale of interactions, the diagrams on the r.h.s. can be sorted into three classes. The first class refers to diagrams of high-energy regime resulting from the quark-gluon interactions, e.g., the first box diagram with two gluon exchanges, which are found to play a dominant role in the regime of the RG scale $k \gtrsim 0.5 \sim 1$ GeV \cite{Fu:2019hdw, Ihssen:2024miv, Fu:2025hcm}. The second class is comprised of diagrams of low-energy interactions including the Yukawa couplings and the four-quark interactions, e.g., the last three diagrams on the right side of the flow equation in \Fig{fig:4quark-eqn}, that is, the box diagram with two meson exchanges, the triangle diagram with one meson exchange and one four-quark vertex, and the fish diagram with two four-quark vertices. The second class diagrams are important in the low-energy region of $k \lesssim 0.5 \sim 1$ GeV, where the chiral symmetry is broken dynamically and the gluons are being decoupled gradually due to the emergence of the finite gluon mass gap. The third class is comprised of the mixing diagrams involving both the quark-gluon couplings and the low-energy interactions, which play a role in between the high and low energy regions. The mixing diagrams include the box diagram with one gluon exchange and one meson exchange, and the triangle diagram with one gluon exchange and one four-quark vertex, as shown by the second and third diagrams on the right side of \Fig{fig:4quark-eqn}, respectively.

The flows for the four-quark couplings in \Eq{eq:Gamma4q} read
\begin{align}
    {\mathrm{Flow}}^{(\alpha)}_{\bar q q \bar q q}& =-k^2\frac{1}{Z_{q}^2}\Tr\Big[\partial_t \Gamma^{(4)}_{\bar{q}q\bar{q}q}\,\mathcal{P}_{\bar{q}q\bar{q}q}^{(\alpha)}\Big]\,,\label{eq:Flow4q}
\end{align}
with
\begin{align}
    &\Gamma^{(4)}_{\bar{q}q\bar{q}q} \equiv \frac{\delta}{\delta\bar{q}}\frac{\delta}{\delta q}\frac{\delta}{\delta\bar{q}}\frac{\delta}{\delta q}\Gamma_{k}[\Phi]\,,\label{}
\end{align}
where the label $\alpha$ denotes the different tensor structures in the Fierz-complete basis in \Eq{eq:set-alpha10tensors}, and $\mathcal{P}_{\bar{q}q\bar{q}q}^{(\alpha)}$ stands for the relevant projection operator. Note that the flows in \Eq{eq:Flow4q} are RG-invariant and dimensionless. Details about the four-quark flows are presented in \app{app:flows-4quark}.

We investigate the four-quark flows at four different representative points in the QCD phase diagram, one in the vacuum with $\mu_B=T=0$, and the other three depicted in \Fig{fig:schematic}: One in the regime of high temperature at vanishing baryon chemical potential, and the other two near the phase boundary, one of which is in the proximity of the critical end point.

In \Fig{fig:flow4pi-sigma} we depict the four-quark flows in the vacuum as functions of the RG scale in the channels of $\alpha=\pi,\,\sigma$, where the contributions from the six loop diagrams on the right side of the flow equation in \Fig{fig:4quark-eqn} are presented separately, labeled in one-by-one correspondence by ``Box-gluon'', ``Box-mixing'', ``Triangle-gluon'', ``Box-meson'', ``Triangle-meson'', ``Fish'', respectively. The results of three classes of diagrams from the ultraviolet to the infrared are colored from the purple to red. It is found that the curves of different colors indeed show dominance at different RG scales as expected. The purple curve, i.e., the flow from the box diagram with two gluon exchanges, is remarkably higher than the other curves in the region of high energy.  With the decrease of RG scale, the cyan curves of mixing diagrams begin to play a role when $k \lesssim 1$ GeV. Subsequently, the red curves of low-energy diagrams take over and play a sizable role when $k \lesssim 0.5$ GeV. Moreover, one finds in each class of diagrams, the box diagram dominates over the triangle and fish diagrams, which indicates that the pion and sigma channels play a dominant role in contrast to the other channels in the four-quark basis in \Eq{eq:set-alpha10tensors}. Comparing the two plots in \Fig{fig:flow4pi-sigma}, one finds the magnitude of the box-meson diagram, i.e., the red solid curve, is larger in the sigma channel than that in the pion channel, that is attributed to the quark mass term of two-pion exchanges, i.e., the third term on the r.h.s. of \Eq{eq:flow4q-box-meson} in the sigma channel, while in the pion channel, this contribution is vanishing, as shown in \Eq{eq:flow4q-box-meson-pi}.

The four-quark flows of the pion and sigma channels at the three representative points in the phase diagram in \Fig{fig:schematic} are presented in \Fig{fig:flow4pi-sigmaT200}, \Fig{fig:flow4pi-sigmaT130mub300}, and \Fig{fig:flow4pi-sigmaT120mub600}, respectively. One can see that in \Fig{fig:flow4pi-sigmaT200} in the regime of high temperature quite above the phase boundary of chiral crossover, the mesonic contributions to the four-quark flows are suppressed significantly due to the decoupling of mesonic degrees of freedom, no matter what is concerned, the mixing diagrams or the purely low-energy diagrams, and the latter is more prominent. The magnitude of the purely gluon box diagram is also decreased a bit due to the increasing Debye screening mass of gluons. When the representative points are close to the phase boundary as shown in \Fig{fig:flow4pi-sigmaT130mub300} and \Fig{fig:flow4pi-sigmaT120mub600}, the mesonic contributions increase a bit, since the sigma mass is lowered down near the chiral crossover. This is more pronounced when it is close to the CEP, at which the sigma mode is the critical mode and thus is massless, as shown by the red solid line in the right panel of \Fig{fig:flow4pi-sigmaT120mub600}.

%
\begin{figure*}[t]
\includegraphics[width=0.45\textwidth]{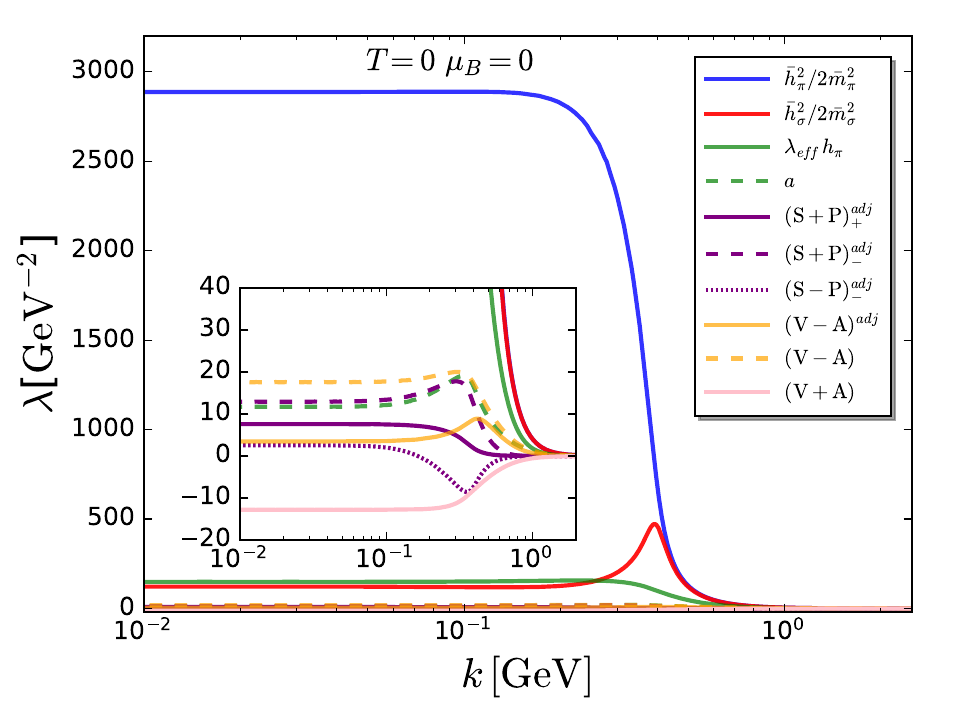}\hspace{0.3cm}
\includegraphics[width=0.45\textwidth]{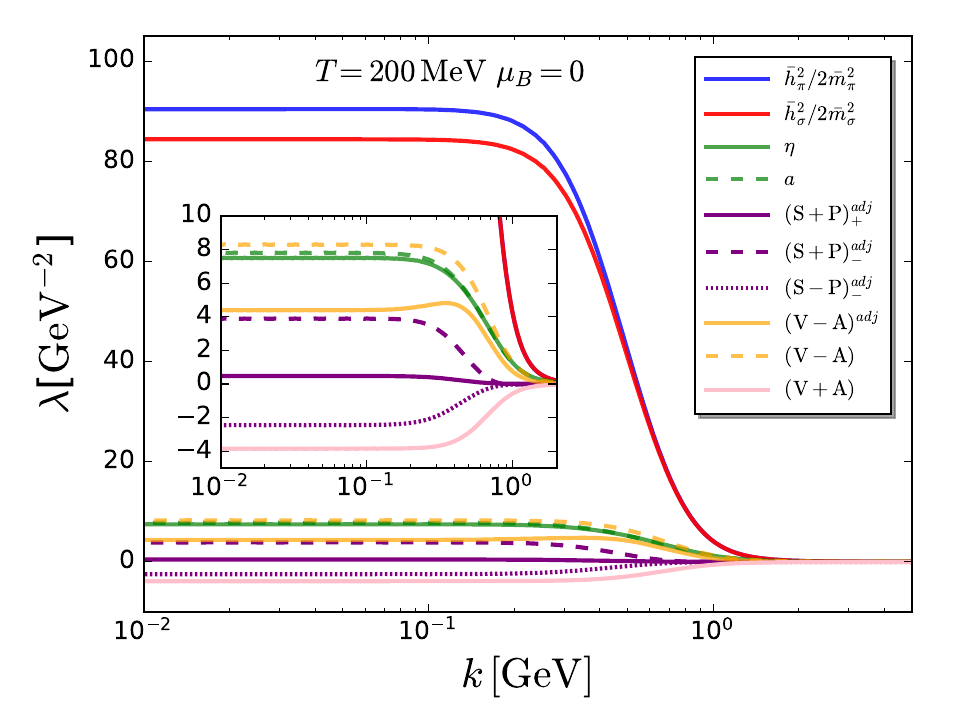}
\includegraphics[width=0.45\textwidth]{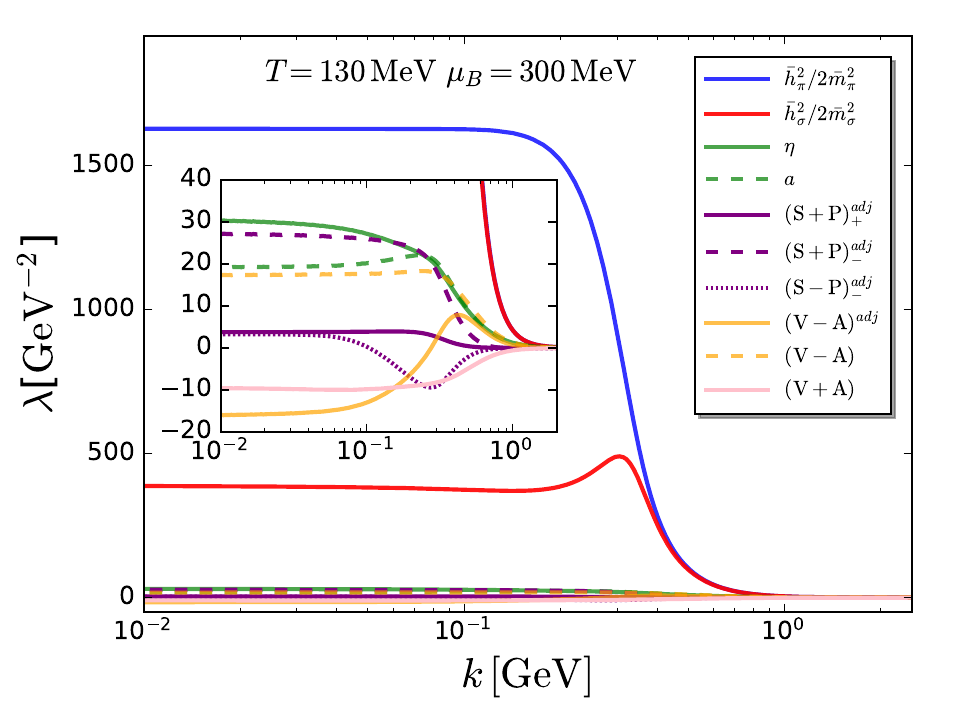}\hspace{0.3cm}
\includegraphics[width=0.45\textwidth]{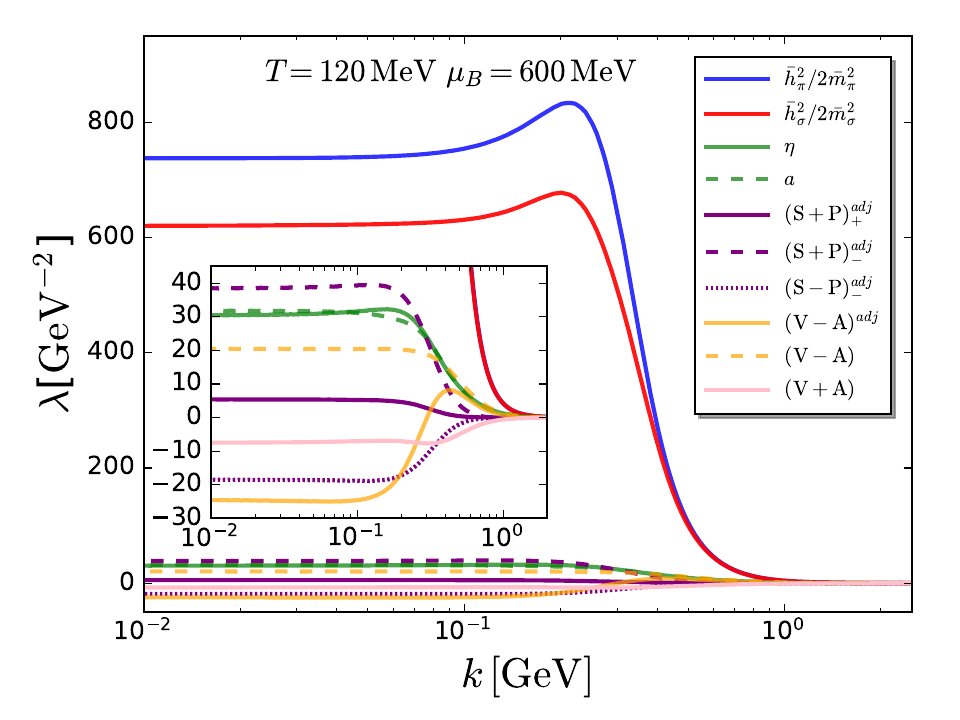}
\caption{Four-quark couplings of Fierz-complete channels as functions of the RG scale in the vacuum (top-left), and at the three representative points in the phase diagram as shown in \Fig{fig:schematic}. The inlays show the zoom-in view for the non-dominant channels.}
\label{fig:lam-k}
\end{figure*}
%

In \Fig{fig:lam-k} we compare all the four-quark couplings of different channels in the Fierz-complete basis in \Eq{eq:set-alpha10tensors}. In the same way, we choose the vacuum as well as the three representative points in the phase diagram in \Fig{fig:schematic}. Since the dynamical hadronization is implemented in the pion and sigma channels, we use the effective four-quark couplings as follows
\begin{align}
    &\frac{\bar h_\pi^2}{2\bar m_\pi^2}\,,\qquad \frac{\bar h_\sigma^2}{2\bar m_\sigma^2}\,,\label{}
\end{align}
instead, where the RG invariant Yukawa couplings and meson masses are presented in \Eq{eq:barh} and \Eq{eq:m-pisig}. Note that in this work we follow the convention in \cite{Fu:2019hdw} that symbols with a bar represent the RG invariant, or renormalized, quantities. The four-quark couplings in \Fig{fig:lam-k} are also RG invariant, i.e.,
\begin{align}
    \bar \lambda_{\alpha}=\frac{\lambda_{\alpha}}{Z_q^2}\,.\label{}
\end{align}

One finds in \Fig{fig:lam-k} that in the vacuum the pion and sigma channels play the overwhelmingly dominant role, and all the other channels are negligible. At $T=130$ MeV and $\mu_B=300$ MeV, which is located in the chiral symmetry broken phase and is near the phase boundary as shown by the square point in \Fig{fig:schematic}, the results are similar with the case in the vacuum, while the magnitude of the pion and sigma channels is suppressed. When it is in the region of high temperature or at larger $\mu_B$ near the CEP, as shown in the right panels in \Fig{fig:lam-k}, the other channels begin to play a role. Specifically, when it is near the CEP, cf., the bottom-right panel, the magnitude of four-quark couplings in the channels $\eta$, $a$, $(S+ P)^{\mathrm{adj}}_-$, $(S- P)^{\mathrm{adj}}_-$, $(V-A)$, $(V-A)^{\mathrm{adj}}$, etc, is increased up to about 10\% of that in the pion and sigma channels.

\subsection{Quark-meson vertices and Yukawa couplings}
\label{subsec:Yukawa}

%
\begin{figure*}[t]
\includegraphics[width=0.9\textwidth]{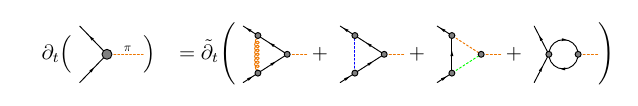}
\includegraphics[width=1.\textwidth]{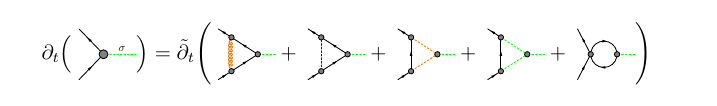}
\caption{Diagrammatic representation of the flow equations of quark-meson vertices, where the orange and green dashed lines denote the $\pi$ and $\sigma$ propagators, respectively, while the blue dashed line stands for both of them.}\label{fig:Yukawa-flow-eq}
\end{figure*}
%

%
\begin{figure*}[t]
\includegraphics[width=0.45\textwidth]{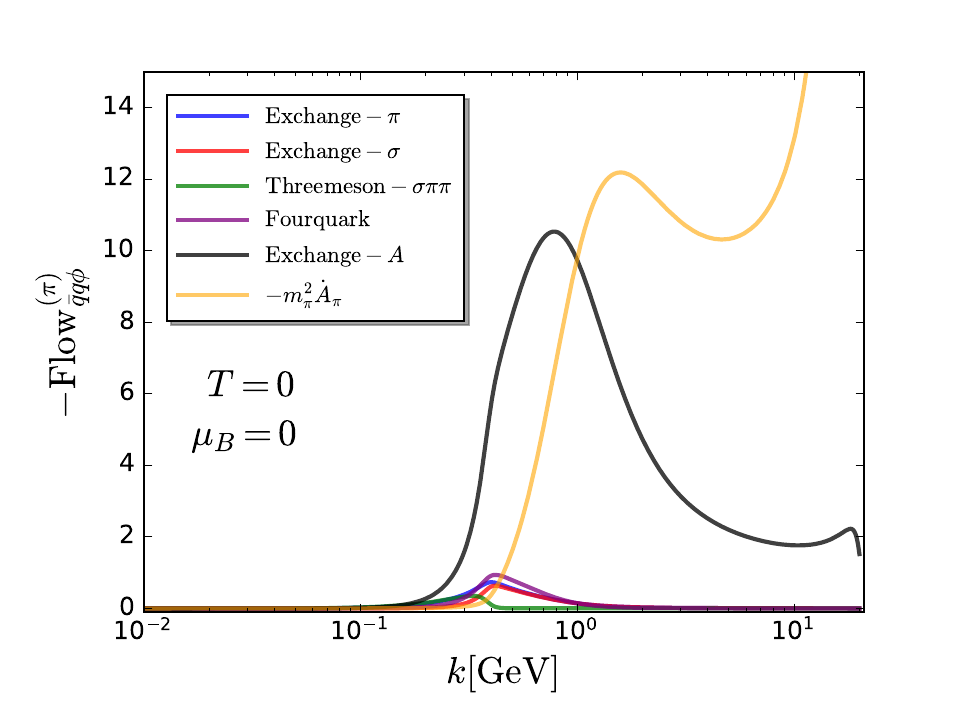}\hspace{0.3cm}
\includegraphics[width=0.45\textwidth]{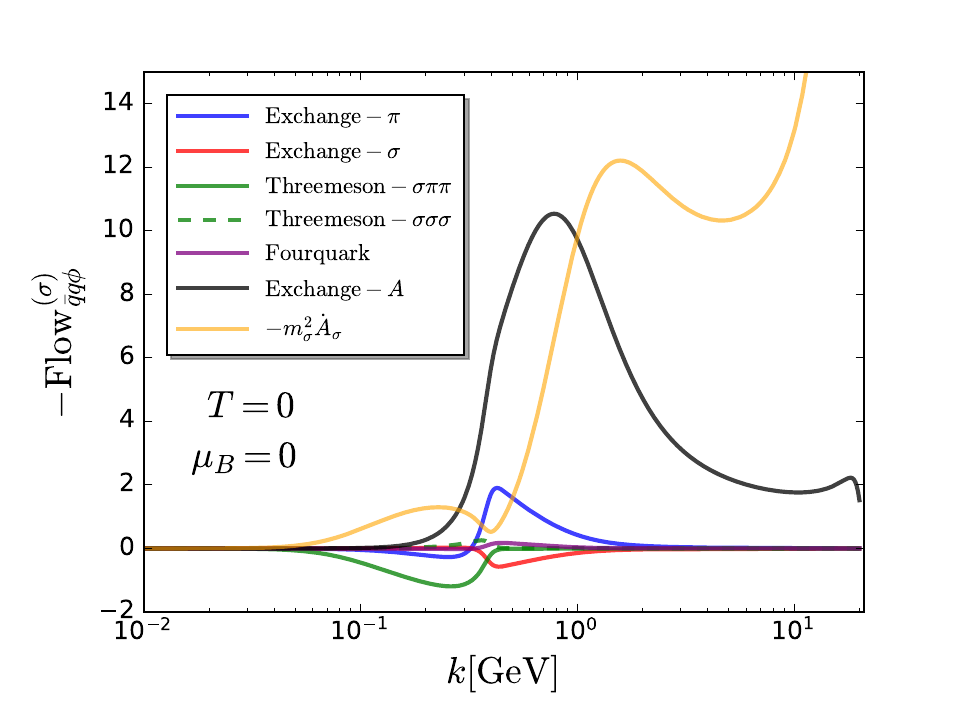}
\caption{Yukawa flows for the pion (left panel) and sigma (right panel) as functions of the RG scale in the vacuum. ``Exchange-A'' denotes the diagram of gluon exchange on the right side of the flow equations in \Fig{fig:Yukawa-flow-eq}, ``Exchange-$\pi/\sigma$'' the diagrams of $\pi/\sigma$ exchange, ``Three-meson-$\sigma\pi\pi/\sigma\sigma\sigma$'' the diagrams from the three-meson vertices, ``Four-quark'' the diagrams from the four-quark vertices. The lines of $\dot{ A}$ represent the contributions from the hadronization function in \Eq{eq:hphiflow}.}\label{fig:flowhpi-hsigma-full-vacuum}
\end{figure*}
%

%
\begin{figure*}[t]
\includegraphics[width=0.45\textwidth]{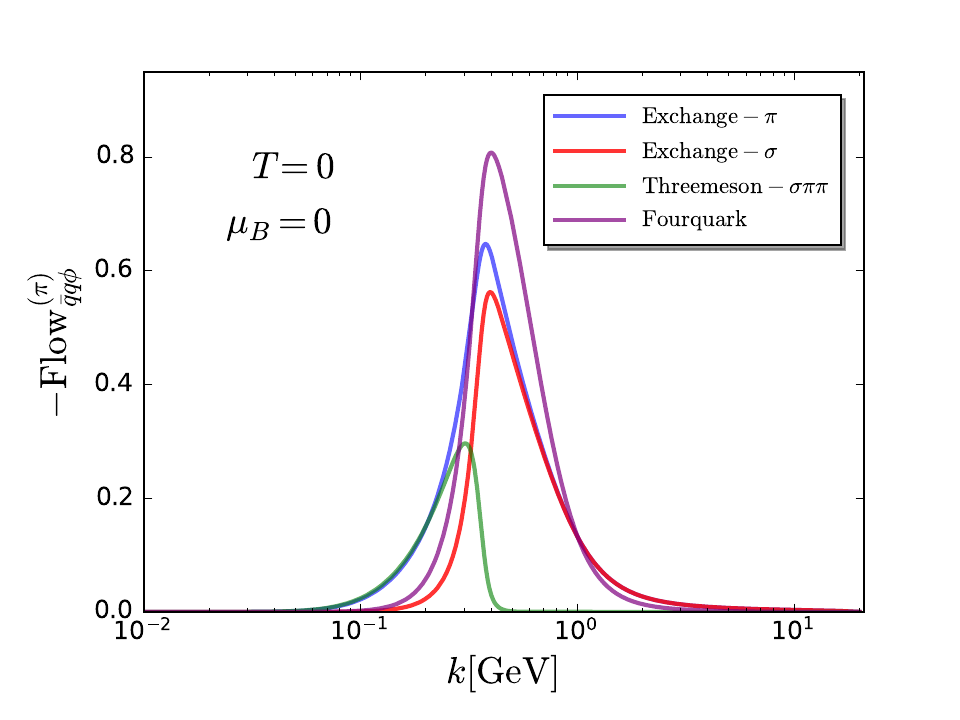}\hspace{0.3cm}
\includegraphics[width=0.45\textwidth]{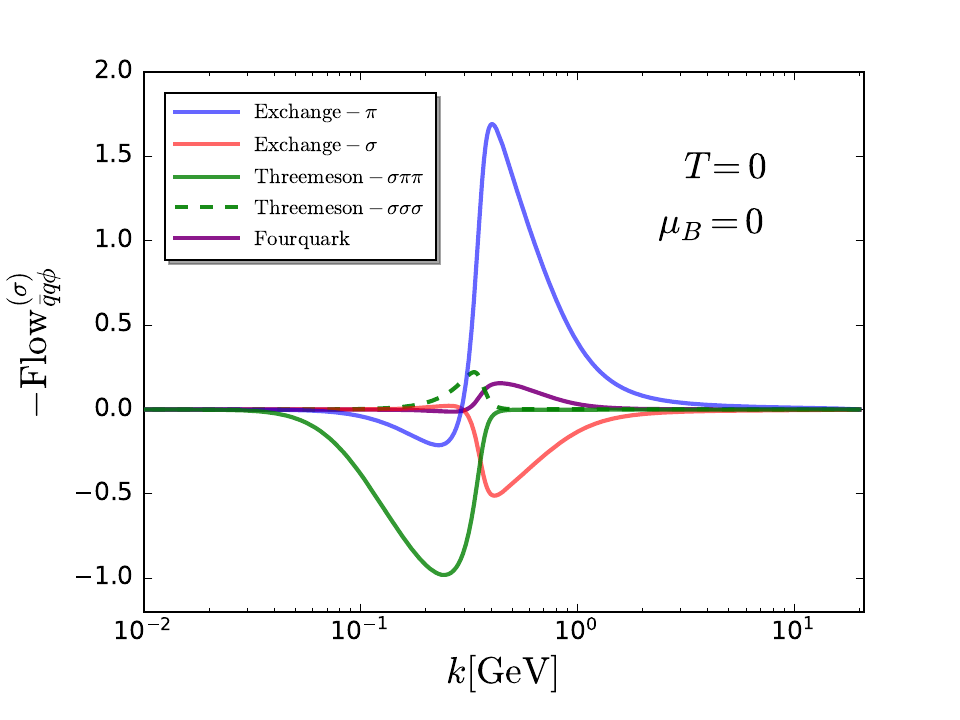}
\caption{Yukawa flows for the pion (left panel) and sigma (right panel) as functions of the RG scale in the vacuum, which are the zoom-in view of \Fig{fig:flowhpi-hsigma-full-vacuum}, without the curves of gluon exchange and hadronization function.}\label{fig:flowhpi-hsigma-vacuum}
\end{figure*}
%

%
\begin{figure*}[t]
\includegraphics[width=0.45\textwidth]{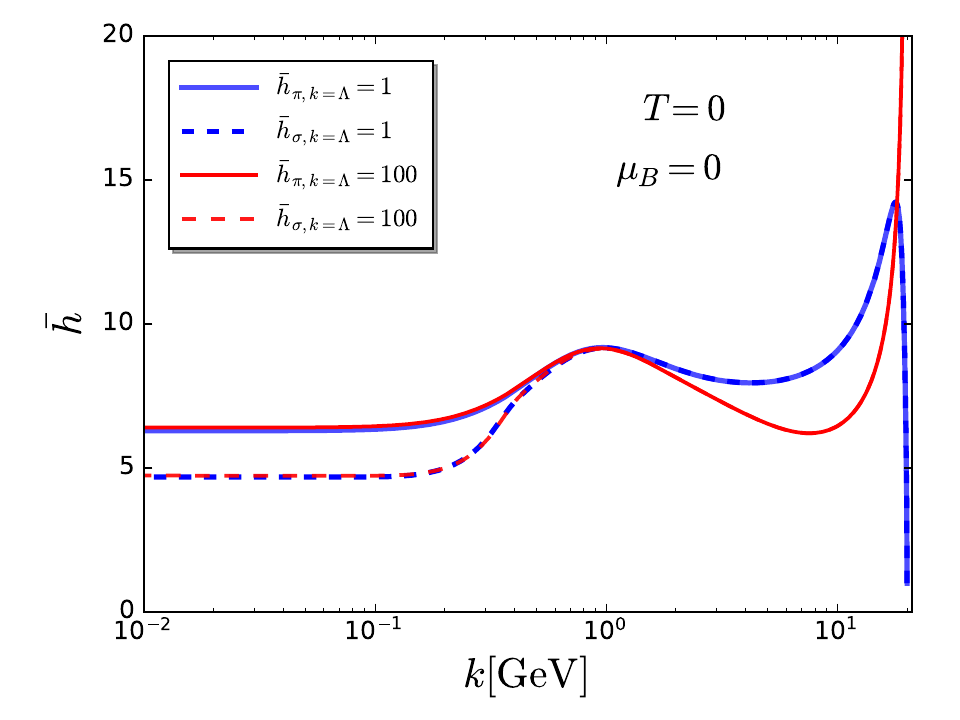}
\includegraphics[width=0.45\textwidth]{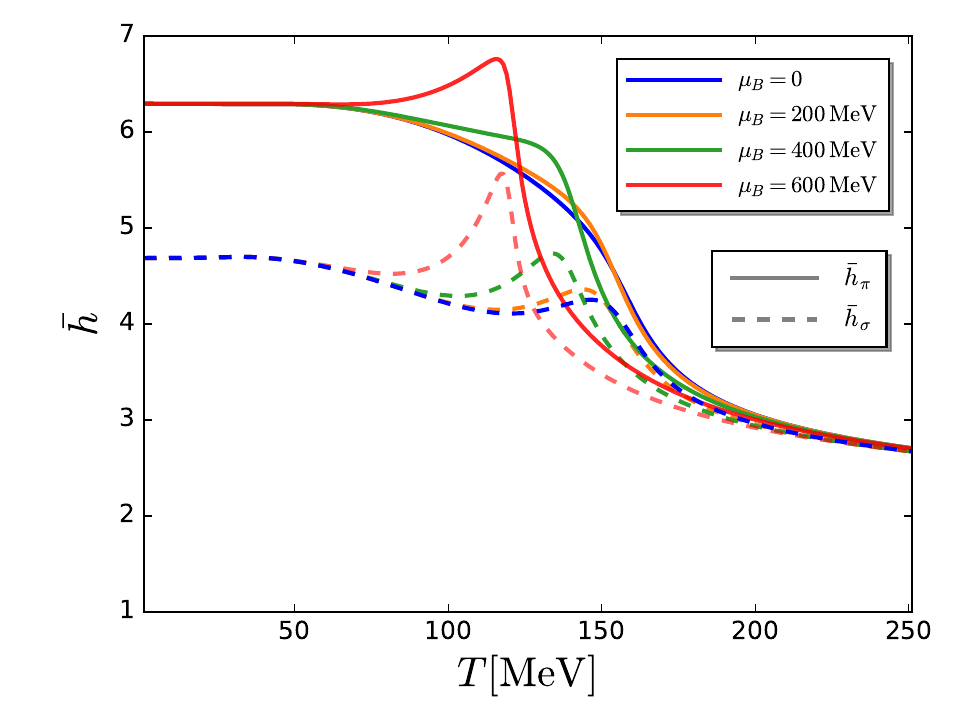}
\caption{Left panel: Yukawa couplings for the pion and sigma as functions of the RG scale in vacuum with different initial values. Right panel: Yukawa couplings at $k=0$ as functions of the temperature with several different values of baryon chemical potentials.}
\label{fig:hphi-k}
\end{figure*}
%

%
\begin{figure*}[t]
\includegraphics[width=0.45\textwidth]{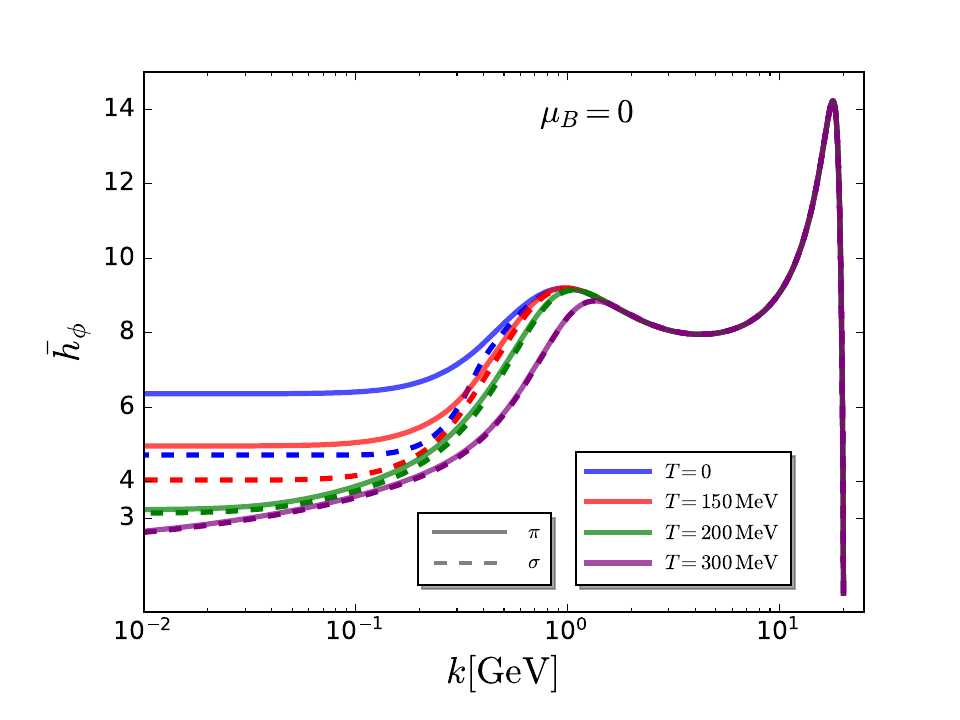}\hspace{0.3cm}
\includegraphics[width=0.45\textwidth]{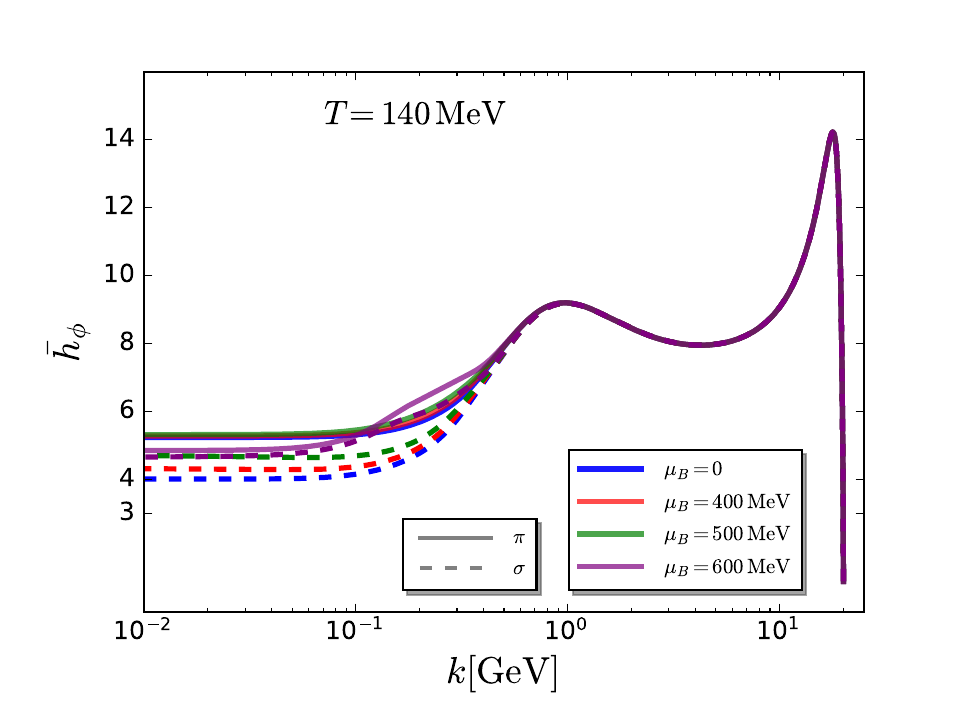}
\caption{Yukawa couplings for the pion and sigma as functions of the RG scale for different values of temperature at $\mu_B=0$ (left panel) and for different values of baryon chemical potential at $T=140$ MeV.}\label{fig:hphi-k-TmuB}
\end{figure*}
%

In \app{app:flow-action} we present the details about the flow equation of QCD effective action with the dynamical hadronization. It is shown there, e.g., \Eq{eq:hphiflow}, that the flows of Yukawa couplings receive contributions not only from the flows of three-point functions in \Fig{fig:Yukawa-flow-eq}, but also from the four-quark flows through the technique of dynamical hadronization, demanding that the four-quark couplings of the $\sigma$ and $\pi$ channels are vanishing at every value of the RG scale $k$, such that the active collective degrees of freedom of chiral dynamics in the low-energy QCD is transferred from the four-quark sector to the quark-meson sector. This explains how the low-energy effective field theory emerges naturally from the fundamental degrees of freedom of QCD with the evolution of RG scale from the ultraviolet to infrared.

In \Fig{fig:flowhpi-hsigma-full-vacuum} and \Fig{fig:flowhpi-hsigma-vacuum} we present the Yukawa flows in \Eq{eq:hphiflow} and \Eq{eq:flowqbarqphi} in the vacuum, arising from the diagrams on the right side of the flow equations for the quark-meson vertices in \Fig{fig:Yukawa-flow-eq}, and from the hadronization function $\dot{A}_{\phi_i}$ in \Eq{eq:dtphi}, respectively, cf., \app{app:flow-action} for more details. It is found that the hadronization function resulting from the four-quark interactions plays a dominant role in the high energy region with the RG scale $k \gtrsim 1$ GeV. Then, it is taken over by the triangular diagram with a gluon exchange. The other triangular diagrams as well as the loop diagram composed of a four-quark vertex and a Yukawa vertex, which are also shown in \Fig{fig:flowhpi-hsigma-vacuum} with a zoom-in view, come into play in the low-energy region. Their magnitude is smaller than that of the gluon exchange diagram, except that in the Yukawa flows of sigma, the magnitude of other diagrams is a bit larger than that of the gluon exchange diagram in the very low energy region.

In the left panel of \Fig{fig:hphi-k}, the Yukawa couplings for the pion and sigma in the vacuum are depicted as functions of the RG scale for different initial values at the UV cutoff $\Lambda=20$ GeV. One can see that the Yukawa couplings are independent of the initial values in the energy region, where the mesonic degrees of freedom begin to play a role, say $k \lesssim 1$ GeV. The Yukawa coupling of pion is found to be relatively larger than that of sigma in the infrared. The dependence of Yukawa couplings at $k=0$ on the temperature with several different fixed values of $\mu_B$ is shown in the right panel of \Fig{fig:hphi-k}. At small values of $\mu_B$, the Yukawa couplings decreases monotonically with the increasing temperature, while at larger $\mu_B$, especially in the proximity of CEP, the Yukawa couplings show a peak structure near the pseudo-critical temperature, resulting from less and less mass of the sigma mode toward the CEP. In \Fig{fig:hphi-k-TmuB} the running of the Yukawa couplings with the RG scale is also investigated at finite temperature and baryon chemical potentials. It is found that the degeneracy between the pion and sigma Yukawa couplings take places as the temperature or chemical potential increases. 

\subsection{Quark-gluon vertex}
\label{subsec:quark-gluon}

%
\begin{figure*}[t]
\includegraphics[width=0.9\linewidth]{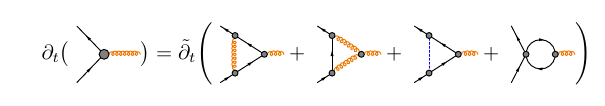}
\caption{Diagrammatic representation of the flow equation of quark-gluon vertex.}
\label{fig:quarkgluon-equ}
\end{figure*}
%

%
\begin{figure}[t]
\includegraphics[width=0.45\textwidth]{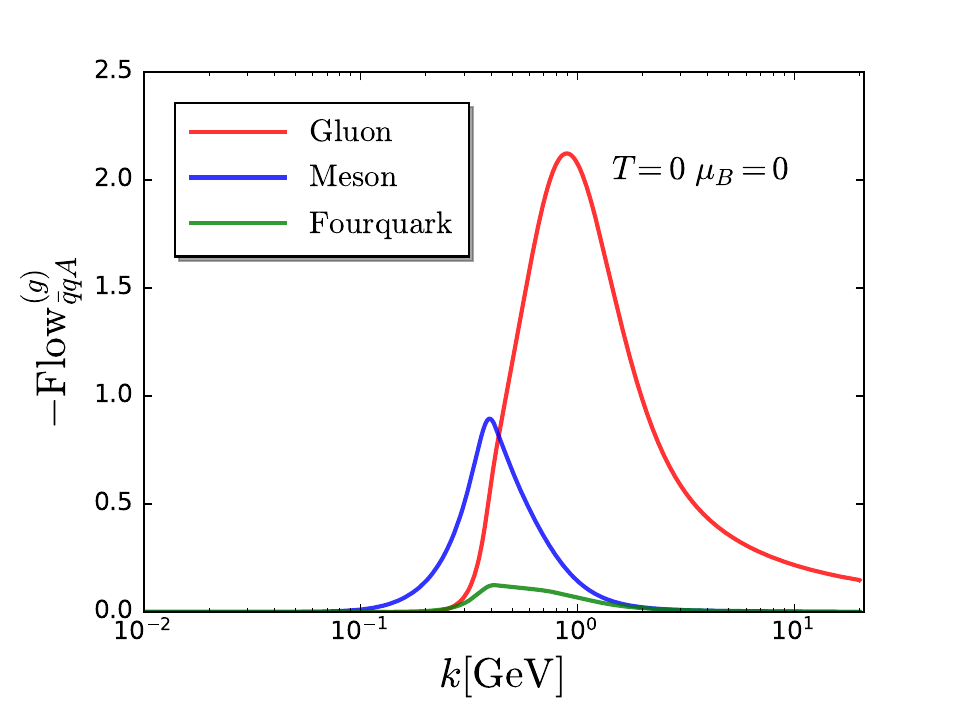}
\caption{Quark-gluon flow in \Eq{eq:dtg} as a function of the RG scale in the vacuum. The red line denotes the total contributions from the first two diagrams on the right side of the flow equation in \Fig{fig:quarkgluon-equ}, i.e., that from the quark and gluon interactions; the blue and green lines represent the last two diagrams, i.e., the meson exchange and four-quark interaction, respectively.}
\label{fig:flowg}
\end{figure}
%

%
\begin{figure}[t]
\includegraphics[width=0.45\textwidth]{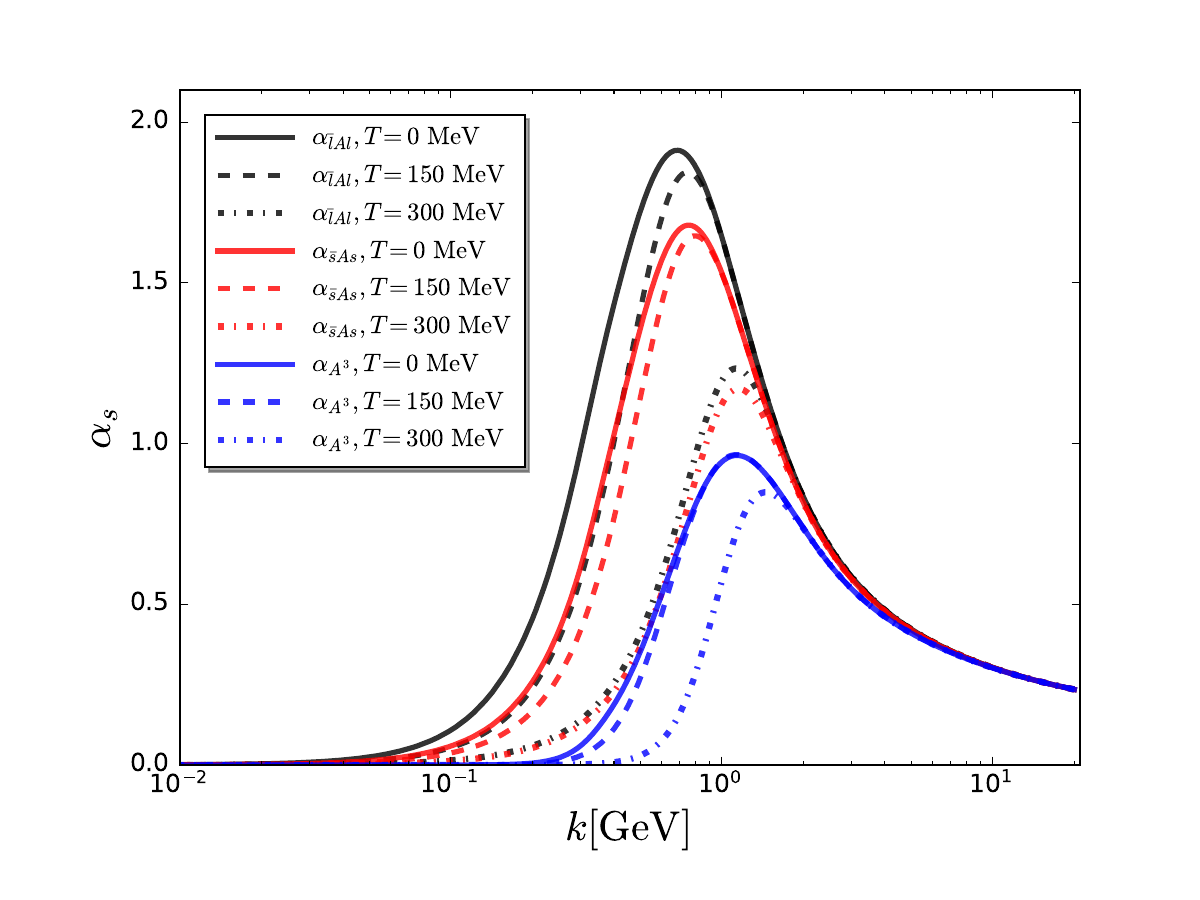}
\caption{Running of the strong couplings for the quark-gluon couplings of $u$ and $d$ light quarks $\alpha_{\bar{l}l A}$ and strange quark $\alpha_{\bar{s} s A}$, and for the three-gluon coupling $\alpha_{A^3}$ with the RG scale at several values of temperature and vanishing baryon chemical potential.}
\label{fig:flowalphas}
\end{figure}
%

%
\begin{figure*}[t]
\includegraphics[width=0.45\textwidth]{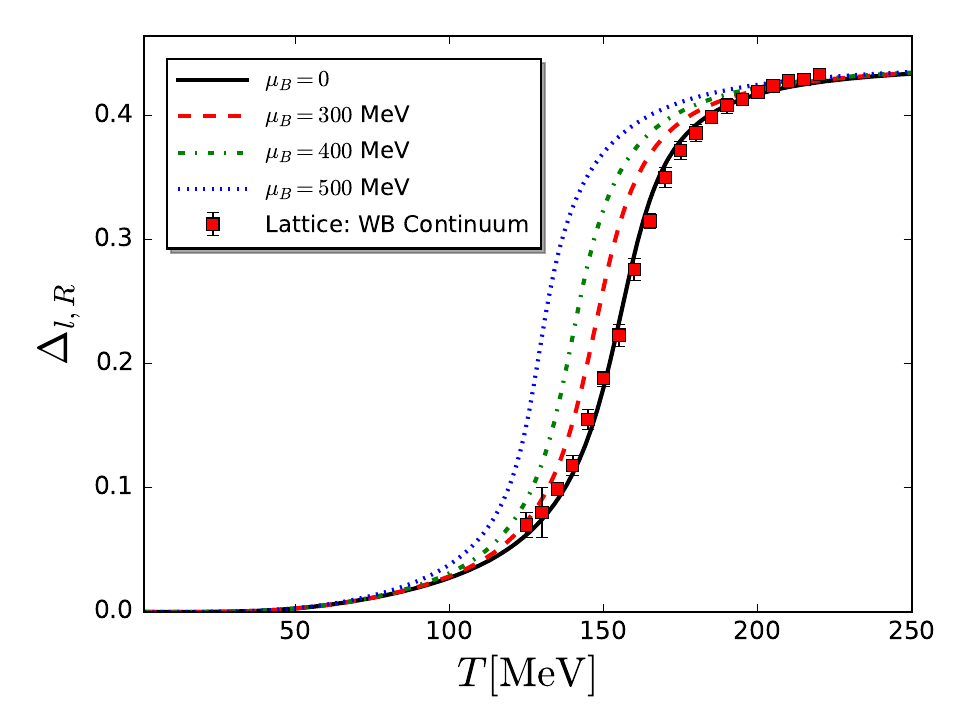}\hspace{0.3cm}
\includegraphics[width=0.45\textwidth]{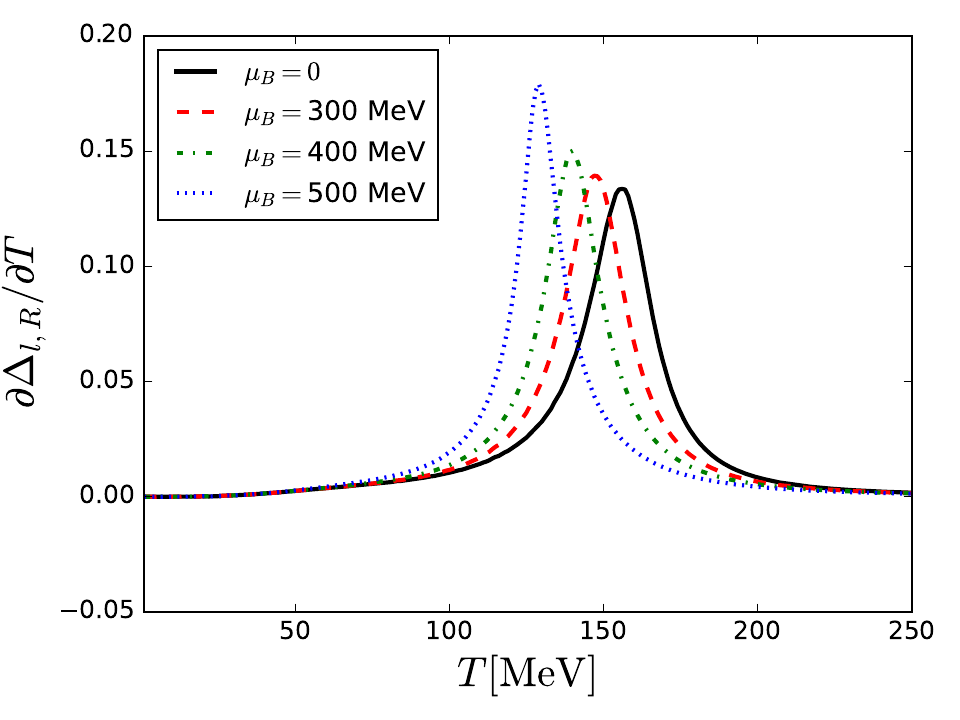}
\caption{Renormalized light quark chiral condensate $\Delta_{l,R}$ (left panel) and its derivative with respect to the temperature $\partial \Delta_{l,R}/ \partial T$ (right panel) as functions of the temperature at several different values of baryon chemical potential. The lattice results at vanishing baryon chemical potential from the Wuppertal-Budapest collaboration (WB) are also presented for comparison \cite{Borsanyi:2010bp}.}\label{fig:deltlR}
\end{figure*}
%

%
\begin{figure}[t]
\includegraphics[width=0.45\textwidth]{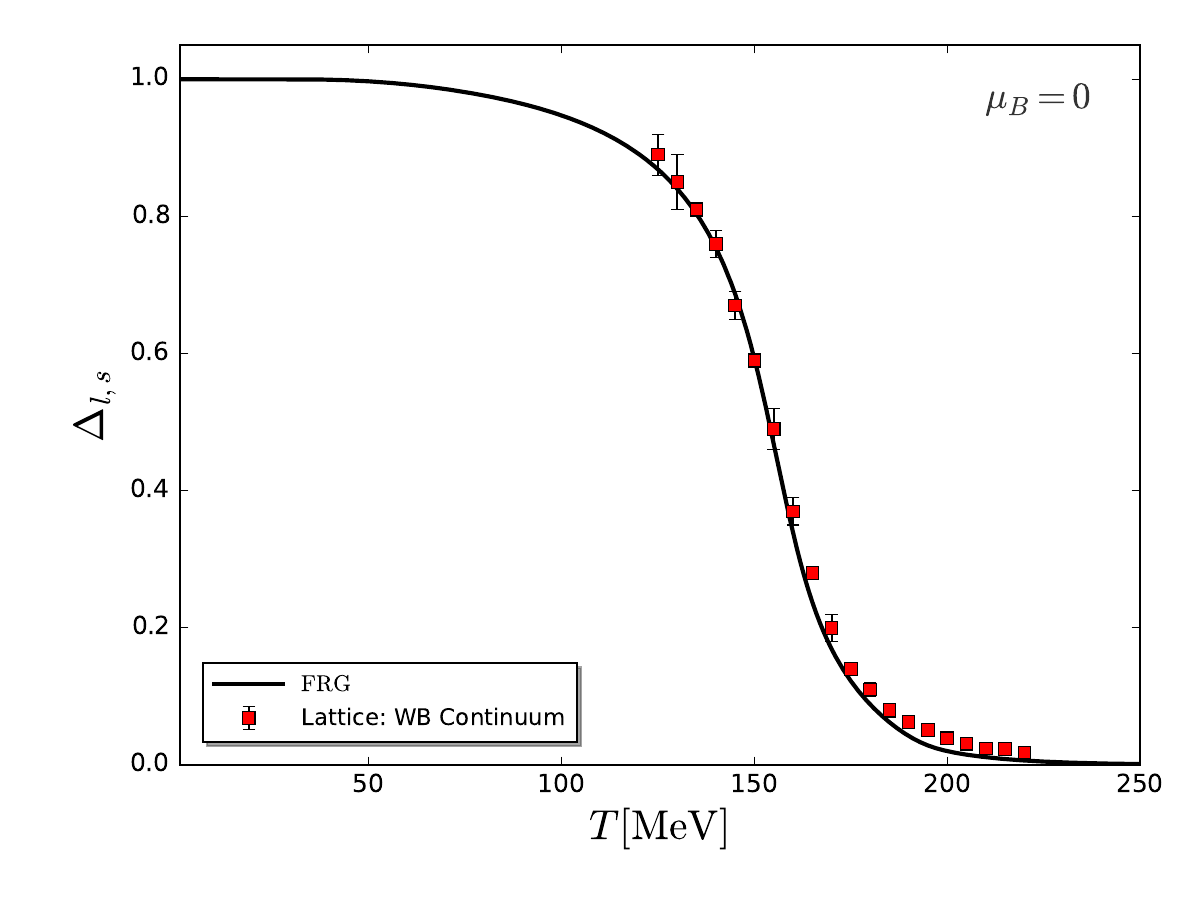}
\caption{Reduced condensate $\Delta_{l,s}$ as a function of the temperature at vanishing baryon chemical potential, in comparison to the lattice results from the Wuppertal-Budapest collaboration (WB) \cite{Borsanyi:2010bp}.}
\label{fig:deltlS}
\end{figure}
%

The four-quark interactions also contribute to the flow of the quark-gluon vertex, as shown by the last diagram in \Fig{fig:quarkgluon-equ}. The other diagrams on the right side of the flow equation correspond to that of the gluon exchange, three-gluon interaction, meson exchange, respectively. From the flow equation of the quark-gluon vertex in \Fig{fig:quarkgluon-equ}, one is able to obtain the flow equation of the renormalized quark-gluon coupling defined in \Eq{eq:gbqqa} after an appropriate projection, which reads
\begin{align}
    \partial_t \bar g_{\bar q  q A}&=\left( \frac{1}{2}\eta_A+\eta_q\right) \bar g_{\bar q  q A}+{\mathrm{Flow}}_{\bar{q} q A}\,,
 \label{eq:dtg}
\end{align}
where the quark-gluon flow is given in \Eq{eq:flowqbarqA}, and $\eta_A$ and $\eta_q$ are the anomalous dimensions for the gluon and quark fields, respectively, defined in \Eq{eq:etaPhia}. In \Fig{fig:flowg} different contributions to the quark-gluon flow ${\mathrm{Flow}}_{\bar{q} q A}$ are presented separately, where the calculation is done in the vacuum. It is found that the four-quark contribution is significantly smaller than the quark-gluon or the meson contributions.

In \Fig{fig:flowalphas} we show the running of the strong couplings obtained from the quark-gluon and three-gluon couplings, to wit, 
\begin{align}
    \alpha_{\bar q  q A}=\frac{\bar{g}_{\bar q  q A}^2}{4\pi}\,,\qquad \alpha_{A^3}=\frac{\bar{g}_{A^3}^2}{4\pi}\,,\label{eq:alphas}
\end{align}
where the renormalized $\bar{g}_{\bar q  q A}$ and $\bar{g}_{A^3}$ are given in \Eq{eq:gbqqa} and \Eq{eq:strongcoupGluon}, respectively. Note that in \Fig{fig:flowalphas} the quark-gluon couplings with respect to the light and strange quarks are differentiated. It is found that all the couplings are identical in the perturbative region, required by the renormalizability and the Slavnov-Taylor identity (STI) of gauge symmetry. They deviate from each other when the RG scale $k \lesssim 3\sim 7 $ GeV, where a gluon mass gap is developed. One finds the ordering $\alpha_{\bar{l}l A} \gtrsim \alpha_{\bar{s} s A} \gtrsim \alpha_{A^3}$ in the regime of low energy scale, which is not only consistent with the results obtained with a relatively simpler truncation with only one single $\sigma$-$\pi$ channel four-quark vertex and Yukawa coupling \cite{Fu:2019hdw}, but also with the results obtained with a much more sophisticated truncation in QCD \cite{Fu:2025hcm}. Moreover, the strong couplings decrease with the increasing temperature.

\section{Chiral phase transition and QCD phase diagram}
\label{sec:phase-diagram}

%
\begin{figure}[t]
\includegraphics[width=0.45\textwidth]{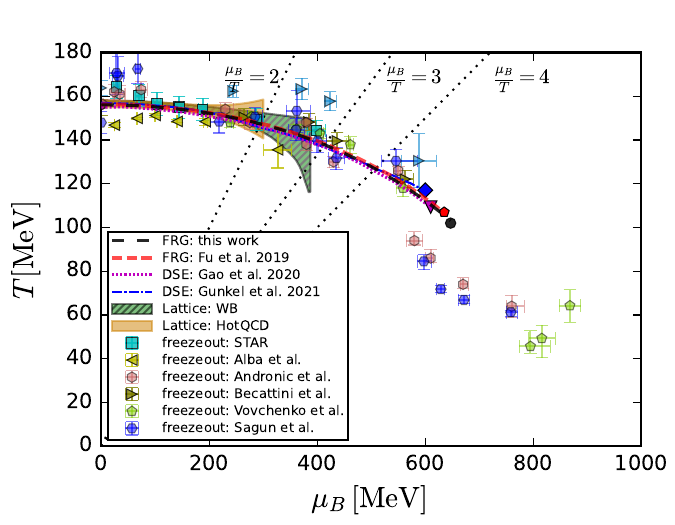}
\caption{QCD phase diagram in the plane of temperature and baryon chemical potential. The phase boundary for the chiral crossover and CEP obtained in this work are depicted in the black dashed line and dot, respectively. Previous results from fRG \cite{Fu:2019hdw} and DSE \cite{Gao:2020fbl, Gunkel:2021oya}, phase boundary in the region of low $\mu_B$ from lattice QCD \cite{HotQCD:2018pds, Borsanyi:2020fev}, chemical freeze-out data extracted from experiments \cite{STAR:2017sal, Alba:2014eba, Andronic:2017pug, Becattini:2016xct, Vovchenko:2015idt, Sagun:2017eye} are also shown in the phase diagram for comparison.}
\label{fig:phase-diagram}
\end{figure}
%

The numerical setup in this work and related parameters, cf. \Tab{tab:Para-Obsers}, are discussed in detail in \app{app:parameters}. In this section we focus on observables relevant for the chiral phase transition and the QCD phase diagram. We begin with the chiral condensate for the quark $q_i=u,d,s$, defined by
\begin{align}
    \Delta_{q_i}=  - m_{q_i}^0 T\sum_{n\in\mathbb{Z}} \int \frac{d^3 q}{(2 \pi)^3} \tr \,G_{q_i\bar q_i} (q)\,,\label{eq:chiralcondG}
\end{align}
up to a renormalization constant, where $G_{q_i\bar q_i}$ denotes the quark propagator and the trace is performed over the Dirac and color spaces. No summation over the flavor index $i$ is assumed. $m_{q_i}^0$ is the current quark mass. Due to the isospin symmetry, one has $\Delta_{l}=\Delta_{u}=\Delta_{d}$ for the light quarks. The renormalized chiral condensate reads
\begin{align}
    \Delta_{q_i,R} = \frac{1}{\mathcal{N}_R}\big[\Delta_{q_i}(T,\mu_B)-\Delta_{q_i}(0,0)\big]\,, \label{eq:chiralcondren}
\end{align}
where the vacuum part is subtracted and it is also made dimensionless with a normalization factor ${\mathcal N}_R\sim f_\pi^4$. In the fRG approach to QCD, the current quark mass is represented by the strength of explicit chiral symmetry breaking $c_\sigma$ as shown in \Eq{eq:QCDaction}, i.e., $m_l^0 \sim c_\sigma$. Consequently, one arrives at
\begin{align}
    \Delta_l =\frac{1}{2} m_l^0\frac{\partial \Omega[\Phi_{\mathrm{EoM}};T,\mu_B]}{\partial {m_l^0}}= \frac{1}{2} c_\sigma \frac{\partial \Omega[\Phi_{\mathrm{EoM}};T,\mu_B]}{ \partial {c_\sigma}}\,,\label{eq:mDer-cDer}
\end{align}
with the thermodynamic potential density
\begin{align}
   \Omega[\Phi_{\mathrm{EoM}};T,\mu_B]= \left(\frac{T}{V}\right) \,\Gamma_{k=0}[\Phi_{\mathrm{EoM}};T,\mu_B]\,,\label{}
\end{align}
i.e., the effective action at the RG scale $k=0$, where all the fields are evaluated at their expectation values $\Phi_{\mathrm{EoM}}$, that is, the solutions of equations of motion (EoM). Substituting \Eq{eq:QCDaction} into \Eq{eq:mDer-cDer}, one finds
\begin{align}
    \Delta_{l}(T,\mu_B) = -\frac{1}{2} c_\sigma \,\sigma_\mathrm{EoM}(T,\mu_B) \,, \label{eq:chiralcondSigma}
\end{align}
as well as 
\begin{align}
    \Delta_{l,R}(T,\mu_B) = -\frac{c_\sigma}{{2 \,\cal N}_{R}}\Big[ \sigma_\mathrm{EoM}(T,\mu_B)-\sigma_\mathrm{EoM}(0,0)\Big]\,. \label{eq:RenCondSigma}
\end{align}

Another related observable is the reduced condensate $\Delta_{l,s}$, which is defined as the weighted difference between the light and strange quark condensates, viz.,
\begin{align}
    \Delta_{l,s}(T,\mu_B) =&\frac{1}{\mathcal{N}_{l,s}}\left[ \Delta_{l}(T,\mu_B) -\left(\frac{m_l^0}{m_s^0}\right)^2\Delta_s(T,\mu_B)\right]\,.\label{eq:Deltals}
\end{align}
The normalization factor $\mathcal{N}_{l,s}$ is chosen as the value in the vacuum, such that one is left with
\begin{align}
    \Delta_{l,s}(T,\mu_B) = \frac{ \Delta_{l}(T,\mu_B) -\left(\frac{m_l^0}{m_s^0}\right)^2\Delta_s(T,\mu_B)}{\Delta_{l}(0,0) - \left(\frac{m_l^0}{m_s^0}\right)^2\Delta_s(0,0)}\,.\label{eq:chiralcondred}
\end{align}
The strange quark condensate is obtained similarly as \Eq{eq:mDer-cDer}:
\begin{align}
    \Delta_s(T,\mu_B)  &=m_s^0\frac{\partial \Omega[\Phi_{\mathrm{EoM}};T,\mu_B]}{\partial {m_s^0}}\nonumber \\[2ex]
    &= c_{\sigma_s}\frac{\partial \Omega[\Phi_{\mathrm{EoM}};T, \mu_B]}{ \partial {c_{\sigma_s}}}\nonumber \\[2ex]
    &= -\frac{1}{\sqrt{2} }c_{\sigma_s}\,\sigma_{s,\textrm{EoM}}(T,\mu_B)\,.\label{eq:CondSigma_s}
\end{align}
Therefore, one obtains
\begin{align}
    \Delta_{l,s}(T,\mu_B) = \frac{\left(\sigma -\sqrt{2}\frac{c_\sigma}{c_{\sigma_s}}\sigma_s\right)_{T,\mu_B}}{\left(\sigma - \sqrt{2}\frac{c_\sigma}{c_{\sigma_s}}\sigma_s\right)_{0,0}}\,, \label{eq:RedCondSigma}
\end{align}
where we have used
\begin{align}
    \frac{m_l^0}{m^0_s}= \frac{c_\sigma}{c_{\sigma_s}} \,. \label{eq:ctomApp}
\end{align}

We show the renormalized light quark condensate and the reduced condensate calculated in this work in \Fig{fig:deltlR} and \Fig{fig:deltlS}, respectively, which are also compared with the lattice results at vanishing $\mu_B$ \cite{Borsanyi:2010bp}. It is found that both our calculated $\Delta_{l,R}$ and $\Delta_{l,s}$ are consistent with the lattice results, except that the reduce condensate is a bit smaller that the lattice result in the region of high temperature. From the inflection point of the curves in the left panel of \Fig{fig:deltlR}, i.e., the peak location of the derivative of the chiral condensate with respect to the temperature as shown in the right panel of \Fig{fig:deltlR}, one is able to obtain the pseudocritical temperature for the continuous chiral crossover for different values of baryon chemical potential, $T_c(\mu_B)$, which provides the information of phase boundary of the chiral crossover. 

The phase boundary and the critical end point (CEP) obtained in this work are plotted in the phase diagram in \Fig{fig:phase-diagram}. To facilitate the comparison, in the same phase diagram we also show the phase boundary lines and CEP obtained from previous functional QCD calculations, including the fRG \cite{Fu:2019hdw} and Dyson-Schwinger equations (DSE) \cite{Gao:2020fbl, Gunkel:2021oya}, the phase boundary in the regime of low baryon chemical potential from lattice QCD \cite{HotQCD:2018pds, Borsanyi:2020fev}, chemical freeze-out data extracted from experiments \cite{STAR:2017sal, Alba:2014eba, Andronic:2017pug, Becattini:2016xct, Vovchenko:2015idt, Sagun:2017eye}. 

The dependence of the pseudocritical temperature of phase boundary on the baryon chemical potential can be expanded around $\mu_B=0$, to wit,
\begin{align}
    \frac{T_c(\mu_B)}{T_c}=1-\kappa_2 \left(\frac{\mu_B}{T_c}\right)^2-\kappa_4 \left(\frac{\mu_B}{T_c}\right)^4+\cdots\,,\label{eq:Tc-muB}
\end{align}
where only the even powers in $\mu_B$ appear due to the charge conjugate symmetry of QCD. The quadratic coefficient $\kappa_2$ represents the curvature of the phase boundary line. We employ the expansion above up to the order of $\mu_B^4$, to fit our calculated $T_c(\mu_B)$ in the ranges of $\mu_B/T \in [0, 3]$ and $[0, 4]$. The obtained coefficients reads
\begin{align}
    \kappa_2=0.0151(1)\,,\qquad \kappa_4=0.00023(2) \,, \label{eq:kappa}
\end{align}
Note that the phase boundary curvature $\kappa_2$ obtained here is a bit larger than $\kappa_2=0.0142(2)$ in \cite{Fu:2019hdw}, which indicates that the inclusion of Fierz-complete four-quark interactions renders the phase boundary less flat slightly. Note that $\kappa_2$ in \Eq{eq:kappa} is consistent within errors with the results from fRG \cite{Fu:2019hdw, Pawlowski:2025jpg, Fu:2026qnl}, DSE \cite{Gao:2020fbl, Gunkel:2021oya}, lattice QCD \cite{Borsanyi:2020fev, Ding:2024sux}. For more relevant discussions, see, e.g., \cite{Fischer:2026uni, Fu:2026qnl}.

Finally, we find that the CEP obtained in this work is located at
\begin{align}
    (T_{_\text{CEP}},\mu_{B_{\text{CEP}}})=(102, 647)\,\text{MeV}\,, \label{eq:CEP}
\end{align}
as shown by the black dot in the phase diagram in \Fig{fig:phase-diagram}. In comparison to the location of CEP, $(T_{_\text{CEP}},\mu_{B_{\text{CEP}}})=(107, 635)$ MeV, obtained with one single channel of four-quark vertex in \cite{Fu:2019hdw} as shown by the red pentagon in \Fig{fig:phase-diagram}, the temperature is a bit smaller and the baryon chemical potential increases slightly. Obviously, this is also consistent with previous estimate of location of CEP from functional QCD \cite{Fu:2019hdw, Gao:2020fbl, Gunkel:2021oya, Fu:2026qnl}, given that the errors arising from truncations of functional QCD computation might increase sizably in the regime of large baryon chemical potential, e.g., $\mu_B/T \gtrsim 4$.

\section{Conclusions and summary}
\label{sec:conclusion}

In this work, we have investigated the dynamics of Fierz-complete four-quark interactions within the fRG approach to QCD at finite temperature and densities, and its influence on the QCD phase diagram. This work could constitute one part of the analysis of systematics for the calculations of QCD phase diagram, from the first-principles QCD at finite temperature and densities within the fRG approach.

In the calculations we have implemented Fierz-complete four-quark interactions for the $u$, $d$ light quarks, which are comprised of ten different tensor channels, of which the sigma and pion channels are replaced with the exchanges of the sigma and pion mesons via the technique of dynamical hadronization. The Yukawa couplings for the sigma and pion are distinguished. 

We have studies the four-quark couplings, Yukawa couplings, and the strong couplings in details across the QCD phase diagram. It is found that in the vacuum the pion and sigma channels play the overwhelmingly dominant role, and all the other channels are negligible. However, when it is near the CEP, the magnitude of four-quark couplings in the channels $\eta$, $a$, $(S+ P)^{\mathrm{adj}}_-$, $(S- P)^{\mathrm{adj}}_-$, $(V-A)$, $(V-A)^{\mathrm{adj}}$, etc, increases sizably and they become more and more important.

In comparison to the single scalar-pseudoscalar channel of four-quark interactions, the dynamics of Fierz-complete four-quark interactions increase a bit the curvature of the phase boundary, and move the CEP to location of larger baryon chemical potential and smaller temperature. Note that the change of magnitude is within the errors arising from the truncations of the fRG approach to QCD at finite temperature and densities.

\section*{Acknowledgements}

We thank Jan M. Pawlowski and Fabian Rennecke for discussions and comments. We also would like to thank the members of the fQCD collaboration \cite{fQCD} for collaborations on related projects. This work is supported by the National Natural Science Foundation of China under Contract No.\ 12447102. Chuang Huang is supported by the Collaborative Research Centre SFB 1225-273811115 (ISOQUANT). Shi Yin is supported by the Alexander von Humboldt foundation.

\appendix 

\section{QCD at finite temperature and densities within the fRG approach}
\label{app:action-QCD}

The QCD effective action used in this work reads
\begin{align}
  &\quad\Gamma_{k}[\Phi]\nonumber\\[2ex]
&=\int_{x} \bigg\{\frac{1}{4}F^a_{\mu\nu}F^a_{\mu\nu}+Z_{c} \big(\partial_{\mu}\bar{c}^a\big)D_{\mu} ^{ab}c^b+\frac{1}{2\xi}\big(\partial_{\mu} A^a_{\mu}\big)^2\nonumber \\[2ex]
  &\quad+Z_{q}\bar{q}\big(\gamma_{\mu}D_{\mu}-\gamma_0 \hat \mu\big) q+m_s(\sigma_s)\bar{q}_s q_s +h_{\sigma}\bar{q}_l T^0\sigma q_l\nonumber  \\[2ex]
  &\quad+h_{\pi} \bar{q}_l i \gamma_5\bm{T}\cdot\bm{\pi}q_l +\frac{1}{2}Z_{\phi}(\partial_{\mu}\phi)^2+V_k(\rho,A_0)\nonumber\\[2ex]
  &\quad-c_\sigma\,\sigma -\frac{1}{\sqrt{2}}\, c_{\sigma_s}\, \sigma_s\bigg\}+\Delta\Gamma_{\mathrm{glue}}+\Gamma_{4q}\,, \label{eq:QCDaction}
\end{align}
with $\int_{x}=\int_0^{1/T}d x_0 \int d^3 x$ and the temperature $T$. This is a RG scale $k$-dependent action. The degrees of freedom (d.o.f) involved are collected in the notation of all fields $\Phi_a\,(a=A, c, \bar{c}, q, \bar{q}, \phi)$, i.e., the gluons, ghosts, anti-ghosts, quarks, anti-quarks, and the composite mesons $\phi_i\,(i=\sigma,\bm{\pi})$, respectively. 

The gluon field strength tensor reads
\begin{align}
    F^a_{\mu\nu}&=Z_{A}^{1/2}\big(\partial_{\mu}A^a_{\nu}-\partial_{\nu}A^a_{\mu}+Z_{A}^{1/2}\bar{g}_{\text{\tiny glue}} f^{abc}A^b_{\mu}A^c_{\nu}\big)\,,\label{eq:Fmunua}
\end{align}
with the gluon wave function $Z_{A}$. In the same way, $Z_{c}$, $Z_{q}$, $Z_{\phi}$ represent the wave functions for the respective fields, and the relevant anomalous dimensions read
\begin{align}
    \eta_{\Phi_a} &=-\frac{\partial_t Z_{\Phi_a}}{Z_{\Phi_a}}\,,  \label{eq:etaPhia}
\end{align}
with the RG time $t=\ln(k/\Lambda)$, where $\Lambda$ denotes the ultraviolet cutoff. The $\bar{g}_{\text{\tiny glue}}$ in \Eq{eq:Fmunua} denotes the renormalized strong coupling, which is given by
\begin{align}
    \bar{g}_{A^3} = \frac{\lambda_{A^3}}{Z_{A}^{3/2}}\,,\quad \bar{g}_{A^4} =\frac{\lambda_{A^4}^{1/2}}{Z_{A}}\,,\quad \bar{g}_{\bar c c A} =\frac{\lambda_{\bar c c A}}{Z_{A}^{1/2}\,Z_{c} }\,, \label{eq:strongcoupGluon}
\end{align}
in the three-gluon, four-gluon, and ghost-gluon vertices. Although they are identical to each other in the perturbative regime, they might deviate due to the emergence of a finite gluon mass gap in the nonperturbative region, see e.g., \cite{Mitter:2014wpa, Cyrol:2016tym, Cyrol:2017ewj, Fu:2025hcm} for more details. The $\lambda$'s in \Eq{eq:strongcoupGluon} are the bare dressing functions for the respective vertices. Moreover, one also finds for the renormalized quark-gluon coupling 
\begin{align}
    \bar{g}_{\bar q  q A} = \frac{\lambda_{\bar q  q A}}{Z_{A}^{1/2}\,Z_{q}}\,. \label{eq:gbqqa}
\end{align}
The covariant derivative in the adjoint and fundamental representations reads
\begin{align}
    D^{ab}_{\mu}&=\partial_{\mu} \delta^{ab}-Z_{A}^{1/2}\bar{g}_{\bar c c A} f^{abc}A^c_{\mu}\,, \label{eq:Dmuab} \\[2ex]
    D_{\mu}&=\partial_{\mu}-i Z_{A}^{1/2}\bar{g}_{\bar q  q A} A^a_{\mu} t^a\,, \label{eq:Dmu}
\end{align}
where one has the color $SU_c(3)$ Lie algebra
\begin{align}
    [t^a,t^b]=i f^{abc} t^c\,,
\end{align}
with the generators $t^a$ ($a=1,\,2,\cdots, 8$) and the structure constant $f^{abc}$. The Landau gauge $\xi=0$ is adopted in the computation. Note that non-classical contributions of the glue sector are collected in $\Delta\Gamma_{\mathrm{glue}}$, and see \cite{Fu:2019hdw} for more details.

The quark field $q=(q_l, q_s)$ includes the $u$ and $d$ light quarks $q_l=(q_u, q_d)$ and the strange quark $q_s$. As we have discussed in the main text, as the RG scale $k$ evolves into the regime of low energy, the chiral symmetry is broken dynamically, the collective d.o.f of bound states, e.g., the pions as the Goldstone bosons of the broken chiral symmetry, plays a dominant role in the low energy dynamics of QCD. The RG evolution provides a tailor-made approach to capture the continuous transition of degrees of freedom, from the fundamental partonic d.o.f in the ultraviolet to the d.o.f of bound states in the infrared. To be more specific, in the fRG the resonant dynamics of four-quark scatterings can be well described by the exchange of emergent bound states as well as their dynamics by themselves, which is known as the dynamical hadronization technique \cite{Gies:2001nw, Gies:2002hq, Pawlowski:2005xe}. Note that in the previous work \cite{Fu:2019hdw} the strange quark related order-parameter potential is obtained from the extension of $N_{f}=2$-flavor effective potential, from which the constituent strange quark mass $m_s(\sigma_s)$ is determined dynamically, see \cite{Fu:2019hdw} for more details. In this work we adopt the same approach for the sector of strange quark. This would be improved on in \cite{Fu:2026qnl}, where a comprehensive 2+1-flavor effective potential with the dynamics of scalar and pseudoscalar mesonic nonets is implemented. The $\hat \mu=\mathrm{diag}(\mu_u, \mu_d, \mu_s)$ in \Eq{eq:QCDaction} denotes the matrix of quark chemical potentials, and $\mu_u=\mu_d=\mu_s=\mu=\mu_B/3$ is used through this work, where $\mu_B$ is the baryon chemical potential.

In \Eq{eq:QCDaction} the light quarks interact with the sigma and pion mesons through the Yukawa couplings $h_{\sigma}$ and $h_{\pi}$, where different couplings are allowed by computing their respective flows, in comparison to the approximation $h_{\sigma}=h_{\pi}$ adopted in \cite{Fu:2019hdw}. Here $T^a$ ($a=1$, 2, 3) stands for the $SU_f(2)$ generators in the flavor space for the light quarks with $\Tr \,T^a T^b=(1/2)\delta^{ab}$ and $T^{0}=(1/2)\mathbb{1}_{2\times 2}$. There is a slight abuse of notations with $t^a$ and $T^a$, which are, however, the generators in the color and two-flavor flavor spaces, respectively, thus would not result in confusion. The RG invariant Yukawa couplings read
\begin{align}
    \bar h_{\pi} = \frac{h_{\pi}}{Z_{\phi}^{1/2}\,Z_{q}}\,, \qquad \bar h_{\sigma} = \frac{h_{\sigma}}{Z_{\phi}^{1/2}\,Z_{q}}\,. \label{eq:barh}
\end{align}

In \Eq{eq:QCDaction} the $V_k$ represents the effective potential, viz.,
\begin{align}
    V_k(\rho,A_0) &=V_{\mathrm{glue}}(A_0) + V_{\mathrm{mat}}(\rho,A_0)\,.  \label{eq:EffPot}
\end{align}
which is a sum of the glue potential $V_{\mathrm{glue}}$ and the mesonic effective potential $V_{\mathrm{mat}}$. $V_{\mathrm{glue}}$ is a functional of the temporal component of the gluon background field $A_0$ or the Polyakov loop $L[A_0]$, thus is also called the Polyakov loop potential. The mesonic effective potential is $O(4)$-invariant with $\rho=\phi^2/2$. The $c_\sigma$ and $c_{\sigma_s}$ in \Eq{eq:QCDaction} break the chiral symmetry explicitly for the light and strange sectors, respectively. Note that in \Eq{eq:QCDaction} all the wave functions, masses, couplings, etc., are RG scale $k$ dependent, and their dependence on $k$ is not shown explicitly for brevity. 

The renormalized, or RG invariant, pion and sigma mass squares read
\begin{align}
    \bar m^{2}_{\pi}=&\frac{1}{Z_{\phi}}\frac{\partial V_{\mathrm{mat}}}{\partial \rho}\,,\quad \bar m^{2}_{\sigma}=\frac{1}{Z_{\phi}}\left(\frac{\partial V_{\mathrm{mat}}}{\partial \rho}+2\rho \frac{\partial^2 V_{\mathrm{mat}}}{\partial \rho^2}\right)\,.\label{eq:m-pisig}
\end{align}
The light and strange quark masses read
\begin{align}
    \bar m_{l}&=\frac{1}{Z_{q}}\frac{h_\pi}{2} \sigma_l, \qquad \bar m_{s}=\frac{1}{Z_{q}}\frac{h_\pi}{\sqrt{2}} \sigma_s\,.\label{}
\end{align}
Note that the quark masses are related to the Yukawa coupling in the pseudo-scalar channel, and here we do not distinguish the Yukawa couplings in the light and strange sectors.

\section{Four-quark interactions of Fierz-complete tensors for the light quarks}
\label{app:action-4quark}

In \Eq{eq:QCDaction} the four-quark interactions for the $u$ and $d$ light quarks $q_l=(q_u, q_d)$ are given by \Eq{eq:Gamma4q}. The Fierz-complete four-quark basis set of light quarks, shown in \Eq{eq:set-alpha10tensors}, is comprised of ten tensor structures for different channels. The first four channels read
\begin{align} 
    &\quad\mathcal{T}^{(V-A)}_{ijmn} \bar {q_l}_i {q_l}_j \bar {q_l}_m {q_l}_n \nonumber \\[2ex]
    &=(\bar{q_l}\gamma_\mu T^0 q_l)^2-(\bar{q_l}i\gamma_\mu\gamma_5 T^0 q_l)^2\,,\\[2ex]
    &\quad\mathcal{T}^{(V+A)}_{ijmn} \bar {q_l}_i {q_l}_j \bar {q_l}_m {q_l}_n \nonumber \\[2ex]
    &=(\bar{q_l}\gamma_\mu T^0 q_l)^2+(\bar{q_l}i\gamma_\mu\gamma_5 T^0 q_l)^2\,,\\[2ex]
    &\quad\mathcal{T}^{(S-P)_{+}}_{ijmn} \bar {q_l}_i {q_l}_j \bar {q_l}_m {q_l}_n \nonumber \\[2ex]
    &=(\bar{q_l}\,T^0 q_l)^2-(\bar{q_l}\,\gamma_5 T^0 q_l)^2\nonumber \\[2ex]
    &\quad+(\bar{q_l}\,T^a q_l)^2-(\bar{q_l}\,\gamma_5 T^a q_l)^2\,,\label{eq:SmPp}\\[2ex]
    &\quad\mathcal{T}^{(V-A)^{\mathrm{adj}}}_{ijmn}\bar {q_l}_i {q_l}_j \bar {q_l}_m {q_l}_n\nonumber \\[2ex]
    &=(\bar{q_l}\gamma_\mu T^0 t^a q_l)^2-(\bar{q_l}i\gamma_\mu\gamma_5 T^0 t^a q_l)^2\,,
\end{align}
which are invariant under the flavor transformations of light quarks: $SU^f_{V}(2)$, $U^f_{V}(1)$, $SU^f_{A}(2)$, and $U^f_{A}(1)$. Here the superscript $f$ denotes the flavor space, on which the groups operate. The second set reads
\begin{align} 
    &\quad\mathcal{T}^{(S+P)_{-}}_{ijmn} \bar {q_l}_i {q_l}_j \bar {q_l}_m {q_l}_n \nonumber \\[2ex]
    &=(\bar{q_l}\,T^0 q_l)^2+(\bar{q_l}\,\gamma_5 T^0 q_l)^2\nonumber \\[2ex]
    &\quad-(\bar{q_l}\,T^a q_l)^2-(\bar{q_l}\,\gamma_5 T^a q_l)^2\,,\label{eq:SpPm}\\[2ex]
    &\quad\mathcal{T}^{(S+P)_{-}^{\mathrm{adj}}}_{ijmn} \bar {q_l}_i {q_l}_j \bar {q_l}_m {q_l}_n \nonumber \\[2ex]
    &=(\bar{q_l}\,T^0t^a q_l)^2+(\bar{q_l}\,\gamma_5 T^0t^a q_l)^2\\[2ex]
    &\quad-(\bar{q_l}\,T^a t^b q_l)^2-(\bar{q_l}\,\gamma_5 T^a t^b q_l)^2\,,
\end{align}
which are invariant under the transformations of $SU^f_{V}(2)$, $U^f_{V}(1)$, $SU^f_{A}(2)$, while break $U^f_{A}(1)$. The third set reads
\begin{align} 
    &\quad\mathcal{T}^{(S-P)_{-}}_{ijmn} \bar {q_l}_i {q_l}_j \bar {q_l}_m {q_l}_n \nonumber \\[2ex]
    &=(\bar{q_l}\,T^0 q_l)^2-(\bar{q_l}\,\gamma_5 T^0 q_l)^2\nonumber \\[2ex]
    &\quad-(\bar{q_l}\,T^a q_l)^2+(\bar{q_l}\,\gamma_5 T^a q_l)^2\,,\label{eq:SmPm}\\[2ex]
    &\quad\mathcal{T}^{(S-P)_{-}^{\mathrm{adj}}}_{ijmn} \bar {q_l}_i {q_l}_j \bar {q_l}_m {q_l}_n \nonumber \\[2ex]
    &=(\bar{q_l}\,T^0t^a q_l)^2-(\bar{q_l}\,\gamma_5 T^0t^a q_l)^2\nonumber \\[2ex]
    &\quad-(\bar{q_l}\,T^a t^b q_l)^2+(\bar{q_l}\,\gamma_5 T^a t^b q_l)^2\,,
\end{align}
which are invariant with respect to the $SU^f_{V}(2)$, $U^f_{V}(1)$, $U^f_{A}(1)$ transformations, while break $SU^f_{A}(2)$. The last set reads
\begin{align} 
    &\quad\mathcal{T}^{(S+P)_{+}}_{ijmn} \bar {q_l}_i {q_l}_j \bar {q_l}_m {q_l}_n \nonumber \\[2ex]
    &=(\bar{q_l}\,T^0 q_l)^2+(\bar{q_l}\,\gamma_5 T^0 q_l)^2\nonumber \\[2ex]
    &\quad+(\bar{q_l}\,T^a q_l)^2+(\bar{q_l}\,\gamma_5 T^a q_l)^2\,,\label{eq:SpPp}\\[2ex]
    &\quad\mathcal{T}^{(S+P)_{+}^{\mathrm{adj}}}_{ijmn} \bar {q_l}_i {q_l}_j \bar {q_l}_m {q_l}_n \nonumber \\[2ex]
    &=(\bar{q_l}\,T^0t^a q_l)^2+(\bar{q_l}\,\gamma_5 T^0 t^a q_l)^2\nonumber \\[2ex]
    &\quad+(\bar{q_l}\,T^a t^b q_l)^2+(\bar{q_l}\,\gamma_5 T^a t^b q_l)^2\,,
\end{align}
which are symmetric with respect to the transformations of $SU^f_{V}(2)$ and $U^f_{V}(1)$, while break $U^f_{A}(1)$ and $SU^f_{A}(2)$. It is more convenient to combine the $\mathcal{T}^{(S-P)_{+}}$ in \Eq{eq:SmPp},  $\mathcal{T}^{(S+P)_{-}}$ in \Eq{eq:SpPm}, $\mathcal{T}^{(S-P)_{-}}$ in \Eq{eq:SmPm}, $\mathcal{T}^{(S+P)_{+}}$ in \Eq{eq:SpPp} to form four new basis vectors as such 
\begin{align} 
    &\quad\mathcal{T}^{\sigma}_{ijmn} \bar {q_l}_i {q_l}_j \bar {q_l}_m {q_l}_n =(\bar{q_l}\,T^0 q_l)^2\,,\\[2ex]
    &\quad\mathcal{T}^{\pi}_{ijmn} \bar {q_l}_i {q_l}_j \bar {q_l}_m {q_l}_n =-(\bar{q_l}\,\gamma_5 T^a q_l)^2\,,\\[2ex]
    &\quad\mathcal{T}^{a}_{ijmn} \bar {q_l}_i {q_l}_j \bar {q_l}_m {q_l}_n =(\bar{q_l}\,T^a q_l)^2\,,\\[2ex]
    &\quad\mathcal{T}^{\eta}_{ijmn} \bar {q_l}_i {q_l}_j \bar {q_l}_m {q_l}_n =-(\bar{q_l}\,\gamma_5 T^0 q_l)^2\,,
\end{align}
which correspond to the scalar and pseudoscalar tensor channels of singlet and triplet, respectively.

\section{Flow equation of the effective action with dynamical hadronization}
\label{app:flow-action}

The flow equation of the effective action in \Eq{eq:QCDaction} reads \cite{Fu:2019hdw, Fu:2022gou}
\begin{align} 
    &\quad\partial_{t}\Gamma_{k}[\Phi]\nonumber\\[2ex]
    &=\frac{1}{2}\mathrm{STr} \big(G[\Phi]\,\partial_t R\big)+\mathrm{Tr}\bigg(G_{\phi\Phi_a}[\Phi] \frac{\delta
     \langle \partial_t \hat \phi \rangle }{\delta \Phi_a} \,R_{\phi}\bigg)\nonumber\\[2ex]
     &\quad-\int \langle   \partial_t \hat \phi_{i}\rangle\, \left( \frac{\delta \Gamma_{k}[\Phi]}{\delta \phi_i}+ c_\sigma \delta_{i\, \sigma}\right)\,, \label{eq:FlowQCD}
\end{align}
where the technique of dynamical hadronization is used for the emergent composite degrees of freedom for the mesons, see e.g., \cite{Gies:2001nw, Gies:2002hq, Pawlowski:2005xe} for more details. In \Eq{eq:FlowQCD} the propagator matrix in the field space reads
\begin{align}
    G[\Phi]= \frac{1}{\Gamma^{(2)}[\Phi]+R}\,, \label{eq:Gk}
\end{align}
with the matrix element
\begin{align}
    G_{\Phi_a \Phi_b}[\Phi]=\left( G[\Phi]\right)_{\Phi_a \Phi_b}\,. \label{}
\end{align}
The two-point correlation function in \Eq{eq:Gk} is given by
\begin{align}
    &\quad(\Gamma^{(2)}[\Phi])_{\Phi_a \Phi_b}\equiv\Gamma^{(2)}_{\Phi_a \Phi_b}[\Phi]\nonumber\\[2ex]
    &= (-1)_{\Phi_a \Phi_b}\frac{\delta^2}{\delta \Phi_a \delta \Phi_b}\Gamma_{k}[\Phi]\,, \label{}
\end{align}
where we have used the notation as follows
\begin{align}
    (-1)_{\Phi_a \Phi_b} \equiv \begin{cases}  -1 & \text{Both $\Phi_a$ and $\Phi_b$ are fermionic} \\ \phantom{-} 1 & \text{Otherwise} \end{cases}. \label{}
\end{align}
Here the $R_k$ in \Eq{eq:FlowQCD} and \Eq{eq:Gk} denotes the infrared regulator in the field space of $\Phi$, and $R_{k,\phi}$ in the subspace of meson fields $\phi$.

We choose the dynamical hadronization for the four-quark interactions of $\sigma$ and $\pi$ channels, that is
\begin{align}
    \langle  \partial_t \hat \phi_i \rangle =&\dot{A}_{\phi_i} \,\bar{q}\tau q\,, \label{eq:dtphi}
\end{align}
with $\tau=(T^0, i \gamma_5\bm{T})$ and $\phi_i\in(\sigma, \pi)$, where $\dot{A}_{\phi_i}$ represents the hadronization function. Substituting \Eq{eq:QCDaction} and \Eq{eq:dtphi} into \Eq{eq:FlowQCD}, and performing the projection onto the four-quark couplings of $\sigma$ and $\pi$ channels, one arrives at
\begin{align}
    \partial_{t} \tilde\lambda_{\phi_i}=2(1+\eta_{q})\tilde\lambda_{\phi_i}+{\mathrm{Flow}}^{(\phi_i)}_{\bar q q \bar q q}+\dot{\tilde A}_{\phi_i}\, \bar h_{\phi_i}\,, \label{eq:flow4q}
\end{align}
where we have adopted the RG invariant dimensionless four-quark couplings and hadronization function
\begin{align}
    \tilde\lambda_{\phi_i}= \frac{k^2\lambda_{\phi_i}}{Z_{q}^2} \,,\qquad
    \dot{\tilde A}_{\phi_i}= \frac{Z_{\phi}^{1/2}}{Z_{q}}k^2 \dot{A}_{\phi_i}\,. \label{eq:tilde-lamA}
\end{align}
Implementing the hadronization conditions for the $\sigma$ and $\pi$ tensor structures of four-quark scatterings
\begin{align}
    \tilde\lambda_{\phi_i}=0,\qquad \text{for}\quad \forall\, k, 
\end{align}
one immediately obtains from \Eq{eq:flow4q}
\begin{align}
    \dot{\tilde A}_{\phi_i}=&-\frac{1}{\bar h_{\phi_i}}{\mathrm{Flow}}^{(\phi_i)}_{\bar q q \bar q q} \,. \label{eq:dottildeA}
\end{align}

In the mean time, doing the projection of \Eq{eq:FlowQCD} onto the Yukawa couplings, one finds for the relevant flows
\begin{align}
    \partial_{t} \bar h_{\phi_i}&=\left(\frac{1}{2}\eta_{\phi}+\eta_{q}\right)\bar h_{\phi_i}+{\mathrm{Flow}}^{(\phi_i)}_{\bar{q} q \phi}-\tilde m^2_{\phi_i}\dot{\tilde A}_{\phi_i}\,, \label{eq:hphiflow}
\end{align}
where the flows of Yukawa couplings as shown in \Fig{fig:Yukawa-flow-eq} read
\begin{align}
    {\mathrm{Flow}}^{(\phi_i)}_{\bar{q} q \phi}& =-\frac{1}{Z_{\phi}^{1/2}Z_{q}}\Tr\Big[\partial_t \Gamma_{\bar{q}q\phi}^{(3)}\mathcal{P}_{\bar{q}q\phi}^{(\phi_i)}\Big]\,,\label{eq:flowqbarqphi}
\end{align}
with
\begin{align}
    \Gamma_{\bar{q}q\phi}^{(3)} \equiv -\frac{\delta}{\delta \phi}\frac{\delta}{\delta\bar{q}}\frac{\delta}{\delta q}\Gamma_{k}[\Phi]\,.\label{}
\end{align}
Here the projection operators for the $\sigma$ and $\pi$ are given by
\begin{align}
    \mathcal{P}_{\bar{q}q\phi}^{(\phi_i=\sigma)}&=-\frac{1}{2N_c}T^0\,,\\[2ex]
    \mathcal{P}_{\bar{q}q\phi}^{(\phi_i=\pi)}&=\frac{1}{6N_c}i\gamma_5\bm{T}\,,\label{}
\end{align}
respectively. In \Eq{eq:hphiflow} the dimensionless and RG invariant sigma and pion masses read
\begin{align}
    \tilde m^2_{\phi_i}=&\frac{m^2_{\phi_i}}{Z_{\phi} k^2}\,. \label{}
\end{align}

The quark-gluon flow in \Eq{eq:dtg} reads
\begin{align}
    {\mathrm{Flow}}_{\bar{q} q A}& =-\frac{1}{Z_{A}^{1/2}Z_{q}}\tr\Big[\left(\partial_t \Gamma_{\bar{q}qA}^{(3)}\right)^{a}_{\mu}\left(\mathcal{P}_{\bar{q}qA}^{(\text{classic})}\right)^{a}_{\mu}\Big]\,,\label{eq:flowqbarqA}
\end{align}
with
\begin{align}
    \left(\Gamma_{\bar{q}qA}^{(3)}\right)^{a}_{\mu} \equiv -\frac{\delta}{\delta A^a_{\mu}}\frac{\delta}{\delta\bar{q}}\frac{\delta}{\delta q}\Gamma_{k}[\Phi]\,,\label{}
\end{align}
where the quark-gluon vertex is projected onto the classical channel, i.e.,
\begin{align}
    \left(\mathcal{P}_{\bar{q}qA}^{(\text{classic})}\right)^{a}_{\mu}= -\frac{1}{8(N_c^2-1)} i \gamma_{\mu}t^{a}\,.\label{}
\end{align}
In this work we have neglected non-classical channels for the quark-gluon vertices, see, e.g., \cite{Mitter:2014wpa, Cyrol:2017ewj, Fu:2025hcm} for relevant discussions. Note that different from the trace $\Tr$ in \Eq{eq:flowqbarqphi}, the lowercase $\tr$ in \Eq{eq:flowqbarqA} indicates that the trace operates only over the Dirac and color spaces, not the flavor space.

\section{Numerical setup and parameters}
\label{app:parameters}

\begin{table}[t]
	\begin{center}
		\begin{tabular}{|c|c||c|}
			\hline & &  \\[-2ex]
			Vacuum Observables  & Value & Parameters in $\Gamma_\Lambda$  \\[1ex]
			\hline & &    \\[-2ex]
			$\bar m_{\pi}$ & \, 138 MeV \, & $c_{\sigma}$ = $3.6\,\mathrm{GeV}^{3}$ \, \\[1ex]
			\hline & &  \\[-2ex] 
			$\bar m_{\sigma}$ & 514 MeV & $c_{\sigma_s}$ = $97.2\,\mathrm{GeV}^{3}$ \\[1ex]
			\hline & &  \\[-2ex] 
			$\bar m_{l}$ & 355 MeV & $ \alpha_{s,\Lambda}$ = 0.235 \\[1ex]
			\hline & &  \\[-2ex] 
			$\bar m_{s}$ & 505 MeV &  \\[1ex]
			\hline
		\end{tabular}
		\caption{Parameters of the strength of explicit chiral symmetry breaking and the strong coupling at the initial ultraviolet cutoff, as well as some observables in the vacuum. The initial UV scale is chosen to be $\Lambda=20$ GeV.}
		\label{tab:Para-Obsers}
	\end{center}\vspace{-0.5cm}
\end{table}

In the numerical calculations we integrate the flow equations starting from an initial scale in the ultraviolet, which is chosen to be $\Lambda=20$ GeV. Quantum and thermal fluctuations are included after the RG scale is evolved toward the infrared. Since QCD is renormalizable and RG invariant, the final results of renormalized observables do not depend on the specific value of $\Lambda$, given that it is large enough such that the flows are integrated beginning from a scale deep in the perturbative regime. At the initial scale, the values of the strength of explicit chiral symmetry breaking, i.e., $c_\sigma$ and $c_{\sigma_s}$ in \Eq{eq:QCDaction}, and the strong coupling are presented in \Tab{tab:Para-Obsers}, see also \Fig{fig:flowalphas}. Note that in \Eq{eq:QCDaction} there is only a mesonic potential for the two-flavor light quarks. Following \cite{Fu:2019hdw} we use the two-flavor mesonic potential to approximate that for the strange quark as well, by employing the relation as follows
\begin{align}
    V_k(\rho,\rho_s) \approx    V_k(\rho) + \frac12 V_k(2 \rho_s)\,.\label{eq:SumofV}
\end{align}
with $\rho_s=\sigma_s^2/2$. Consequently, one is able to obtain several observables in the vacuum as shown in \Tab{tab:Para-Obsers}, which include the masses of the pion, sigma, constituent light and strange quark masses. The symbols with a bar represent that there are renormalized quantities.

As we have discussed in \sec{subsec:quark-gluon}, only the classical tensor structure of the quark-gluon vertex is included in the calculations. In order to compensate for the missing non-classical tensor structures, we use the same approach in \cite{Braun:2014ata, Fu:2019hdw} and introduce a phenomenological infrared enhancement for the quark-gluon coupling of classical channel. This is implemented via the replacement as follows
\begin{align}
    \partial_{t} \bar{g}_{\bar q q A} \rightarrow \bar{g}_{\bar q q A}\, \partial_{t} \varsigma_{a,b}(k)+\varsigma_{a,b}(k)  \partial_{t} \bar{g}_{\bar q q A}\,,\label{eq:IR-fct}
\end{align}
where the infrared enhancement function reads
\begin{align}
    \varsigma_{a,b}(k)&=1+a\frac{(k/b)^\delta}{\exp[ (k/b)^\delta]-1}\,. \label{eq:IR-enhance}
\end{align}
Obviously, one has $\varsigma_{a,b}(k) \to 1$ with $k\gtrsim b$ and $\varsigma_{a,b}(k) \to 1+a$ with $k \lesssim b$. Thus, this enhancement works only as the RG scale $k$ is below some scale value $b$, and the parameter $a$ determine the strength of the enhancement. In this work, we choose the same values of $b=$ 2 GeV and $\delta=2$ as those in \cite{Fu:2019hdw}. We find that the strength of enhancement $a = 0.013$ is sufficient to guarantee the dynamical chiral symmetry breaking, i.e., only 1.3\% larger quark-gluon coupling required, smaller than the value $a=0.034$ in \cite{Fu:2019hdw}. This is reasonable, since such difference is compensated by the Fierz-complete four quark interactions.

Furthermore, we also need a phenomenological glue potential $V_{\mathrm{glue}}(A_0)$ in \Eq{eq:EffPot}. In this work, we employ the same parametrization for the glue potential as in \cite{Fu:2019hdw}, which reads
\begin{align}
    \frac{V_\text{glue}(L,\bar{L})}{T^4}&= -\frac{a(T)}{2} \bar L L + b(T)\ln M_H(L,\bar{L}) \nonumber\\[2ex]
    &\quad + \frac{c(T)}{2} (L^3+\bar L^3) + d(T) (\bar{L} L)^2\,,\label{eq:polpot}
\end{align}
with the Polyakov loop $L(A_0)$ and its conjugate $\bar{L}(A_0)$. Thus, the glue potential is also called the Polyakov loop potential. The Haar measure in \Eq{eq:polpot} reads
\begin{align}
    M_H (L, \bar{L})&= 1 -6 \bar{L}L + 4 (L^3+\bar{L}^3) - 3  (\bar{L}L)^2\,.
\end{align}
The temperature dependence of coefficients $a,c,d$ is parameterized as 
\begin{align}
  x(T)&= \frac{x_1 + x_2/(t_{\text{\tiny{YM}}}+1) + x_3/(t_{\text{\tiny{YM}}}+1)^2}{1 + x_4/(t_{\text{\tiny{YM}}}+1) + x_5/(t_{\text{\tiny{YM}}}+1)^2}\,,\label{eq:xT}
\end{align}
for $x\in \{a,c,d\}$. That for the coefficient $b$ reads
\begin{align}
    b(T )&= b_1 (t_{\text{\tiny{YM}}}+1)^{-b_4}\left (1 -e^{b_2/(t_{\text{\tiny{YM}}}+1)^{b_3}} \right)\,.\label{eq:bT}
\end{align}
The constants in \ref{eq:xT} and \ref{eq:bT} are taken from \cite{Lo:2013hla}. The relation between the Polyakov loop potential in pure gauge theory and QCD is well captured by a linear rescaling for their respective reduced temperature \cite{Haas:2013qwp, Herbst:2013ufa}, i.e.,
\begin{align}
    t_{\text{\tiny{YM}}}&\rightarrow \alpha\,t_{\text{\tiny{glue}}} \quad \text{with}\quad t_{\text{\tiny{glue}}}= (T-T_c^\text{\tiny{glue}})/T_c^\text{\tiny{glue}},\label{}
\end{align}
where $\alpha=0.57$, $T_c^\text{\tiny{glue}}$=225 MeV are used in this work.

%
\begin{figure*}[t]
\includegraphics[width=0.45\textwidth]{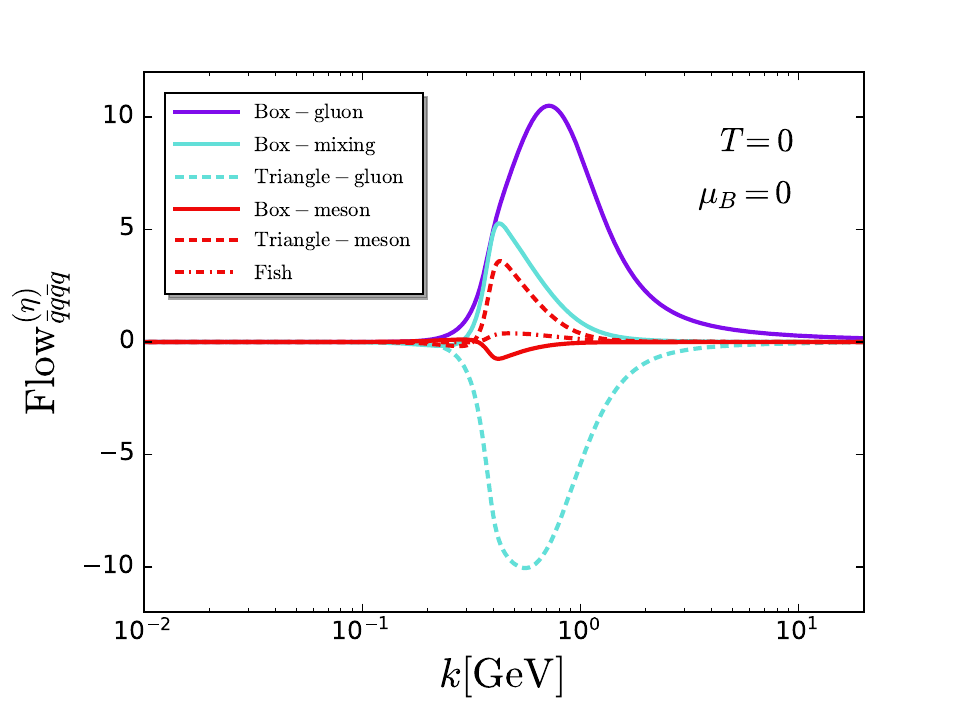}\hspace{0.3cm}
\includegraphics[width=0.45\textwidth]{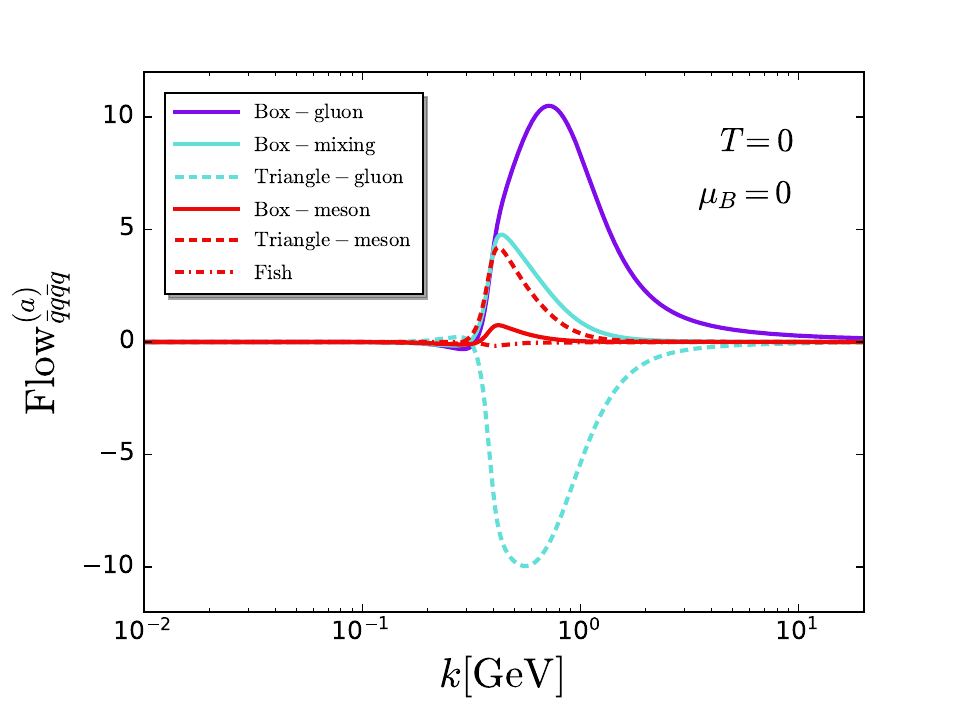}
\caption{Four-quark flows for the $\eta$ (left panel) and $a$ (right panel) channels as functions of the RG scale in the vacuum, where different lines in the legend correspond one by one to the six loop diagrams on the right side of the flow equation in \Fig{fig:4quark-eqn}.}\label{fig:flow4etaA}
\end{figure*}
%

%
\begin{figure*}[t]
\includegraphics[width=0.45\textwidth]{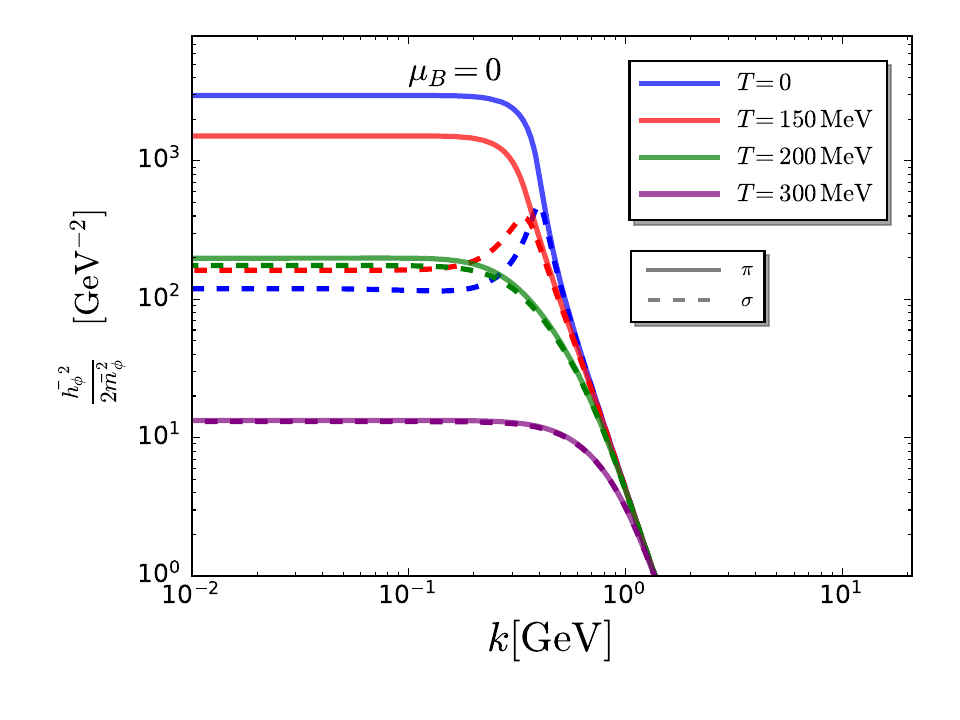}\hspace{0.3cm}
\includegraphics[width=0.45\textwidth]{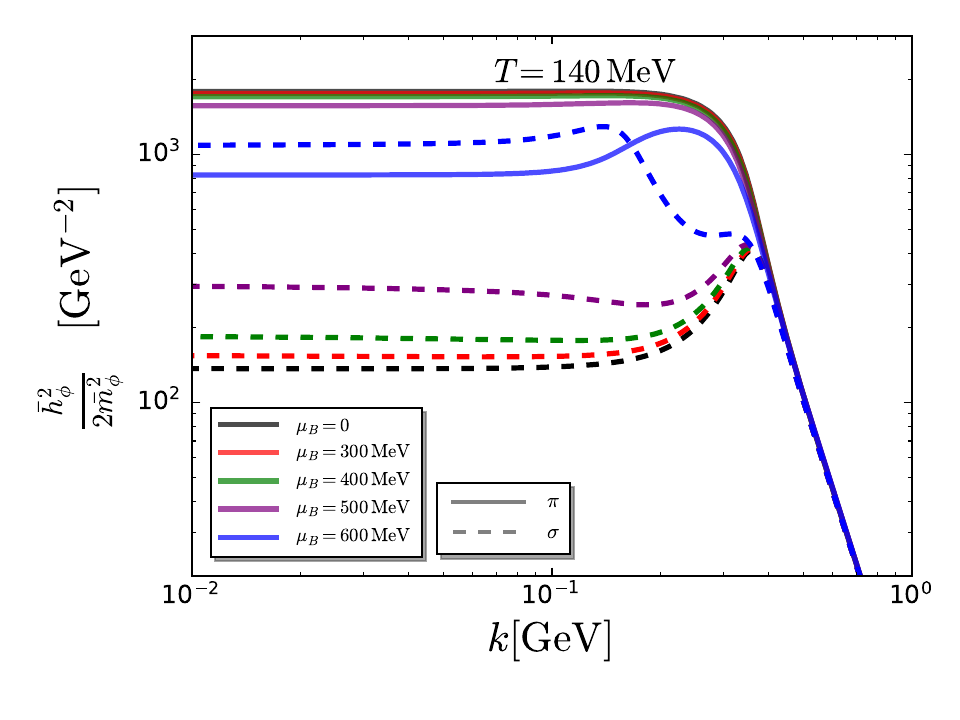}
\caption{Effective four-quark couplings of the pion and sigma channels as functions of the RG scale at different temperature and baryon chemical potentials. The left panel shows the temperature dependence at vanishing baryon chemical potential. The right panel show the dependence of $\mu_B$ at a fixed value of temperature $T=140$ MeV.}\label{fig:lam-TmuB-pisig}
\end{figure*}
%

%
\begin{figure*}[t]
\includegraphics[width=0.45\textwidth]{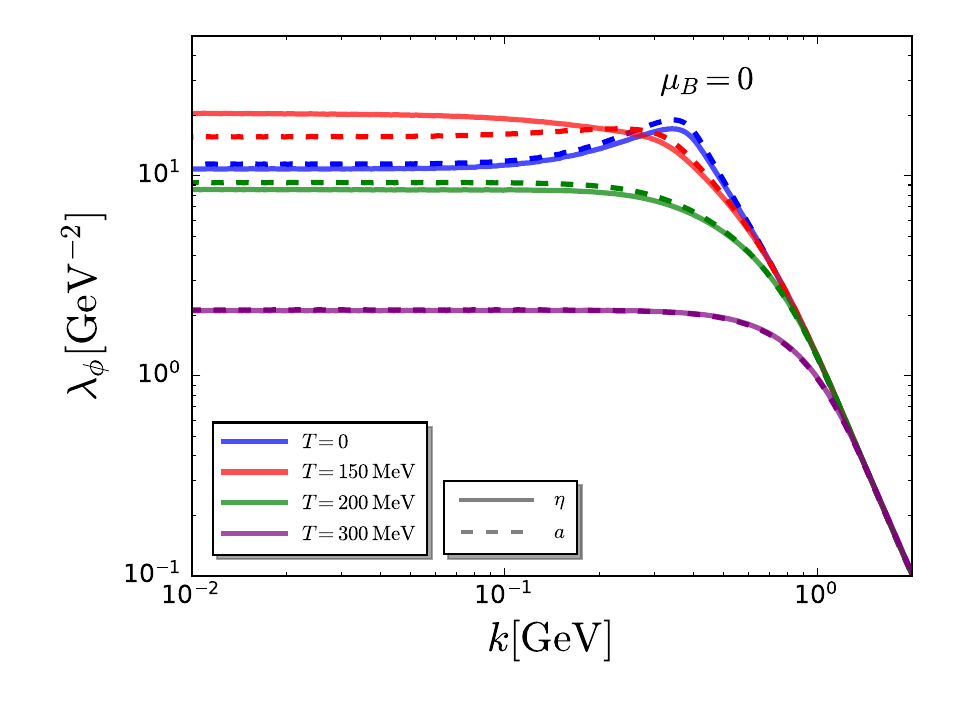}\hspace{0.3cm}
\includegraphics[width=0.45\textwidth]{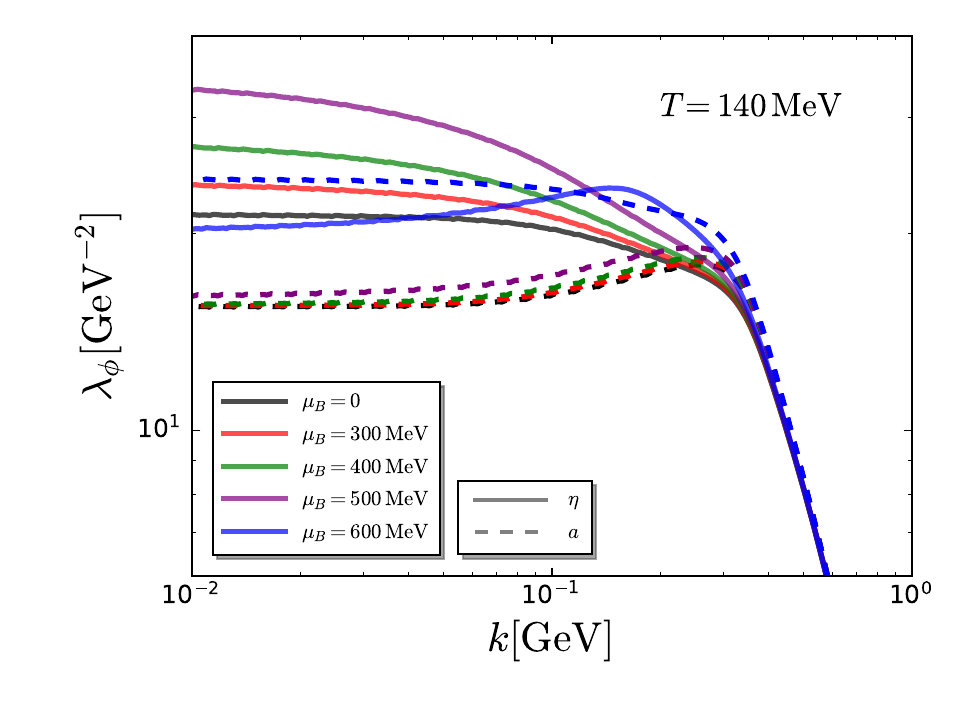}
\caption{Four-quark couplings of the $\eta$ and $a$ channels as functions of the RG scale at different temperature and baryon chemical potentials. The left panel shows the temperature dependence at vanishing baryon chemical potential. The right panel show the dependence of $\mu_B$ at a fixed value of temperature $T=140$ MeV.}\label{fig:lam-TmuB-etaA}
\end{figure*}
%

%
\begin{figure*}[t]
\includegraphics[width=0.45\textwidth]{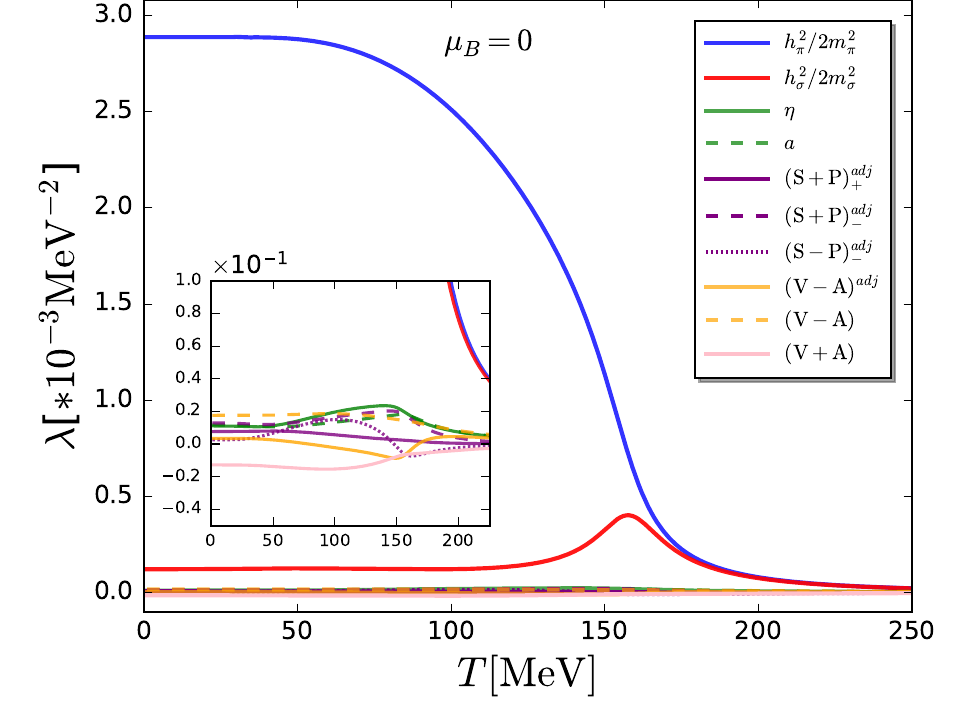}\hspace{0.3cm}
\includegraphics[width=0.45\textwidth]{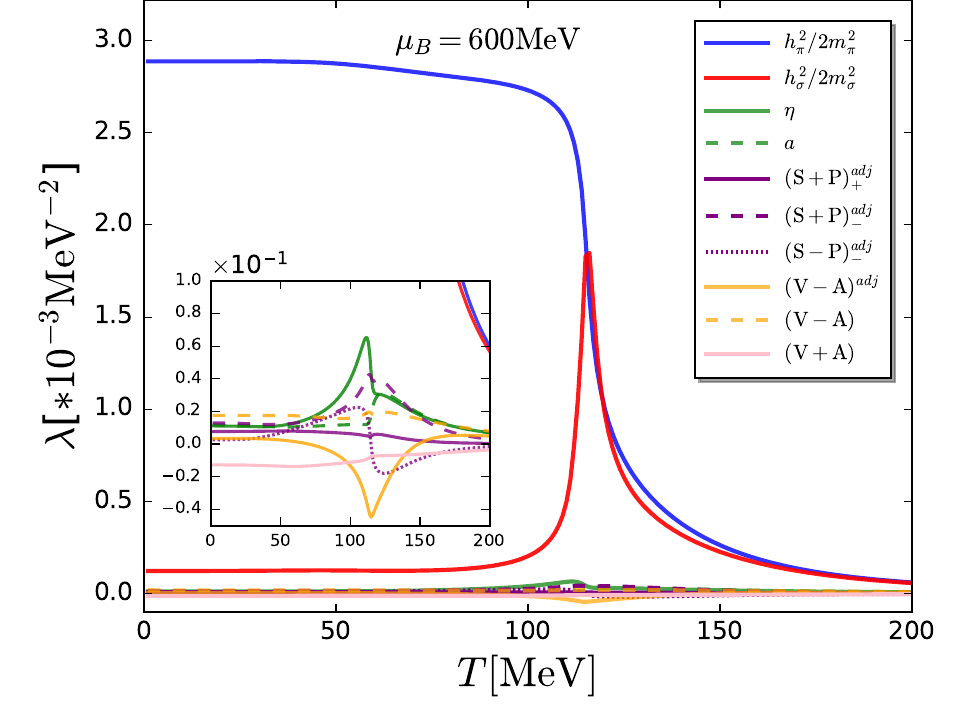}
\caption{Four-quark couplings of all Fierz-complete channels at $k=0$ as functions of the temperature with $\mu_B=0$ (left panel) and $\mu_B=570$ MeV (right panel). The inlays show the zoom-in view for the non-dominant channels.}\label{fig:lam-T}
\end{figure*}
%

%
\begin{figure*}[t]
\includegraphics[width=0.45\textwidth]{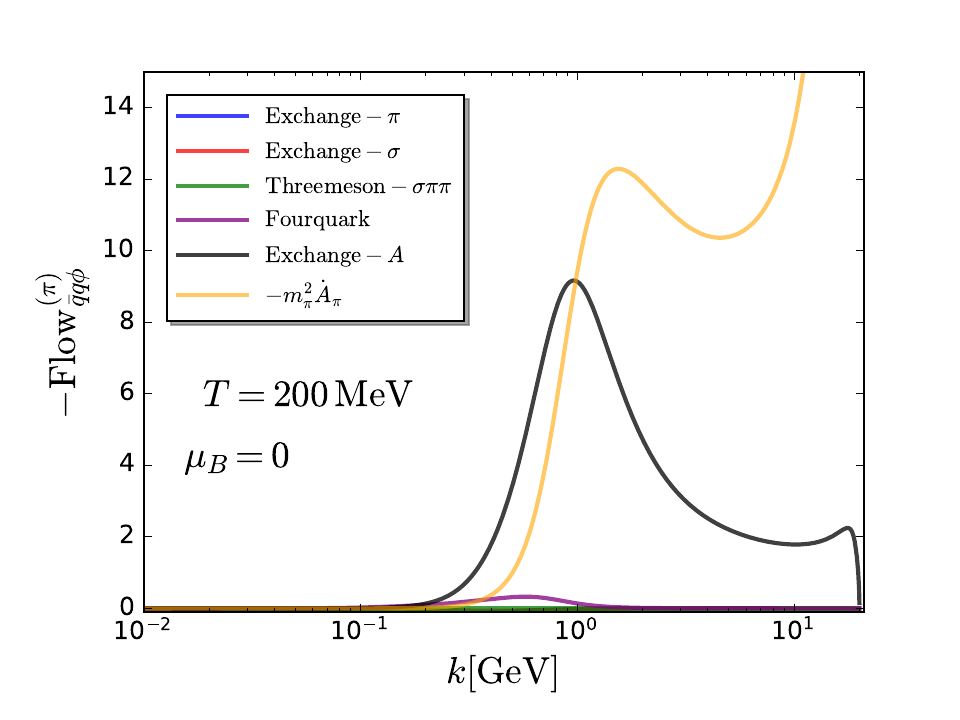}\hspace{0.3cm}
\includegraphics[width=0.45\textwidth]{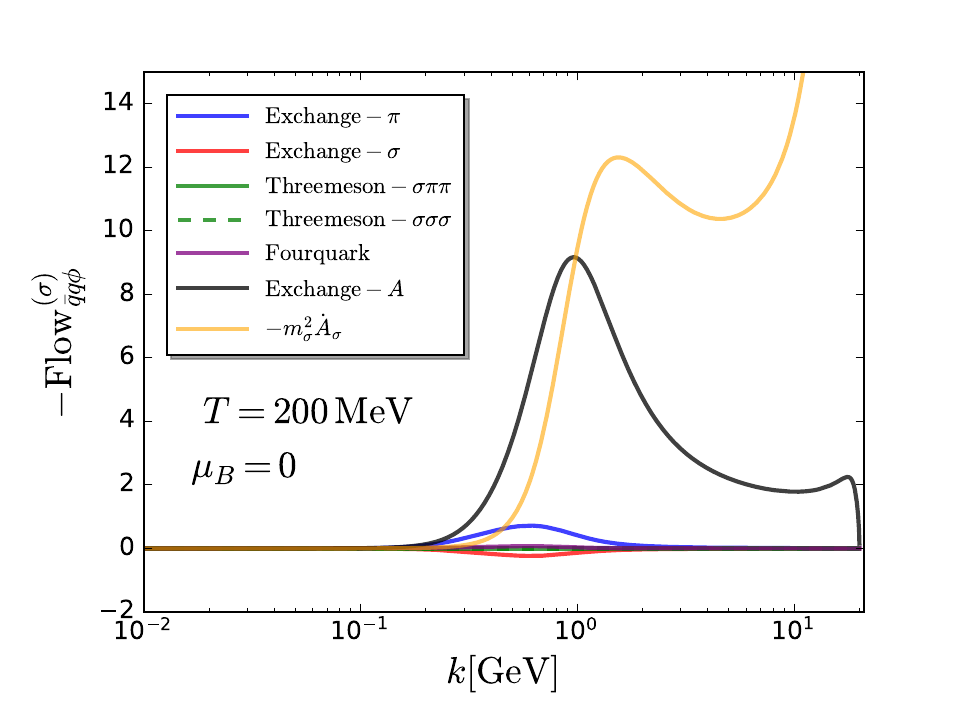}
\caption{Yukawa flows for the pion (left panel) and sigma (right panel) as functions of the RG scale with $T=200$ MeV and $\mu_B=0$. ``Exchange-A'' denotes the diagram of gluon exchange on the right side of the flow equations in \Fig{fig:Yukawa-flow-eq}, ``Exchange-$\pi/\sigma$'' the diagrams of $\pi/\sigma$ exchange, ``Three-meson-$\sigma\pi\pi/\sigma\sigma\sigma$'' the diagrams from the three-meson vertices, ``Four-quark'' the diagrams from the four-quark vertices. The lines of $\dot{ A}$ represent the contributions from the hadronization function in \Eq{eq:hphiflow}.}\label{fig:flowhpi-hsigmaT200}
\end{figure*}
%

%
\begin{figure*}[t]
\includegraphics[width=0.45\textwidth]{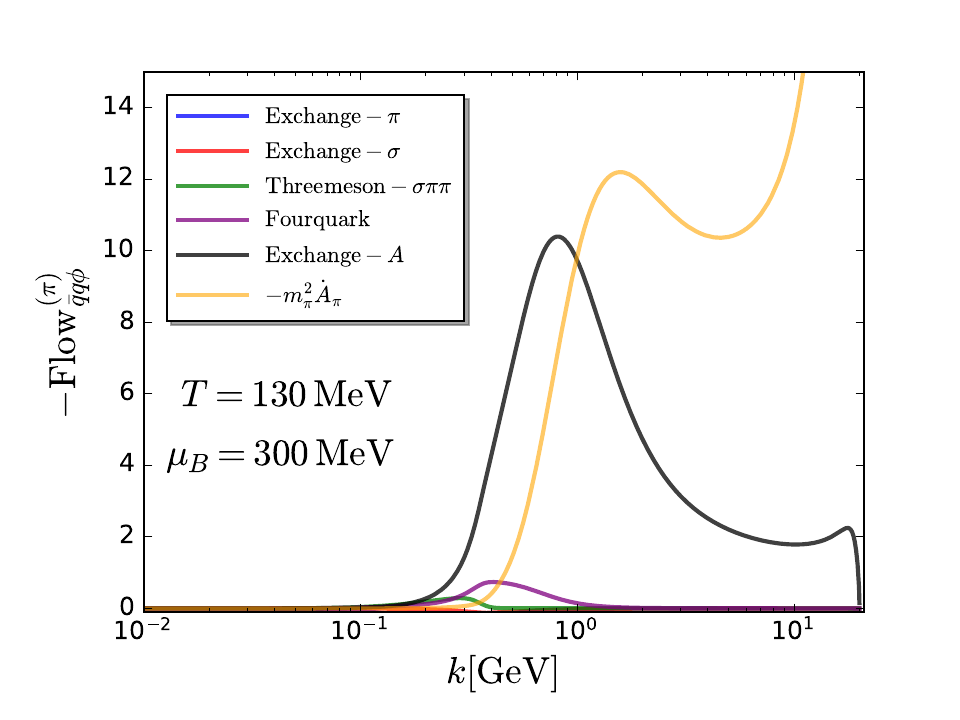}\hspace{0.3cm}
\includegraphics[width=0.45\textwidth]{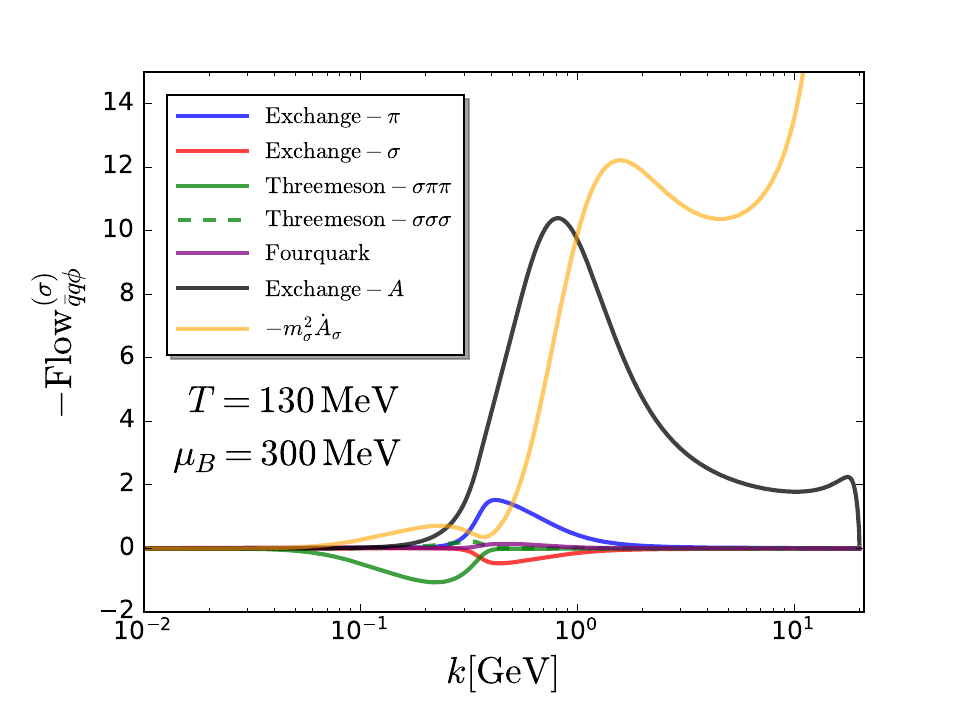}
\caption{Yukawa flows for the pion (left panel) and sigma (right panel) as functions of the RG scale with $T=130$ MeV and $\mu_B=300$ MeV. The labels of curves are the same as \Fig{fig:flowhpi-hsigmaT200}.}\label{fig:flowhpi-hsigmaT130muB300}
\end{figure*}
%

%
\begin{figure*}[t]
\includegraphics[width=0.45\textwidth]{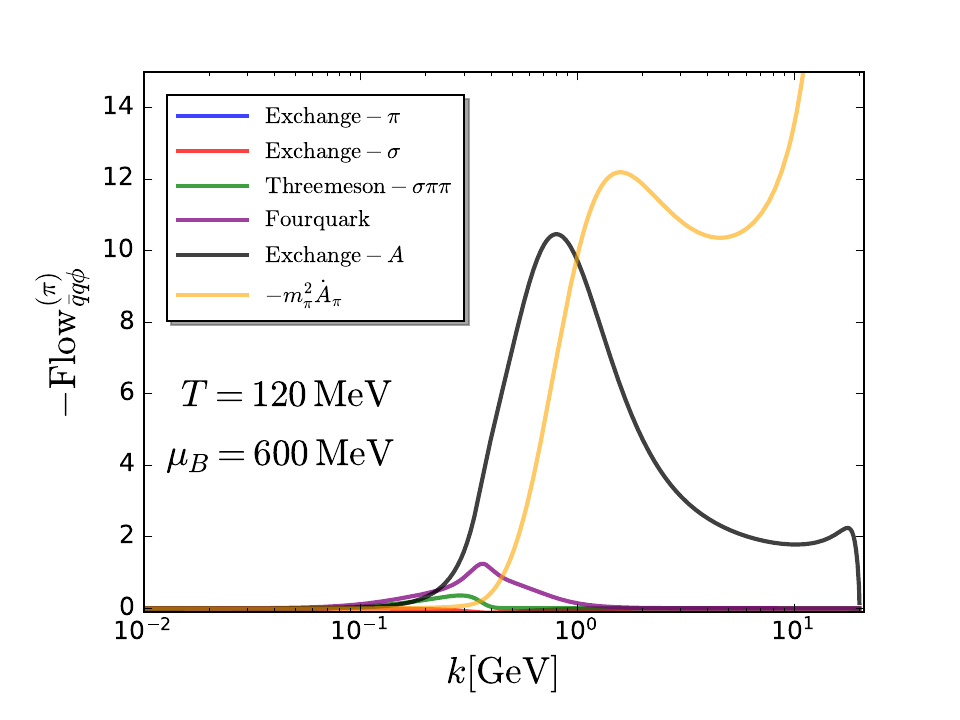}\hspace{0.3cm}
\includegraphics[width=0.45\textwidth]{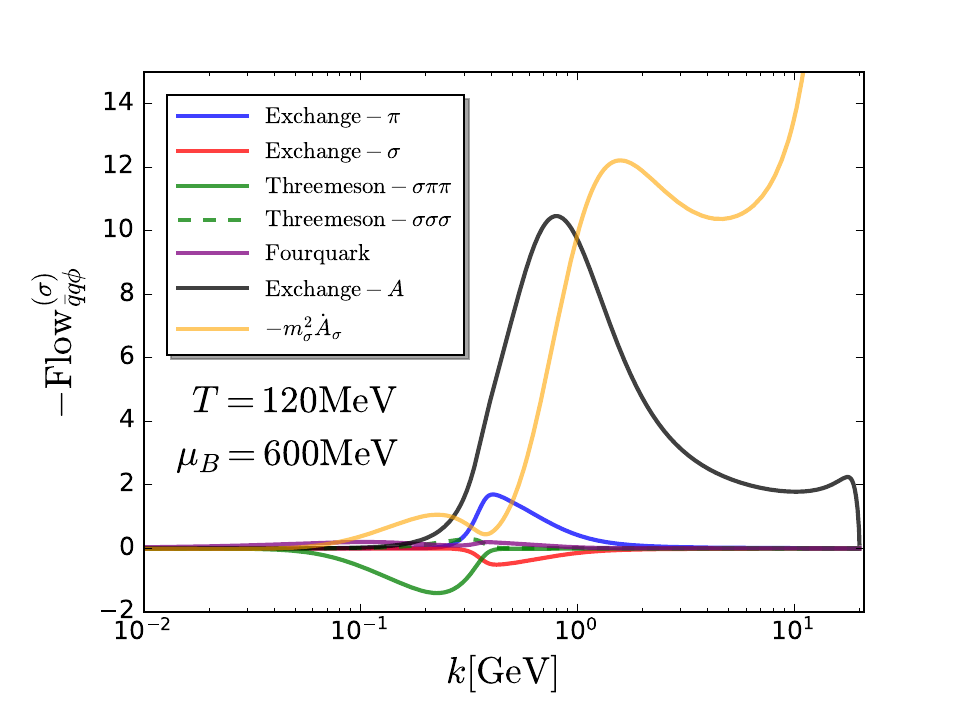}
\caption{Yukawa flows for the pion (left panel) and sigma (right panel) as functions of the RG scale with $T=120$ MeV and $\mu_B=600$ MeV. The labels of curves are the same as \Fig{fig:flowhpi-hsigmaT200}.}\label{fig:flowhpi-hsigmaT120muB600}
\end{figure*}
%

\section{Some further numerical results}
\label{app:further-nums}

\subsection{Four-quark couplings and flows}
\label{app:4quark-couplings}

In \Fig{fig:flow4etaA} we show the four-quark flows for the $\eta$ and $a$ channels in the vacuum. In comparison to the four-quark flows for the pion and sigma channels in \Fig{fig:flow4pi-sigma}, the triangle-gluon diagram here contributes a sizable minus value. The magnitude of box-meson diagram in the $\eta$ and $a$ channels is smaller than its magnitude in the pion and sigma channels. The dependence of the effective four-quark couplings in the pion and sigma channels on the temperature and baryon chemical potential are investigated in \Fig{fig:lam-TmuB-pisig}. It is found that the four-quark coupling of pion channel decreases monotonically with the increasing temperature or baryon chemical potential, while the four-quark coupling of sigma channel increases firstly and then decreases. In \Fig{fig:lam-TmuB-etaA} we show the dependence of the four-quark couplings in the $\eta$ and $a$ channels on the temperature and baryon chemical potential. The four-quark couplings of all Fierz-complete channels at $k=0$ are depicted in \Fig{fig:lam-T} as functions of the temperature with $\mu_B=0$ and $\mu_B=600$ MeV.

\subsection{Yukawa couplings and flows}
\label{app:Yukawa-couplings}

The Yukawa flows for the pion and sigma as functions of the RG scale at the three representative points in the QCD phase diagram as shown in \Fig{fig:schematic} are investigated in \Fig{fig:flowhpi-hsigmaT200}, \Fig{fig:flowhpi-hsigmaT130muB300}, and \Fig{fig:flowhpi-hsigmaT120muB600}, respectively.

\section{Flow equations of the Yukawa couplings}
\label{app:Yukawa-flows}

In this section, we provide the computational details of the flow equations for Yukawa couplings. For convenience, we first introduce notations for the propagators after all the indices have been traced over. The renormalized gluon propagator reads
\begin{align}
    G_A=\frac{1}{(q_0-p_0)^{2}+\bm{q}^{2}(1+r_{B}(\bm{q}^2/k^2))}\,, \label{eq:GA}
\end{align}
with the flat regulator for bosons
\begin{align}
    r_{B}(x)&= \Big(\frac{1}{x}-1\Big)\Theta(1-x)\,,\label{eq:regulatorOpt2}
\end{align}
where $\Theta(x)$ is the Heaviside step function. The pion and sigma propagators read
\begin{align}
    G_{\pi/\sigma}=\frac{1}{(q_0-p_0)^2+\bm{q}^{2}(1+r_{B}(\bm{q}^2/k^2))+\bar{m}_{\pi/\sigma}^2}\,. \label{eq:Gphi}
\end{align}
The renormalized quark propagator is given by
\begin{align}
    &\quad G_{q \bar q}(q)\nonumber\\[2ex]
    &=\frac{1}{i\big[q_0 \gamma_0+ (1+r_{F}(\bm{q}^2/k^2))\bm{q} \cdot \bm{\gamma}\big]+\bar m_q}\nonumber\\[2ex]
    &=\frac{-i\big[q_0 \gamma_0+ (1+r_{F}(\bm{q}^2/k^2))\bm{q} \cdot \bm{\gamma}\big]+\bar m_q}{q_0^2+ \bm{q}^2 (1+r_{F}(\bm{q}^2/k^2))^2+\bar m_q^2}\nonumber\\[2ex]
    &=G_{q}(q)\Big(-i\big[q_0 \gamma_0+ (1+r_{F}(\bm{q}^2/k^2))\bm{q} \cdot \bm{\gamma}\big]+\bar m_q\Big)\,.\label{eq:Gqbarq}
\end{align}
with 
\begin{align}
    G_{q}(q)\equiv \frac{1}{q_0^2+ \bm{q}^2 (1+r_{F}(\bm{q}^2/k^2))^2+\bar m_q^2}\,,
\end{align}
where the flat regulator for fermions reads
\begin{align}
    r_{F}(x)=\left(\frac{1}{\sqrt{x}}-1\right)\Theta(1-x)\,.\label{eq:rFopt}
\end{align}
Furthermore, it is useful to employ the shorthand notation as follows
\begin{align}
    q_{f}^2=q_0^2+ \bm{q}^2 (1+r_{F}(\bm{q}^2/k^2))^2\,.
\end{align}
Note that in \Eq{eq:GA}, \Eq{eq:Gphi} and \Eq{eq:Gqbarq}, $q=(q_0, \bm q)$ and $p=(p_0, \bm p)$ denote the momenta for the inner and external quark propagators, respectively. The Matsubara frequency summation is performed for the inner quark line, i.e.,
\begin{align}
    q_0=2\pi T \left(n+\frac{1}{2}\right)\,, \qquad n \in \mathbb{Z} \,.
\end{align}
while for the external propagators of both fermions and bosons, the momentum dependence are neglected, i.e., $\bm p=0$, and the lowest mode for the fermionic Matsubara frequency is adopted, that is
\begin{align}
    p_0=\pi T \,.
\end{align}

With these preparations, one is able to proceed with the flows of Yukawa couplings defined in \Eq{eq:flowqbarqphi}. One finds for the sigma 
\begin{align}
    &\quad \textrm{Flow}^{(\sigma)}_{\bar{q}q\phi}\nonumber\\[2ex]
    &=T\sum_{n}\int\frac{d^{3}q}{(2\pi)^3}\tilde{\partial}_t\Bigg[\frac{3}{2}G^2_\pi G_q \bar h^2_\pi \bar m_q S_{2\pi \sigma}  \nonumber\\[2ex]
    &\quad+\frac{3}{4}G_\pi G^2_q \bar h^2_\pi \bar h_\sigma(\bar m^2_q-q^2_f)-\frac{1}{4}G_\sigma G^2_q \bar h^3_\sigma (\bar m^2_q-q^2_f)\nonumber\\[2ex]
    &\quad+4G_A G^2_q \bar h_\sigma \bar g_{\bar{q}qA}^2(\bar m^2_q-q^2_f)-\frac{1}{2}G^2_\sigma G_q \bar h^2_\sigma \bar m_q S_{3\sigma}\nonumber\\[2ex]
    &\quad-\frac{1}{24}G^2_q \bar h_\sigma (\bar m^2_q-q^2_f) \sum_{i}a_i \bar \lambda_i\Bigg] \,.\label{eq:Flow-qqsig}
\end{align}
with 
\begin{align}
    \sum_{i} a_i \bar \lambda_i &=3 \bar \lambda_a +3 \bar\lambda_\eta -75 \bar\lambda_\pi-3 \bar\lambda_\sigma -16 \bar \lambda_{(S+P)_-^{\text{adj}}}\nonumber\\[2ex]
    &\quad +24 \bar \lambda_{(V+A)} \,,
\end{align}
and the three-meson vertices
\begin{align}
    S_{3\sigma}&=\sqrt{2 \bar\rho}V'(\bar \rho)+(2 \bar \rho)^{\frac{3}{2}}V^{(3)}(\bar \rho)\,,\\[2ex]
    S_{2\pi\sigma}&=\sqrt{2 \bar\rho}V''(\bar \rho)\,.
\end{align}
where we have simplified the notation of the matter sector of potential $V=V_{\mathrm{mat}}$ in \Eq{eq:EffPot} and used the renormalized $\bar\rho=Z_\phi \rho$. Note that $\tilde{\partial}_t$ in \Eq{eq:Flow-qqsig} only acts on the regulators in propagators. The flow of pion Yukawa coupling is given by
\begin{align}
    &\quad \textrm{Flow}^{(\pi)}_{\bar{q}q\phi}\nonumber\\[2ex]
    &=T\sum_{n}\int\frac{d^{3}q}{(2\pi)^3}\tilde{\partial}_t \Bigg[-G_\pi G_\sigma G_q \bar h_\pi \bar h_{\sigma}S_{2\pi \sigma}\nonumber\\[2ex]
    &\quad-\frac{1}{4}G_\sigma G_q^2 \bar h_\pi \bar h_\sigma^2 (\bar m^2_q+q^2_f)-\frac{1}{4}G_\pi G^2_q \bar h^3_\pi (\bar m^2_q+q^2_f)\nonumber\\[2ex]
    &\quad-4G_A G^2_q \bar g_{\bar{q}qA}^2 \bar h_\pi(\bar m^2_q+q^2_f)\nonumber\\[2ex]
    &\quad+\frac{1}{6}G^2_q \bar h_\pi (\bar m^2_q+q^2_f)\sum_{i} b_i \bar \lambda_i\Bigg] \,,\label{eq:Flow-qqpion}
\end{align}
with
\begin{align}
    \sum_{i} b_i\lambda_i &=(9\lambda_a -3\lambda_\eta +9\lambda_\pi-69\lambda_\sigma -16\lambda_{(S+P)_-^{\text{adj}}}\nonumber\\[2ex]
    &\quad+32\lambda_{(S+P)_+^{\text{adj}}}+24\lambda_{(V+A)}) \,.
\end{align}

\section{Flow equations of the four-quark interactions}
\label{app:flows-4quark}

As shown in \Eq{eq:Flow4q}, the four-quark couplings of different tensor structures in the Fierz-complete basis can be obtained via the corresponding projection as follow,
\begin{align}
    \lambda_{\alpha} =-\Tr\Big[\Gamma^{(4)}_{\bar{q}q\bar{q}q}\,\mathcal{P}_{\bar{q}q\bar{q}q}^{(\alpha)}\Big]\,.\label{}
\end{align}
In the following we present contributions of different diagrams in \Fig{fig:4quark-eqn} to the four-quark flows. We begin with the ``Box-gluon'' diagram
\begin{align}
    &\quad \left({\mathrm{Flow}}^{(\sigma)}_{\bar q q \bar q q}\right)_{\text{Box-gluon}}\nonumber\\[2ex]
    &=-T\sum_{n}\int\frac{d^{3}q}{(2\pi)^3}\tilde{\partial}_t \bigg[\frac{1}{16}\bar g_{\bar q qA}^4 G_A^2G_q^2(18 \bar m_q^2+19 q_f^2)\bigg]\,,\label{}
\end{align}
for the $\sigma$ channel. One finds for the $\pi$ and $\eta$ channels
\begin{align}
    &\quad\left({\mathrm{Flow}}^{(\pi)}_{\bar q q \bar q q}\right)_{\text{Box-gluon}}=\left({\mathrm{Flow}}^{(\eta)}_{\bar q q \bar q q}\right)_{\text{Box-gluon}}\nonumber\\[2ex]
    &=\left({\mathrm{Flow}}^{(\sigma)}_{\bar q q \bar q q}\right)_{\text{Box-gluon}}\,.\label{}
\end{align}
The $a$ channel reads
\begin{align}
    &\quad \left({\mathrm{Flow}}^{(a)}_{\bar q q \bar q q}\right)_{\text{Box-gluon}}\nonumber\\[2ex]
    &=-T\sum_{n}\int\frac{d^{3}q}{(2\pi)^3}\tilde{\partial}_t\bigg[\frac{1}{48}\bar g_{\bar q qA}^4 G_A^2G_q^2(-74 \bar m_q^2+57 q_f^2)\bigg]\,.\label{}
\end{align}
The $(V+A)$ channel reads
\begin{align}
    &\quad \left({\mathrm{Flow}}^{(V+A)}_{\bar q q \bar q q}\right)_{\text{Box-gluon}}\nonumber\\[2ex]
    &=T\sum_{n}\int\frac{d^{3}q}{(2\pi)^3}\tilde{\partial}_t\bigg[\frac{1}{48}\bar g_{\bar q qA}^4 G_A^2 G_q^2(30 \bar m_q^2+13 q_f^2)\bigg]\,.\label{}
\end{align}
The $(V-A)$ channel reads
\begin{align}
    &\quad \left({\mathrm{Flow}}^{(V-A)}_{\bar q q \bar q q}\right)_{\text{Box-gluon}}\nonumber\\[2ex]
    &=-T\sum_{n}\int\frac{d^{3}q}{(2\pi)^3}\tilde{\partial}_t\bigg[\frac{2}{3}\bar g_{\bar q qA}^4 G_A^2 G_q^2 q_f^2\bigg]\,.\label{}
\end{align}
The $(V-A)^{\text{adj}}$ channel reads
\begin{align}
    &\quad \left({\mathrm{Flow}}^{(V-A)^{\text{adj}}}_{\bar q q \bar q q}\right)_{\text{Box-gluon}}\nonumber\\[2ex]
    &=-T\sum_{n}\int\frac{d^{3}q}{(2\pi)^3}\tilde{\partial}_t\bigg[\frac{1}{8}\bar g_{\bar q qA}^4 G_A^2G_q^2 q_f^2\bigg]\,.\label{}
\end{align}
The $(S+P)^{\text{adj}}_+$ channel reads
\begin{align}
    &\quad \left({\mathrm{Flow}}^{(S+P)^{\text{adj}}_+}_{\bar q q \bar q q}\right)_{\text{Box-gluon}}\nonumber\\[2ex]
    &=-T\sum_{n}\int\frac{d^{3}q}{(2\pi)^3}\tilde{\partial}_t\bigg[\frac{3}{4}\bar g_{\bar q qA}^4 G_A^2G_q^2 \bar m_q^2\bigg]\,.\label{}
\end{align}
The $(S+P)^{\text{adj}}_-$ channel reads
\begin{align}
    &\quad \left({\mathrm{Flow}}^{(S+P)^{\text{adj}}_-}_{\bar q q \bar q q}\right)_{\text{Box-gluon}}\nonumber\\[2ex]
    &=T\sum_{n}\int\frac{d^{3}q}{(2\pi)^3}\tilde{\partial}_t\bigg[\frac{1}{4}\bar g_{\bar q qA}^4 G_A^2G_q^2 \bar m_q^2\bigg]\,.\label{}
\end{align}
The $(S-P)^{\text{adj}}_-$ channel reads
\begin{align}
    &\quad \left({\mathrm{Flow}}^{(S-P)^{\text{adj}}_-}_{\bar q q \bar q q}\right)_{\text{Box-gluon}}\nonumber\\[2ex]
    &=-T\sum_{n}\int\frac{d^{3}q}{(2\pi)^3}\tilde{\partial}_t \bigg[\frac{5}{4}\bar g_{\bar q qA}^4 G_A^2G_q^2 \bar m_q^2\bigg]\,.\label{}
\end{align}

We move on to the ``Box-mixing'' diagram in \Fig{fig:4quark-eqn}. The $\sigma$ channel reads
\begin{align}
    &\quad \left({\mathrm{Flow}}^{(\sigma)}_{\bar q q \bar q q}\right)_{\text{Box-mixing}}\nonumber\\[2ex]
    &=-T\sum_{n}\int\frac{d^{3}q}{(2\pi)^3}\tilde{\partial}_t\bigg[\frac{1}{72}\bar g_{\bar q qA}^2 G_AG_q^2\Big(96 G_\pi \bar h_\pi^2 q_f^2\nonumber\\[2ex]
    &\quad+G_{\sigma}\bar h_\sigma^2(27 \bar m_q^2-32q_f^2)\Big)\bigg]\,.\label{}
\end{align}
The $\pi$ channel reads
\begin{align}
    &\quad \left({\mathrm{Flow}}^{(\pi)}_{\bar q q \bar q q}\right)_{\text{Box-mixing}}\nonumber\\[2ex]
    &=-T\sum_{n}\int\frac{d^{3}q}{(2\pi)^3}\tilde{\partial}_t\bigg[\frac{1}{72}\bar g_{\bar q qA}^2 G_A G_q^2\Big(32G_\pi \bar h_\pi^2 q_f^2\nonumber\\[2ex]
    &\quad+G_{\sigma} \bar h_\sigma^2(27 \bar m_q^2+32q_f^2)\Big)\bigg]\,.\label{}
\end{align}
The $\eta$ channel reads
\begin{align}
    &\quad \left({\mathrm{Flow}}^{(\eta)}_{\bar q q \bar q q}\right)_{\text{Box-mixing}}\nonumber\\[2ex]
    &=T\sum_{n}\int\frac{d^{3}q}{(2\pi)^3}\tilde{\partial}_t\bigg[\frac{1}{72}\bar g_{\bar q qA}^2 G_AG_q^2\Big(-96G_\pi \bar h_\pi^2 q_f^2\nonumber\\[2ex]
    &\quad+G_{\sigma}\bar h_\sigma^2(27 \bar m_q^2+32q_f^2)\Big)\bigg]\,.\label{}
\end{align}
The $a$ channel reads
\begin{align}
    &\quad \left({\mathrm{Flow}}^{(a)}_{\bar q q \bar q q}\right)_{\text{Box-mixing}}\nonumber\\[2ex]
    &=T\sum_{n}\int\frac{d^{3}q}{(2\pi)^3}\tilde{\partial}_t\bigg[\frac{1}{72}\bar g_{\bar q qA}^2 G_AG_q^2\Big(-32 G_\pi \bar h_\pi^2 q_f^2\nonumber\\[2ex]
    &\quad+G_{\sigma}\bar h_\sigma^2(27 \bar m_q^2-32q_f^2)\Big)\bigg]\,.\label{}
\end{align}
The $(V+A)$ channel reads
\begin{align}
    &\quad \left({\mathrm{Flow}}^{(V+A)}_{\bar q q \bar q q}\right)_{\text{Box-mixing}}\nonumber\\[2ex]
    &=T\sum_{n}\int\frac{d^{3}q}{(2\pi)^3}\tilde{\partial}_t\bigg[\frac{1}{8}\bar g_{\bar q qA}^2 G_A G_q^2 G_{\sigma}\bar h_\sigma^2 \bar m_q^2\bigg]\,.\label{}
\end{align}
The $(V-A)^{\text{adj}}$ channel reads
\begin{align}
    &\quad \left({\mathrm{Flow}}^{(V-A)^{\text{adj}}}_{\bar q q \bar q q}\right)_{\text{Box-mixing}}\nonumber\\[2ex]
    &=-T\sum_{n}\int\frac{d^{3}q}{(2\pi)^3}\tilde{\partial}_t\bigg[\frac{3}{4}\bar g_{\bar q qA}^2 G_A G_q^2 G_{\sigma}\bar h_\sigma^2 \bar m_q^2\bigg]\,.\label{}
\end{align}
The $(S+P)^{\text{adj}}_{-}$ channel reads
\begin{align}
    &\quad \left({\mathrm{Flow}}^{(S+P)^{\text{adj}}_{-}}_{\bar q q \bar q q}\right)_{\text{Box-mixing}}\nonumber\\[2ex]
    &=-T\sum_{n}\int\frac{d^{3}q}{(2\pi)^3}\tilde{\partial}_t\bigg[\frac{1}{6}\bar g_{\bar q qA}^2 G_AG_q^2\Big(7G_\pi \bar h_\pi^2 q_f^2\nonumber\\[2ex]
    &\quad+3G_{\sigma} \bar h_\sigma^2q_f^2\Big)\bigg]\,.\label{}
\end{align}
The $(S+P)^{\text{adj}}_{+}$ channel reads
\begin{align}
    &\quad \left({\mathrm{Flow}}^{(S+P)^{\text{adj}}_{+}}_{\bar q q \bar q q}\right)_{\text{Box-mixing}}\nonumber\\[2ex]
    &=T\sum_{n}\int\frac{d^{3}q}{(2\pi)^3}\tilde{\partial}_t\bigg[\frac{1}{6}\bar g_{\bar q qA}^2 G_AG_q^2\Big(G_\pi \bar h_\pi^2 q_f^2\nonumber\\[2ex]
    &\quad-G_{\sigma}\bar h_\sigma^2 q_f^2\Big)\bigg]\,.\label{}
\end{align}
The other two channels of the gluon-meson mixing diagram are vanishing, i.e.,
\begin{align}
    \left({\mathrm{Flow}}^{(V-A)}_{\bar q q \bar q q}\right)_{\text{Box-mixing}}=\left({\mathrm{Flow}}^{(S-P)^{\text{adj}}_{-}}_{\bar q q \bar q q}\right)_{\text{Box-mixing}}=0\,.\label{}
\end{align}

We proceed with the ``Box-meson'' diagram in \Fig{fig:4quark-eqn}. The $\sigma$ channel reads
\begin{align}
    &\quad \left({\mathrm{Flow}}^{(\sigma)}_{\bar q q \bar q q}\right)_{\text{Box-meson}}\nonumber\\[2ex]
    &=-T\sum_{n}\int\frac{d^{3}q}{(2\pi)^3}\tilde{\partial}_t\bigg[\frac{1}{24}\Big(G_q^2G_\pi^2\bar h_\pi^4q_f^2-G_q^2G_\pi G_\sigma \bar h_\pi^2 \bar h_\sigma^2q_f^2\nonumber\\[2ex]
    &\quad+18G_q^2G_\pi^2  \bar h_\pi^4  \bar m_q^2+6G_\sigma^2\bar h_\sigma^4G_q^2 \bar m_q^2\Big)\bigg]\,.\label{eq:flow4q-box-meson}
\end{align}
The $\pi$ channel reads
\begin{align}
    &\quad \left({\mathrm{Flow}}^{(\pi)}_{\bar q q \bar q q}\right)_{\text{Box-meson}}\nonumber\\[2ex]
    &=T\sum_{n}\int\frac{d^{3}q}{(2\pi)^3}\tilde{\partial}_t\bigg[\frac{1}{24}\Big(G_q^2G_\pi^2 \bar h_\pi^4q_f^2-G_q^2G_\pi G_\sigma \bar h_\pi^2 \bar h_\sigma^2 q_f^2\nonumber\\[2ex]
    &\quad-12G_q^2G_\pi G_\sigma \bar h_\pi^2 \bar h_\sigma^2 \bar m_q^2\Big)\bigg]\,.\label{eq:flow4q-box-meson-pi}
\end{align}
It is found that the quark-meson box diagram with two sigma exchanges only contributes to the sigma channel of four-quark vertices in \Eq{eq:flow4q-box-meson}, and it is vanishing for other channels. We proceed to the $\eta$ channel of the quark-meson box diagram:
\begin{align}
    &\quad \left({\mathrm{Flow}}^{(\eta)}_{\bar q q \bar q q}\right)_{\text{Box-meson}}\nonumber\\[2ex]
    &=-T\sum_{n}\int\frac{d^{3}q}{(2\pi)^3}\tilde{\partial}_t\bigg[\frac{1}{24}\Big(G_q^2G_\pi^2 \bar h_\pi^4 q_f^2\nonumber\\[2ex]
    &\quad-G_q^2G_\pi G_\sigma \bar h_\pi^2 \bar h_\sigma^2 q_f^2\Big)\bigg]\,.\label{}
\end{align}
The $a$ channel reads
\begin{align}
    &\quad \left({\mathrm{Flow}}^{(a)}_{\bar q q \bar q q}\right)_{\text{Box-meson}}\nonumber\\[2ex]
    &=T\sum_{n}\int\frac{d^{3}q}{(2\pi)^3}\tilde{\partial}_t\bigg[\frac{1}{24}\Big(G_q^2 G_\pi^2 \bar h_\pi^4 q_f^2\nonumber\\[2ex]
    &\quad-G_q^2G_\pi G_\sigma \bar h_\pi^2 \bar h_\sigma^2 q_f^2\Big)\bigg]\,.\label{}
\end{align}
The $(V-A)$ channel reads
\begin{align}
    &\quad \left({\mathrm{Flow}}^{(V-A)}_{\bar q q \bar q q}\right)_{\text{Box-meson}}\nonumber\\[2ex]
    &=-T\sum_{n}\int\frac{d^{3}q}{(2\pi)^3}\tilde{\partial}_t\bigg[\frac{1}{48}\Big(G_q^2 G_\pi^2 \bar h_\pi^4 q_f^2\nonumber\\[2ex]
    &\quad-G_q^2 G_\pi G_\sigma \bar h_\pi^2 \bar h_\sigma^2 q_f^2\Big)\bigg]\,.\label{}
\end{align}
The $(V+A)$ channel reads
\begin{align}
    &\quad \left({\mathrm{Flow}}^{(V+A)}_{\bar q q \bar q q}\right)_{\text{Box-meson}}\nonumber\\[2ex]
    &=T\sum_{n}\int\frac{d^{3}q}{(2\pi)^3}\tilde{\partial}_t\bigg[\frac{1}{16}\Big(G_q^2G_\pi^2 \bar h_\pi^4 q_f^2\nonumber\\[2ex]
    &\quad-G_q^2G_\pi G_\sigma \bar h_\pi^2 \bar h_\sigma^2 q_f^2\Big)\bigg]\,.\label{}
\end{align}
The $(V-A)^{\text{adj}}$ channel reads
\begin{align}
    &\quad \left({\mathrm{Flow}}^{(V-A)^{\text{adj}}}_{\bar q q \bar q q}\right)_{\text{Box-meson}}\nonumber\\[2ex]
    &=T\sum_{n}\int\frac{d^{3}q}{(2\pi)^3}\tilde{\partial}_t\bigg[\frac{1}{4}\Big(G_q^2G_\pi^2 \bar h_\pi^4q_f^2+G_q^2G_\pi G_\sigma \bar h_\pi^2 \bar h_\sigma^2 q_f^2\Big)\bigg]\,.\label{}
\end{align}
The $(S-P)^{\text{adj}}_-$ channel reads
\begin{align}
    &\quad \left({\mathrm{Flow}}^{(S-P)^{\text{adj}}_-}_{\bar q q \bar q q}\right)_{\text{Box-meson}}\nonumber\\[2ex]
    &=-T\sum_{n}\int\frac{d^{3}q}{(2\pi)^3}\tilde{\partial}_t\bigg[\frac{1}{4}\Big(G_q^2 G_\pi^2 \bar h_\pi^4q_f^2\nonumber\\[2ex]
    &\quad-G_q^2G_\pi G_\sigma \bar h_\pi^2 \bar h_\sigma^2 q_f^2\Big)\bigg]\,.\label{}
\end{align}
The other two channels are vanishing, i.e.,
\begin{align}
    &\quad \left({\mathrm{Flow}}^{(S+P)^{\text{adj}}_+}_{\bar q q \bar q q}\right)_{\text{Box-meson}}\nonumber\\[2ex]
    &=\left({\mathrm{Flow}}^{(S+P)^{\text{adj}}_-}_{\bar q q \bar q q}\right)_{\text{Box-meson}}=0\,.\label{}
\end{align}

We proceed with the ``Fish'' diagram, whose flows can be generally expressed as the form: 
\begin{align}
    &\quad \left(\mathrm{Flow}^{(\alpha)}_{\bar{q}q\bar{q}q}\right)_{\text{Fish}}\nonumber\\[2ex]
    &=-\tilde{\partial}_t\bigg[T\sum_{n}\int\frac{d^{3}q}{(2\pi)^3}G_{q}^2 \,L^{(\alpha)}\bigg]\,.\label{}
\end{align}
\begin{widetext}

The $\sigma$ channel reads
\begin{align}
    &\quad L^{(\sigma)}\nonumber\\[2ex]
    &= -\frac{1}{162} \bigg[ q_f^2 \Big(-18 \leta^2 + 52 \lSm^2 - 39 \lSp^2 - 206 \lSpp^2 + 266 \lSp \lSpp \nonumber\\[2ex]
    &\quad+ 192 \lSp \lVA - 384 \lSpp \lVA + 66 \lSm \lVAadj - 128 \lSp \lVAadj  \nonumber\\[2ex]
    &\quad+ 72 \leta \lSm + 42 \leta \lSp - 171 \leta \lSpp - 108 \leta \lVA - 72 \leta \lVAadj \nonumber\\[2ex]
    &\quad- 108 \leta \lVpA - 18 \la \leta + 24 \la \lSm + 24 \la \lSp - 69 \la \lSpp - 32 \lSpp \lVAadj \nonumber\\[2ex]
    &\quad - 360 \la \lVA - 66 \la \lVAadj + 72 \la \lVpA + 192 \lSp \lVpA - 384 \lSpp \lVpA \Big) \nonumber\\[2ex]
    &\quad +\bar m_q^2 \Big(-216 \la^2 - 630 \leta^2 - 352 \lSm^2 - 345 \lSp^2 - 178 \lSpp^2 \nonumber\\[2ex]
    &\quad- 120 \lSm \lSp - 24 \lSm \lSpp + 118 \lSp \lSpp - 192 \lSp \lVA  \nonumber\\[2ex]
    &\quad- 2880 \lVA^2 - 66 \lSm \lVAadj + 128 \lSp \lVAadj + 32 \lSpp \lVAadj  \nonumber\\[2ex]
    &\quad- 572 \lVAadj^2 - 72 \leta \lSm - 42 \leta \lSp + 171 \leta \lSpp + 108 \leta \lVA \nonumber\\[2ex]
    &\quad+ 72 \leta \lVAadj + 396 \leta \lVpA + 18 \la \leta - 24 \la \lSm - 24 \la \lSp + 384 \lSpp \lVA \nonumber\\[2ex]
    &\quad+ 69 \la \lSpp + 360 \la \lVA + 66 \la \lVAadj - 72 \la \lVpA - 384 \lSm \lVpA \nonumber\\[2ex]
    &\quad- 576 \lSp \lVpA + 1488 \lSpp \lVpA - 576 \lVpA^2- 1056 \lVA \lVAadj \Big)\bigg]\,.\label{}
\end{align}
The $\pi$ channel reads
\begin{align}
    &\quad L^{(\pi)}\nonumber\\[2ex]
    &= \frac{1}{162} \bigg[ q_f^2 \Big(18 \leta^2 + 20 \lSm^2 - 39 \lSp^2 + 10 \lSp \lSpp + 50 \lSpp^2- 72 \leta \lSm \nonumber\\[2ex]
    &\quad+ 192 \lSp \lVA - 30 \lSm \lVAadj - 128 \lSp \lVAadj + 96 \lSpp \lVAadj  \nonumber\\[2ex]
    &\quad- 6 \leta \lSp - 27 \leta \lSpp + 288 \leta \lVA + 24 \leta \lVAadj - 144 \leta \lVpA \nonumber\\[2ex]
    &\quad+ 192 \lSp \lVpA + 18 \la \leta - 24 \la \lSm - 24 \la \lSp - 21 \la \lSpp \nonumber\\[2ex]
    &\quad+ 36 \la \lVA + 30 \la \lVAadj + 36 \la \lVpA \Big) \nonumber\\[2ex]
    &\quad+ \bar m_q^2 \Big(-18 \leta^2 + 256 \lSm^2 - 153 \lSp^2 + 142 \lSpp^2 - 120 \lSm \lSp + 128 \lSp \lVAadj \nonumber\\[2ex]
    &\quad - 24 \lSm \lSpp - 10 \lSp \lSpp - 192 \lSp \lVA + 30 \lSm \lVAadj  \nonumber\\[2ex]
    &\quad - 96 \lSpp \lVAadj + 72 \leta \lSm + 6 \leta \lSp + 27 \leta \lSpp - 288 \leta \lVA \nonumber\\[2ex]
    &\quad - 24 \leta \lVAadj - 144 \leta \lVpA + 192 \lSp \lVpA + 414 \la \leta + 24 \la \lSm - 288 \lVpA^2 \nonumber\\[2ex]
    &\quad + 24 \la \lSp + 21 \la \lSpp - 36 \la \lVA - 30 \la \lVAadj - 36 \la \lVpA \nonumber\\[2ex]
    &\quad + 288 \lVA^2 + 480 \lVA \lVAadj - 124 \lVAadj^2 + 384 \lSm \lVpA + 336 \lSpp \lVpA  \Big) \bigg]\,.\label{}
\end{align}
The $\eta$ channel reads
\begin{align}
    &\quad L^{(\eta)}\nonumber\\[2ex]
    &= \frac{1}{162} \bigg[ q_f^2 \Big(-1386 \leta^2 + 20 \lSm^2 - 39 \lSp^2 + 10 \lSp \lSpp + 50 \lSpp^2 \nonumber\\[2ex]
    &\quad- 192 \lSp \lVA - 30 \lSm \lVAadj + 128 \lSp \lVAadj - 96 \lSpp \lVAadj  \nonumber\\[2ex]
    &\quad - 24 \la \lSm - 24 \la \lSp - 21 \la \lSpp - 36 \la \lVA - 66 \la \lVAadj \nonumber\\[2ex]
    &\quad - 36 \la \lVpA - 192 \lSp \lVpA - 72 \leta \lSm + 570 \leta \lSp - 27 \leta \lSpp \nonumber\\[2ex]
    &\quad - 72 \leta \lVA + 120 \leta \lVAadj + 792 \leta \lVpA + 126 \la \leta \Big) \nonumber\\[2ex]
    &\quad+ \bar m_q^2 \Big(4194 \leta^2 + 256 \lSm^2 + 231 \lSp^2 - 242 \lSpp^2 - 120 \lSm \lSp \nonumber\\[2ex]
    &\quad - 24 \lSm \lSpp - 10 \lSp \lSpp + 192 \lSp \lVA + 30 \lSm \lVAadj \nonumber\\[2ex]
    &\quad + 96 \lSpp \lVAadj - 342 \la \leta + 24 \la \lSm + 24 \la \lSp + 21 \la \lSpp \nonumber\\[2ex]
    &\quad + 36 \la \lVA + 66 \la \lVAadj + 36 \la \lVpA + 576 \lSp \lVpA + 336 \lSpp \lVpA \nonumber\\[2ex]
    &\quad + 72 \leta \lSm - 1722 \leta \lSp + 27 \leta \lSpp + 72 \leta \lVA - 120 \leta \lVAadj \nonumber\\[2ex]
    &\quad - 2808 \leta \lVpA - 288 \lVA^2 - 1056 \lVA \lVAadj + 4 \lVAadj^2 + 384 \lSm \lVpA \nonumber\\[2ex]
    &\quad + 288 \lVpA^2  - 128 \lSp \lVAadj\Big) \bigg]\,.\label{}
\end{align}
The $a$ channel reads
\begin{align}
    &\quad L^{(a)}\nonumber\\[2ex]
    &= \frac{1}{162} \bigg[ q_f^2 \Big(1188 \la^2 + 18 \leta^2 - 52 \lSm^2 + 39 \lSp^2 - 266 \lSp \lSpp- 128 \lSp \lVAadj \nonumber\\[2ex]
    &\quad+ 206 \lSpp^2 + 192 \lSp \lVA - 384 \lSpp \lVA - 66 \lSm \lVAadj- 108 \leta \lVpA  \nonumber\\[2ex]
    &\quad - 32 \lSpp \lVAadj - 306 \la \leta - 24 \la \lSm + 552 \la \lSp - 1083 \la \lSpp \nonumber\\[2ex]
    &\quad - 144 \la \lVA - 30 \la \lVAadj + 720 \la \lVpA + 192 \lSp \lVpA - 384 \lSpp \lVpA \nonumber\\[2ex]
    &\quad - 72 \leta \lSm - 42 \leta \lSp + 171 \leta \lSpp - 108 \leta \lVA - 216 \leta \lVAadj  \Big) \nonumber\\[2ex]
    &\quad+\bar m_q^2 \Big(1188 \la^2 - 18 \leta^2 + 352 \lSm^2 - 423 \lSp^2 - 590 \lSpp^2 \nonumber\\[2ex]
    &\quad + 120 \lSm \lSp + 24 \lSm \lSpp + 650 \lSp \lSpp - 192 \lSp \lVA  \nonumber\\[2ex]
    &\quad - 1152 \lVA^2 + 66 \lSm \lVAadj + 128 \lSp \lVAadj + 32 \lSpp \lVAadj  \nonumber\\[2ex]
    &\quad - 68 \lVAadj^2 - 342 \la \leta + 24 \la \lSm + 600 \la \lSp - 1221 \la \lSpp \nonumber\\[2ex]
    &\quad + 144 \la \lVA + 30 \la \lVAadj + 1008 \la \lVpA + 192 \lSp \lVpA - 720 \lSpp \lVpA \nonumber\\[2ex]
    &\quad + 72 \leta \lSm + 42 \leta \lSp - 171 \leta \lSpp + 108 \leta \lVA + 216 \leta \lVAadj \nonumber\\[2ex]
    &\quad - 180 \leta \lVpA + 384 \lSm \lVpA + 1152 \lVpA^2+ 384 \lSpp \lVA - 480 \lVA \lVAadj\Big) \bigg]\,.\label{}
\end{align}
The $(V-A)$ channel reads
\begin{align}
    &\quad L^{(V-A)}\nonumber\\[2ex]
    &= \frac{1}{162} \bigg[ q_f^2 \Big(-9 \leta^2 + 24 \lSm^2 - 64 \lSp^2 - 24 \lSpp^2 + 64 \lSp \lSpp \nonumber\\[2ex]
    &\quad+ 540 \lVA^2 + 288 \lVA \lVAadj - 72 \lVAadj^2 - 24 \leta \lSm - 72 \leta \lSp \nonumber\\[2ex]
    &\quad+ 24 \leta \lSpp - 162 \leta \lVpA + 288 \lSm \lVpA + 648 \lVpA^2 + 9 \la \leta \nonumber\\[2ex]
    &\quad+ 24 \la \lSm + 24 \la \lSp - 24 \la \lSpp + 54 \la \lVpA \Big) \nonumber\\[2ex]
    &\quad+\bar m_q^2 \Big(9 \leta^2 - 24 \lSm^2 + 64 \lSp^2 + 24 \lSpp^2 - 64 \lSp \lSpp + 216 \la \lVA \nonumber\\[2ex]
    &\quad- 540 \lVA^2 - 288 \lVA \lVAadj + 72 \lVAadj^2 + 24 \leta \lSm + 72 \leta \lSp \nonumber\\[2ex]
    &\quad- 24 \leta \lSpp - 504 \leta \lVA - 384 \leta \lVAadj + 162 \leta \lVpA + 1536 \lSm \lVA \nonumber\\[2ex]
    &\quad+ 320 \lSm \lVAadj - 288 \lSm \lVpA + 384 \lSp \lVA - 256 \lSp \lVAadj  \nonumber\\[2ex]
    &\quad+ 64 \lSpp \lVAadj + 6048 \lVA \lVpA + 1152 \lVAadj \lVpA - 648 \lVpA^2 - 9 \la \leta \nonumber\\[2ex]
    &\quad - 24 \la \lSm - 24 \la \lSp + 24 \la \lSpp  - 54 \la \lVpA - 384 \lSpp \lVA \Big) \bigg]\,.\label{}
\end{align}
The $(V+A)$ channel reads
\begin{align}
    &\quad L^{(V+A)}\nonumber\\[2ex]
    &= \frac{1}{54} \bigg[ q_f^2 \Big(9 \leta^2 + 36 \lSm^2 + 25 \lSp^2 + 10 \lSp \lSpp + 50 \lSpp^2 \nonumber\\[2ex]
    &\quad+ 96 \lSm \lVA - 30 \leta \lSp - 27 \leta \lSpp - 54 \leta \lVA + 2 \lSm \lVAadj \nonumber\\[2ex]
    &\quad+ 9 \la \leta - 21 \la \lSpp + 18 \la \lVA + 6 \la \lVAadj + 504 \lVA \lVpA \nonumber\\[2ex]
    &\quad+ 96 \lVAadj \lVpA + 108 \lVpA^2 \Big) \nonumber\\[2ex]
    &\quad+\bar m_q^2 \Big(-9 \leta^2 + 48 \lSm^2 - 25 \lSp^2 - 50 \lSpp^2 - 120 \lSm \lSp+ 1152 \lVA^2 \nonumber\\[2ex]
    &\quad- 24 \lSm \lSpp - 10 \lSp \lSpp - 96 \lSm \lVA  - 2 \lSm \lVAadj \nonumber\\[2ex]
    &\quad+ 480 \lVA \lVAadj + 68 \lVAadj^2 + 30 \leta \lSp + 27 \leta \lSpp + 54 \leta \lVA \nonumber\\[2ex]
    &\quad- 360 \leta \lVpA - 9 \la \leta + 21 \la \lSpp - 18 \la \lVA - 6 \la \lVAadj + 72 \la \lVpA \nonumber\\[2ex]
    &\quad+ 384 \lSm \lVpA + 336 \lSpp \lVpA - 504 \lVA \lVpA - 96 \lVAadj \lVpA + 756 \lVpA^2 \Big) \bigg]\,.\label{}
\end{align}
The $(V-A)^{\text{adj}}$ channel reads
\begin{align}
    &\quad L^{(V-A)^{\text{adj}}}\nonumber\\[2ex]
    &= -\frac{1}{27} \bigg[ q_f^2 \Big(18 \leta^2 + 21 \lSm^2 + 32 \lSp^2 + 6 \lSpp^2 - 32 \lSp \lSpp \nonumber\\[2ex]
    &\quad - 72 \lVA \lVAadj + 57 \lVAadj^2 - 18 \la \leta - 12 \la \lSm - 12 \la \lSp \nonumber\\[2ex]
    &\quad+ 3 \la \lSpp - 6 \leta \lSm + 36 \leta \lSp - 21 \leta \lSpp \Big) \nonumber\\[2ex]
    &\quad+ \bar m_q^2 \Big(-18 \leta^2 - 21 \lSm^2 - 32 \lSp^2 - 6 \lSpp^2 + 32 \lSp \lSpp \nonumber\\[2ex]
    &\quad + 72 \lVA \lVAadj - 57 \lVAadj^2 + 18 \la \leta + 12 \la \lSm + 12 \la \lSp \nonumber\\[2ex]
    &\quad - 3 \la \lSpp + 6 \leta \lSm - 36 \leta \lSp + 21 \leta \lSpp - 192 \lSm \lVA \nonumber\\[2ex]
    &\quad - 192 \lSp \lVA + 48 \lSpp \lVA - 184 \lSm \lVAadj  - 288 \leta \lVA \nonumber\\[2ex]
    &\quad + 128 \lSp \lVAadj + 12 \leta \lVAadj + 36 \la \lVAadj- 80 \lSpp \lVAadj \Big) \bigg]\,.\label{}
\end{align}
The $(S-P)_{-}^{\text{adj}}$ channel reads
\begin{align}
    &\quad L^{(S-P)_{-}^{\text{adj}}}\nonumber\\[2ex]
    &= -\frac{1}{27} \bigg[ q_f^2 \Big(-18 \leta^2 - 20 \lSm^2 - 14 \lSpp^2 - 70 \lSp \lSpp + 36 \lSm \lVA \nonumber\\[2ex]
    &\quad+ 42 \lSm \lVAadj - 9 \leta \lSm + 15 \leta \lSp + 45 \leta \lSpp - 9 \leta \lVAadj \nonumber\\[2ex]
    &\quad - 18 \la \leta - 3 \la \lSm + 15 \la \lSp + 3 \la \lSpp - 9 \la \lVAadj \nonumber\\[2ex]
    &\quad+ 108 \lSm \lVpA \Big) \nonumber\\[2ex]
    &\quad+ \bar m_q^2 \Big(18 \leta^2 - 16 \lSm^2 + 14 \lSpp^2 + 168 \lSm \lSpp + 70 \lSp \lSpp \nonumber\\[2ex]
    &\quad - 36 \lSm \lVA - 42 \lSm \lVAadj - 144 \lVA \lVAadj - 84 \lVAadj^2 + 18 \la \leta \nonumber\\[2ex]
    &\quad+ 39 \la \lSm - 15 \la \lSp - 3 \la \lSpp + 9 \la \lVAadj - 60 \lSm \lVpA \nonumber\\[2ex]
    &\quad - 240 \lSp \lVpA - 48 \lSpp \lVpA + 117 \leta \lSm - 15 \leta \lSp - 45 \leta \lSpp \nonumber\\[2ex]
    &\quad + 9 \leta \lVAadj + 288 \leta \lVpA \Big) \bigg]\,.\label{}
\end{align}
The $(S+P)_{-}^{\text{adj}}$ channel reads
\begin{align}
    &\quad L^{(S+P)_{-}^{\text{adj}}}\nonumber\\[2ex]
    &= \frac{2}{27} \bigg[ q_f^2 \Big(-42 \lSm \lSpp - 48 \lSp \lVA + 12 \lSpp \lVA + 9 \la \lSm - 9 \la \lVAadj \nonumber\\[2ex]
    &\quad + 32 \lSp \lVAadj - 20 \lSpp \lVAadj + 9 \leta \lSm - 36 \leta \lVA + 15 \leta \lVAadj \nonumber\\[2ex]
    &\quad - 36 \leta \lVpA + 60 \lSp \lVpA + 12 \lSpp \lVpA \Big) \nonumber\\[2ex]
    &\quad+ \bar m_q^2 \Big(-18 \lSp^2 + 24 \lSpp^2 + 42 \lSm \lSpp + 48 \lSp \lSpp + 48 \lSp \lVA \nonumber\\[2ex]
    &\quad - 12 \lSpp \lVA - 9 \la \lSm - 54 \la \lSp + 9 \la \lVAadj - 32 \lSp \lVAadj \nonumber\\[2ex]
    &\quad + 20 \lSpp \lVAadj - 9 \leta \lSm + 54 \leta \lSp - 72 \leta \lSpp + 36 \leta \lVA \nonumber\\[2ex]
    &\quad - 15 \leta \lVAadj + 36 \leta \lVpA - 60 \lSp \lVpA - 12 \lSpp \lVpA - 144 \lVA \lVAadj \nonumber\\[2ex]
    &\quad+ 24 \lVAadj^2 - 144 \lSm \lVpA \Big) \bigg]\,.\label{}
\end{align}
The $(S+P)_{+}^{\text{adj}}$ channel reads
\begin{align}
    &\quad L^{(S+P)_{+}^{\text{adj}}}\nonumber\\[2ex]
    &= -\frac{4}{27} \bigg[ q_f^2 \Big(-30 \lSm \lSp - 6 \lSm \lSpp - 18 \la \lVA - 24 \lSpp \lVA - 6 \la \lVAadj \nonumber\\[2ex]
    &\quad - 8 \lSpp \lVAadj - 18 \la \lVpA + 84 \lSpp \lVpA + 18 \leta \lSm - 18 \leta \lVA \nonumber\\[2ex]
    &\quad - 6 \leta \lVAadj - 18 \leta \lVpA \Big) \nonumber\\[2ex]
    &\quad +\bar m_q^2 \Big(36 \lSm^2 - 18 \lSp^2 - 24 \lSpp^2 + 30 \lSm \lSp + 6 \lSm \lSpp \nonumber\\[2ex]
    &\quad+ 96 \lSp \lSpp - 54 \la \lSpp + 18 \la \lVA + 24 \lSpp \lVA + 6 \la \lVAadj \nonumber\\[2ex]
    &\quad+ 8 \lSpp \lVAadj + 18 \la \lVpA - 84 \lSpp \lVpA - 18 \leta \lSm - 90 \leta \lSpp \nonumber\\[2ex]
    &\quad+ 18 \leta \lVA + 6 \leta \lVAadj + 18 \leta \lVpA - 144 \lVA^2 - 96 \lVA \lVAadj \nonumber\\[2ex]
    &\quad - 16 \lVAadj^2 + 144 \lVpA^2 \Big) \bigg]\,.\label{}
\end{align}

The flows of two triangle diagrams, ``Triangle-gluon'', ``Triangle-meson'', can combined together, denoted collectively by ``Triangle''. The $\sigma$ channel reads
\begin{align}
    &\quad \left({\mathrm{Flow}}^{(\sigma)}_{\bar q q \bar q q}\right)_{\text{Triangle}}\nonumber\\[2ex]
    &= T\sum_n \int \frac{d^3 q}{(2\pi)^3} \tilde{\partial}_t \bigg[\frac{1}{72} G_q^2 G_\sigma \bar h_\sigma^2 q_f^2 \big( -8 \lSm + 8 \lSp - 23 \lSpp - 48 \lVA - 10 \lVAadj - 48 \lVpA \big) \nonumber\\[2ex]
    &\quad + \frac{1}{72} G_\pi G_q^2 \bar h_\pi^2 \Big( -24 m_q^2 \lSm + q_f^2 \big( -24 \lSm + 14 \lSp - 57 \lSpp - 36 \lVA - 72 \lVAadj \nonumber\\[2ex]
    &\quad - 36 \lVpA \big) \Big)  - \frac{1}{108} G_A G_q^2 \bar{g}_{qqA}^2 \Big( 3 m_q^2 \big( 75 \la + 8 \lSm - 64 \lSp + 121 \lSpp - 78 \lVA - 98 \lVAadj \nonumber\\[2ex]
    &\quad + 114 \lVpA \big)  - 4 q_f^2 \big( 72 \la - 24 \leta - 32 \lSp + 136 \lSpp + 81 \lVpA \big) \Big) \bigg]\,.\label{}
\end{align}
The $\pi$ channel reads
\begin{align}
    &\quad \left({\mathrm{Flow}}^{(\pi)}_{\bar q q \bar q q}\right)_{\text{Triangle}}\nonumber\\[2ex]
    &= T\sum_n \int \frac{d^3 q}{(2\pi)^3} \tilde{\partial}_t \bigg[-\frac{1}{72} G_q^2 G_\sigma \bar h_\sigma^2 q_f^2 \big( -8 \lSm + 8 \lSp + 7 \lSpp - 12 \lVA - 22 \lVAadj - 12 \lVpA \big) \nonumber\\[2ex]
    &\quad -\frac{1}{72} G_\pi G_q^2 \bar h_\pi^2 \Big( 24 m_q^2 \lSm + q_f^2 \big( 24 \lSm - 2 \lSp - 9 \lSpp + 24 \lVA - 40 \lVAadj + 24 \lVpA \big) \Big) \nonumber\\[2ex]
    &\quad + \frac{1}{108} G_A G_q^2 \bar{g}_{\bar{q}qA}^2 \Big( 3 m_q^2 \big( 21 \la + 8 \lSm - 64 \lSp + 7 \lSpp - 114 \lVA - 14 \lVAadj + 78 \lVpA \big) \nonumber\\[2ex]
    &\quad + 4 q_f^2 \big( 24 \la + 24 \leta - 32 \lSp - 24 \lSpp + 81 \lVpA \big) \Big) \bigg]\,.\label{}
\end{align}
The $\eta$ channel reads
\begin{align}
    &\quad \left({\mathrm{Flow}}^{(\eta)}_{\bar q q \bar q q}\right)_{\text{Triangle}}\nonumber\\[2ex]
    &= T\sum_n \int \frac{d^3 q}{(2\pi)^3} \tilde{\partial}_t \bigg[\frac{1}{72} G_q^2 G_\sigma \bar h_\sigma^2 \Big( 108 m_q^2 \leta + q_f^2 \big( 36 \leta + 8 \lSm - 8 \lSp + 23 \lSpp \nonumber\\[2ex]
    &\quad - 120 \lVA - 22 \lVAadj + 24 \lVpA \big) \Big) \nonumber\\[2ex]
    &\quad - \frac{1}{72} G_\pi G_q^2 \bar h_\pi^2 \Big( 12 m_q^2 \big( 18 \la - 9 \leta + 2 \lSm \big) - q_f^2 \big( 108 \leta + 24 \lSm - 14 \lSp + 57 \lSpp\nonumber\\[2ex]
    &\quad - 36 \lVA - 24 \lVAadj  - 36 \lVpA \big) \Big) \nonumber\\[2ex]
    &\quad + \frac{1}{108} G_A G_q^2 \bar{g}_{\bar{q}qA}^2 \Big( 3 m_q^2 \big( 75 \la + 288 \leta + 8 \lSm - 64 \lSp + 121 \lSpp + 114 \lVA \nonumber\\[2ex]
    &\quad + 62 \lVAadj - 78 \lVpA \big) + 4 q_f^2 \big( 72 \la + 192 \leta - 32 \lSp + 136 \lSpp - 81 \lVpA \big) \Big) \bigg]\,.\label{}
\end{align}
The $a$ channel reads
\begin{align}
    &\quad \left({\mathrm{Flow}}^{(a)}_{\bar q q \bar q q}\right)_{\text{Triangle}}\nonumber\\[2ex]
    &= T\sum_n \int \frac{d^3 q}{(2\pi)^3} \tilde{\partial}_t \bigg[\frac{1}{72} G_q^2 G_\sigma \bar h_\sigma^2 \Big( 108 m_q^2 \la - q_f^2 \big( 36 \la + 8 \lSm - 8 \lSp - 7 \lSpp \nonumber\\[2ex]
    &\quad - 12 \lVA - 10 \lVAadj - 12 \lVpA \big) \Big) \nonumber\\[2ex]
    &\quad + \frac{1}{72} G_\pi G_q^2 \bar h_\pi^2 \Big( 12 m_q^2 \big( 3 \la + 6 \leta + 2 \lSm \big) - q_f^2 \big( 36 \la - 24 \lSm + 2 \lSp + 9 \lSpp \nonumber\\[2ex]
    &\quad+ 96 \lVA + 8 \lVAadj - 48 \lVpA \big) \Big) \nonumber\\[2ex]
    &\quad - \frac{1}{108} G_A G_q^2 \bar{g}_{\bar{q}qA}^2 \Big( 3 m_q^2 \big( 309 \la + 8 \lSm - 64 \lSp + 7 \lSpp + 78 \lVA - 46 \lVAadj - 114 \lVpA \big) \nonumber\\[2ex]
    &\quad - 4 q_f^2 \big( 240 \la + 24 \leta - 32 \lSp - 24 \lSpp - 81 \lVpA \big) \Big) \bigg]\,.\label{}
\end{align}
The $(V-A)$ channel reads
\begin{align}
    &\quad \left({\mathrm{Flow}}^{(V-A)}_{\bar q q \bar q q}\right)_{\text{Triangle}}\nonumber\\[2ex]
    &= T\sum_n \int \frac{d^3 q}{(2\pi)^3} \tilde{\partial}_t \bigg[\frac{1}{72} G_q^2 G_\sigma \bar h_\sigma^2 \Big( 108 m_q^2 \lVA - q_f^2 \big( 9 \leta - 8 \lSm + 8 \lSp - 8 \lSpp - 18 \lVpA \big) \Big) \nonumber\\[2ex]
    &\quad + \frac{1}{72} G_\pi G_q^2 \bar h_\pi^2 \Big( 108 m_q^2 \lVA - q_f^2 \big( 27 \la - 8 \lSm + 24 \lSp - 8 \lSpp - 54 \lVpA \big) \Big) \nonumber\\[2ex]
    &\quad - \frac{2}{9} G_A G_q^2 \bar{g}_{\bar{q}qA}^2 \Big( m_q^2 \big( 6 \la - 5 \lSm + 4 \lSp - \lSpp - 18 \lVpA \big) + 6 q_f^2 \lVAadj \Big) \bigg]\,.\label{}
\end{align}
The $(V+A)$ channel reads
\begin{align}
    &\quad \left({\mathrm{Flow}}^{(V+A)}_{\bar q q \bar q q}\right)_{\text{Triangle}}\nonumber\\[2ex]
    &= T\sum_n \int \frac{d^3 q}{(2\pi)^3} \tilde{\partial}_t \bigg[\frac{1}{24} G_q^2 G_\sigma \bar h_\sigma^2 \Big( 36 m_q^2 \lVpA + q_f^2 \big( 3 \leta + 7 \lSpp + 6 \lVA + 2 \lVAadj \big) \Big) \nonumber\\[2ex]
    &\quad + \frac{1}{24} G_\pi G_q^2 \bar h_\pi^2 \Big( 12 m_q^2 \big( 2 \lSm + 3 \lVpA \big) + q_f^2 \big( 9 \la - 10 \lSp - 9 \lSpp + 18 \lVA \big) \Big) \nonumber\\[2ex]
    &\quad + \frac{1}{12} G_A G_q^2 \bar{g}_{\bar{q}qA}^2 \Big( m_q^2 \big( 3 \la - 7 \lSpp + 66 \lVA + 22 \lVAadj + 18 \lVpA \big) - 12 q_f^2 \lVpA \Big) \bigg]\,.\label{}
\end{align}
The $(V-A)^{\text{adj}}$ channel reads
\begin{align}
    &\quad \left({\mathrm{Flow}}^{(V-A)^{\text{adj}}}_{\bar q q \bar q q}\right)_{\text{Triangle}}\nonumber\\[2ex]
    &= T\sum_n \int \frac{d^3 q}{(2\pi)^3} \tilde{\partial}_t \bigg[\frac{1}{12} G_q^2 G_\sigma \bar h_\sigma^2 \Big( 18 m_q^2 \lVAadj - q_f^2 \big( 4 \lSm - 4 \lSp + \lSpp \big) \Big) \nonumber\\[2ex]
    &\quad + \frac{1}{12} G_\pi G_q^2 \bar h_\pi^2 \Big( 18 m_q^2 \lVAadj - q_f^2 \big( -2 \lSm - 12 \lSp + 7 \lSpp \big) \Big) \nonumber\\[2ex]
    &\quad + \frac{1}{6} G_A G_q^2 \bar{g}_{\bar{q}qA}^2 \Big( m_q^2 \big( 15 \la - 20 \lSm + 16 \lSp - 10 \lSpp \big) - 12 q_f^2 \big( 3 \lVA - \lVAadj \big) \Big) \bigg]\,.\label{}
\end{align}
The $(S+P)^{\text{adj}}_{+}$ channel reads
\begin{align}
    &\quad \left({\mathrm{Flow}}^{(S+P)^{\text{adj}}_{+}}_{\bar q q \bar q q}\right)_{\text{Triangle}}\nonumber\\[2ex]
    &= T\sum_n \int \frac{d^3 q}{(2\pi)^3} \tilde{\partial}_t \bigg[\frac{1}{6} G_q^2 G_\sigma \bar h_\sigma^2 \Big( 9 m_q^2 \lSpp + 2 q_f^2 \big( 3 \lVA + \lVAadj + 3 \lVpA \big) \Big) \nonumber\\[2ex]
    &\quad - \frac{1}{6} G_\pi G_q^2 \bar h_\pi^2 \Big( 15 m_q^2 \lSpp - 2 q_f^2 \big( 3 \lSm - 3 \lVA - \lVAadj - 3 \lVpA \big) \Big) \nonumber\\[2ex]
    &\quad - \frac{1}{9} G_A G_q^2 \bar{g}_{\bar{q}qA}^2 \Big( 6 m_q^2 \big( 6 \lVA + 2 \lVAadj + 3 \lVpA \big) + q_f^2 \big( 3 \la - 3 \leta + 46 \lSpp \big) \Big) \bigg]\,.\label{}
\end{align}
The $(S-P)^{\text{adj}}_{-}$ channel reads
\begin{align}
    &\quad \left({\mathrm{Flow}}^{(S-P)^{\text{adj}}_{-}}_{\bar q q \bar q q}\right)_{\text{Triangle}}\nonumber\\[2ex]
    &= T\sum_n \int \frac{d^3 q}{(2\pi)^3} \tilde{\partial}_t \bigg[\frac{1}{12} G_q^2 G_\sigma \bar h_\sigma^2 \Big( 18 m_q^2 \lSm - q_f^2 \big( \lSm + 5 \lSp + \lSpp + 3 \lVAadj \big) \Big) \nonumber\\[2ex]
    &\quad + \frac{1}{12} G_\pi G_q^2 \bar h_\pi^2 \Big( -30 m_q^2 \lSm + q_f^2 \big( 3 \lSm + 5 \lSp + 15 \lSpp + 3 \lVAadj \big) \Big) \nonumber\\[2ex]
    &\quad - \frac{1}{6} G_A G_q^2 \bar{g}_{\bar{q}qA}^2 \Big( m_q^2 \big( 6 \la + 10 \lSm - 5 \lSp - \lSpp + 18 \lVA + 21 \lVAadj + 18 \lVpA \big)\nonumber\\[2ex]
    &\quad + 6 q_f^2 \lSm \Big) \bigg]\,.\label{}
\end{align}
The $(S+P)^{\text{adj}}_{-}$ channel reads
\begin{align}
    &\quad \left({\mathrm{Flow}}^{(S+P)^{\text{adj}}_{-}}_{\bar q q \bar q q}\right)_{\text{Triangle}}\nonumber\\[2ex]
    &= T\sum_n \int \frac{d^3 q}{(2\pi)^3} \tilde{\partial}_t \bigg[\frac{1}{2} G_q^2 G_\sigma \bar h_\sigma^2 \Big( 3 m_q^2 \lSp - q_f^2 \big( \lSm - \lVAadj \big) \Big) \nonumber\\[2ex]
    &\quad + \frac{1}{6} G_\pi G_q^2 \bar h_\pi^2 \Big( 3 m_q^2 \big( 3 \lSp - 4 \lSpp \big) + q_f^2 \big( 3 \lSm - 12 \lVA + 5 \lVAadj - 12 \lVpA \big) \Big) \nonumber\\[2ex]
    &\quad + \frac{1}{9} G_A G_q^2 \bar{g}_{\bar{q}qA}^2 \Big( 9 m_q^2 \big( \lSm - 6 \lVA + 2 \lVAadj \big) + q_f^2 \big( 21 \la + 9 \leta - 2 \lSp - 22 \lSpp \big) \Big) \bigg]\,.\label{}
\end{align}

\end{widetext}

\vfill 

\bibliography{ref-lib}%

\end{document}